%% file: mobile_guards.tex
\documentclass[letterpaper,11pt]{article}

\usepackage{fancyhdr}
\usepackage{xspace}
\usepackage{graphicx}
\usepackage{subfigure}
\usepackage{psfrag}
\usepackage{amsmath,amssymb,amsthm}
\usepackage{fullpage}
\usepackage[usenames]{color}
\usepackage{multirow,tabularx,array,ctable}

\usepackage[unicode,final,colorlinks=true,bookmarksopen=true]{hyperref} 
\hypersetup{linkcolor=red,urlcolor=blue,citecolor=blue}

\input{preamble}

\newcommand{\sep}{,\xspace}

\begin{document}

\invalidatenew

\phantomsection
\addcontentsline{toc}{section}{Title \& Abstract}

\title{Guarding curvilinear art galleries with edge or mobile guards via
  2-dominance of triangulation graphs}

\author{Menelaos I. Karavelas$^{\dagger,\ddagger}$\\[5pt]
\it{}$^\dagger$Department of Applied Mathematics,\\
\it{}University of Crete,\\
\it{}GR-714 09 Heraklion, Greece\\
\texttt{mkaravel@tem.uoc.gr}\\[5pt]
\it{}$^\ddagger$Institute of Applied and Computational Mathematics,\\
\it{}Foundation for Research and Technology - Hellas,\\
\it{}P.O. Box 1385, GR-711 10 Heraklion, Greece}

\maketitle


\begin{abstract}
\input{abstract}

\bigskip\noindent
\textit{Key\;words:}\/
art gallery\sep curvilinear polygons\sep triangulation graphs\sep
2-dominance\sep edge guards\sep diagonal guards\sep mobile guards\sep
\pconvex polygons\sep monotone \pconvex polygons
\smallskip\par\noindent
\end{abstract}

\clearpage


\input{maintext}


\paragraph*{Acknowledgements}
%
The work in this paper has been partially supported by the IST
Programme of the EU (FET Open) Project under Contract No
\acsProjectNr\ -- (\acsAcronym\ - \acsFullTitle).

\phantomsection
\addcontentsline{toc}{section}{References}

\bibliographystyle{abbrv}
\bibliography{mobile_guards}


\clearpage

\appendix

\phantomsection
\addcontentsline{toc}{section}{Appendix}

\section*{\Large{}Appendix}

\input{appendix}

\end{document}

%% file: preamble.tex
\newtheorem{theorem}{Theorem}
\newtheorem{lemma}[theorem]{Lemma}

\newtheorem{remark}{Remark}

\newenvironment{mathdescription}%
  {\begin{list}{}%
    {\renewcommand{\makelabel}[1]%
      {\textbf{$\boldsymbol{##1}$\hfil}}}}
  {\end{list}}

\newcommand{\hide}[1]{}

\newcommand{\trg}[1]{T_{{#1}}}
\newcommand{\lseg}[1]{\overline{#1}}

\newcommand{\guard}[1][]{monitor{#1}\xspace}
\newcommand{\gset}{guard set\xspace}

\newcommand{\ie}{i.e.,\xspace}
\newcommand{\resp}{resp.,\xspace}
\newcommand{\pconvex}[1][p]{{#1}iece\-wise-convex\xspace}
\newcommand{\pconcave}[1][p]{{#1}iecewise-concave\xspace}

\newcommand{\nin}{\not\in}

\newcommand{\bm}{\lfloor\frac{n+1}{3}\rfloor}
\newcommand{\lbmg}{\lfloor\frac{n}{3}\rfloor}

\newcommand{\be}{\lfloor\frac{2n+1}{5}\rfloor}
\newcommand{\lbeg}{\lceil\frac{n}{3}\rceil}
\newcommand{\ubew}{\lfloor\frac{3n}{7}\rfloor}

\newcommand{\mbd}{\lceil\frac{n+1}{4}\rceil}

\newcommand{\Chvatal}{Chv{\'a}tal\xspace}
\newcommand{\Gyori}{Gy{\"o}ri\xspace}

\newcommand{\acsProjectNr}{IST-006413}

\newcommand{\acsAcronym}{ACS}
\newcommand{\acsTitle}{Algorithms for Complex Shapes}
\newcommand{\acsSubTitle}{with Certified Numerics and Topology}
\newcommand{\acsFullTitle}{\acsTitle\ \acsSubTitle}

\newcommand{\new}[1]{\textcolor{BlueGreen}{#1}}
\newcommand{\newbegin}{\color{BlueGreen}}
\newcommand{\newend}{\color{black}}

\newcommand{\invalidatenew}{%
  \renewcommand{\new}[1]{##1}
  \renewcommand{\newbegin}{}
  \renewcommand{\newend}{}
}

\long\def\symbolfootnote[#1]#2{\begingroup%
  \def\thefootnote{\fnsymbol{footnote}}\footnote[#1]{#2}\endgroup} 
\long\def\symbolfootnotemark[#1]{\begingroup%
  \def\thefootnote{\fnsymbol{footnote}}\footnotemark[#1]\endgroup}
\long\def\symbolfootnotetext[#1]#2{\begingroup%
  \def\thefootnote{\fnsymbol{footnote}}\footnotetext[#1]{#2}\endgroup}

%% file: abstract.tex
In this paper we consider the problem of monitoring an art gallery
modeled as a polygon, the edges of which are arcs of curves, with edge
or mobile guards. Our focus is on piecewise-convex polygons, i.e.,
polygons that are locally convex, except possibly at the vertices, and
their edges are convex arcs.

We transform the problem of monitoring a piecewise-convex polygon to the
problem of \emph{2-dominating} a properly defined triangulation graph
with edges or diagonals, where 2-dominance requires that every
triangle in the triangulation graph has at least two of its vertices
in its 2-dominating set. We show that:
(1) $\lfloor\frac{n+1}{3}\rfloor$ diagonal guards are always
sufficient and sometimes necessary, and
(2) $\lfloor\frac{2n+1}{5}\rfloor$ edge guards are always sufficient
and sometimes necessary,
in order to 2-dominate a triangulation graph.
Furthermore, we show how to compute:
(1) a diagonal 2-dominating set of size $\lfloor\frac{n+1}{3}\rfloor$
in linear time and space,
(2) an edge 2-dominating set of size $\lfloor\frac{2n+1}{5}\rfloor$ in
$O(n^2)$ time and $O(n)$ space, and
(3) an edge 2-dominating set of size $\lfloor\frac{3n}{7}\rfloor$ in
$O(n)$ time and space.

Based on the above-mentioned results, we prove that, for
piecewise-convex polygons, we can compute:
(1) a mobile guard set of size $\lfloor\frac{n+1}{3}\rfloor$ in
$O(n\log{}n)$ time,
(2) an edge guard set of size $\lfloor\frac{2n+1}{5}\rfloor$ in
$O(n^2)$ time, and
(3) an edge guard set of size $\lfloor\frac{3n}{7}\rfloor$ in
$O(n\log{}n)$ time.
All space requirements are linear.
Finally, we show that $\lfloor\frac{n}{3}\rfloor$ mobile or 
$\lceil\frac{n}{3}\rceil$ edge guards are sometimes necessary.

When restricting our attention to monotone piecewise-convex polygons,
the bounds mentioned above drop: $\lceil\frac{n+1}{4}\rceil$ edge or
mobile guards are always sufficient and sometimes necessary; such an
edge or mobile guard set, of size at most $\lceil\frac{n+1}{4}\rceil$,
can be computed in $O(n)$ time and space.

%% file: maintext.tex

\section{Introduction}
\label{sec:intro}

In recent years Computational Geometry has made a shift towards
curvilinear objects. Recent works have addressed both combinatorial
properties and algorithmic aspects of such problems, as well as the
necessary algebraic techniques required to tackle the predicates
used in the algorithms involving these objects. The pertinent
literature is quite extensive; the interested reader may consult the
recent book edited by Boissonnat and Teillaud \cite{BoiTeil:ECG:book}
for a collection of recent results for various classical Computational
Geometry problems involving curvilinear objects. Despite the apparent
shift towards the curvilinear world, and despite the vast range of
application areas for art gallery problems, including robotics 
\cite{ks-errss,xcb-pvicb-86}, motion planning
\cite{lw-apcfp-79,m-aaspt-88}, computer vision
\cite{sc-bwfmr-86,y-3damv-86,ae-caps-83,t-prgc-80}, graphics
\cite{m-wcohs-87,ci-tsc-84}, CAD/CAM \cite{b-beddc-88,ek-hstsr-89} and
wireless networks \cite{egs-gpepp-07},
there are very few works dealing with the well-known art gallery and
illumination class of problems when the objects involved are curvilinear
\cite{uz-ics-89,cgl-iscd-89,cruz-pcs-94,cgru-ihcs-95,kt-gcagv-08,
ktt-gcagv-09,clu-vgac-09}.

The original art gallery problem was posted by Klee to \Chvatal:
given a simple polygon $P$ with $n$ vertices, what is the minimum
number of point guards that are required in order to \guard the interior of 
$P$? \Chvatal \cite{c-ctpg-75} proved that $\lfloor\frac{n}{3}\rfloor$
vertex guards are always sufficient and sometimes necessary, while Fisk
\cite{f-spcwt-78}, a few years later, gave exactly the same result using
a much simpler proof technique based on polygon triangulation and
coloring the vertices of the triangulated polygon with three colors.
Lee and Lin \cite{ll-ccagp-86} showed that computing the minimum
number of vertex guards for a simple polygon is NP-hard, which is also
the case for point guards as shown by Aggarwal \cite{a-agpiv-84}.
In the context of curvilinear polygons, \ie polygons the edges of which
may be linear segments or arcs of curves, Karavelas, T{\'o}th and
Tsigaridas \cite{ktt-gcagv-09} have shown that
$\lfloor\frac{2n}{3}\rfloor$ vertex guards are always sufficient and
sometimes necessary in order to \guard \pconvex polygons (\ie locally
convex polygons, except possibly at the vertices, the edges of which
are convex arcs), whereas $\lceil\frac{n}{2}\rceil$ point guards are
sometimes necessary.
In the same paper it is also shown that $2n-4$ point guards are always
sufficient and sometimes necessary in order to \guard \pconcave
polygons, \ie locally concave polygons, except possibly at the
vertices, the edges of which are convex arcs. In the special case of
monotone \pconvex polygons, \ie polygons for which there exists a line
$L$ such that every line $L^\perp$ perpendicular to $L$ intersects the
polygon at at most two connected components, then
$\lfloor\frac{n}{2}\rfloor+1$ vertex or $\lfloor\frac{n}{2}\rfloor$
point guards are always sufficient and sometimes necessary
\cite{kt-gcagv-08}.
Cano-Vila, Longi and Urrutia \cite{clu-vgac-09} have also studied
the problem of \guard[ing] \pconvex polygons with vertex or point
guards. More precisely, they have indicated an alternative way for
proving the upper bound in \cite{ktt-gcagv-09} for the case of
vertex guards, and have improved the upper bound for the case of
point guards to $\lfloor\frac{5n}{8}\rfloor$.

Soon after the first results on \guard[ing] polygons with vertex or point
guards, other types of guarding models where considered. Toussaint
introduced in 1981 the notion of \emph{edge guards}. A point $p$ in the
interior of the polygon is considered to be \guard[ed] if it is visible
from at least one point of an edge in the \gset. Edge guards
where introduced as a guarding model in which guards where allowed to
move along the edges of the polygon. Another variation, dating back to
1983, is due to O'Rourke: guards are allowed to move along any diagonal
of the polygon. This type of guards has been called \emph{mobile guards}.
Toussaint conjectured that, except
for a few polygons, $\lfloor\frac{n}{4}\rfloor$ edge guards are always
sufficient. There are only two known counterexamples to this
conjecture, with $n=7,11$, due to Paige and Shermer
(cf. \cite{s-rrag-92}), requiring $\lfloor\frac{n+1}{4}\rfloor$ edge
guards. The first step towards Toussaint's conjecture was made by
O'Rourke \cite{o-gnfmg-83,o-agta-87} who proved that
$\lfloor\frac{n}{4}\rfloor$ mobile guards are always sufficient and
occasionally necessary in order to \guard any polygon with $n$
vertices. The technique by O'Rourke amounts to reducing the
problem of \guard[ing] a simple polygon to that of dominating a
\emph{triangulation graph} of the polygon. A triangulation graph is a
maximal outerplanar graph, all internal faces of which are
triangles. Dominance in this context means that at least one of the
vertices of each triangle in the triangulation graph is an endpoint of
a mobile guard. Shermer \cite{s-rrag-92} settled the problem of
\guard[ing] triangulation graphs with edge guards by showing that
$\lfloor\frac{3n}{10}\rfloor$ edge guards are always sufficient and
sometimes necessary, except for $n=3,6$ or $13$, in which case one
extra edge guard may be necessary. When considering orthogonal
polygons, \ie polygons the edges of which are axes-aligned, the
afore-mentioned upper and lower bounds drop. Aggarwal
\cite{a-agpiv-84} showed that $\lfloor\frac{3n+4}{16}\rfloor$ mobile
guards are sufficient and sometimes necessary in order to \guard
orthogonal polygons with $n$ vertices, a bound that was later on
matched for edge guards by Bjorling-Sachs \cite{b-egrc-98}. Finally,
\Gyori, Hoffmann, Kriegel and Shermer \cite{ghks-ggprp-96} showed
that when an orthogonal polygon with $n$ vertices contains $h$ holes,
$\lfloor\frac{3n+4h+4}{16}\rfloor$ mobile guards are sufficient and
sometimes necessary in order to \guard it.

In this paper we consider the problem of \guard[ing] \pconvex
polygons with edge or mobile guards. In our context an edge guard is
an edge of the polygon, whereas a mobile guard is an edge or a
diagonal of the polygon (a diagonal is a straight-line segment inside
the polygon connecting two polygon vertices).
Our proof technique capitalizes
on the technique used by O'Rourke to prove tight bounds on the number
of mobile guards that are necessary and sufficient for \guard[ing]
linear polygons \cite{o-agta-87}. As we have already mentioned above,
O'Rourke's paradigm reduces the geometric guarding problem to a
problem of diagonal dominance for the triangulation graph of the
linear polygon; the solution for the dominance problem is also a
solution for the original geometric mobile guarding problem. In our
case, the paradigm involves two steps: firstly the reduction of the
geometric problem to an appropriately defined combinatorial problem,
and secondly mapping the solution of the combinatorial problem to a
solution for the geometric problem.
More precisely, in order to \guard \pconvex polygons with
mobile or edge guards, we first reduce the problem of \guard[ing] our
\pconvex polygon $P$ to the problem of 2-dominating an appropriately
defined triangulation graph. Given a triangulation graph $\trg{P}$ of
a polygon $P$, a set of edges/diagonals of $\trg{P}$ is a 2-dominating
set of $\trg{P}$ if every triangle in $\trg{P}$ has at least two of
its vertices incident to an edge/diagonal in the 2-dominating set.
\new{We prove that $\bm$ diagonal guards or $\be$ edge guards are
always sufficient and sometimes necessary in order to 2-dominate
$\trg{P}$}. The proofs of sufficiency are inductive on the number of
vertices of $P$. In the case of diagonal 2-dominance, our proof yields
a linear time and space algorithm.

In the case of edge 2-dominance, the inductive step incorporates edge
contraction operations, thus yielding an $O(n^2)$ \new{time and $O(n)$
space algorithm, where $n$ is the number of vertices of $P$}. A linear
time and space algorithm can be attained by slightly relaxing the size
of the edge 2-dominating set. More precisely, we show inductively that
we can 2-dominate $\trg{P}$ with $\ubew$ edges; the proof is similar,
though more complicated, to the proof presented for the case of diagonal 
2-dominance. As in the \new{diagonal} 2-dominance case, it does not make use 
of edge contraction operations, thus permitting us to transform it to a
linear time and space algorithm. As a final note, the proof of
sufficiency for the diagonal 2-dominance problem is not the simplest
possible; in Section \ref{sec:trg-diag-guard-alt} of the Appendix we
present a much simpler alternate proof. The drawback of this alternate
proof is that it makes use of edge contractions, rendering it
unsuitable as the basis for a time-efficient algorithm; we present it,
however, for the sake of completeness.

\newcommand{\resetrowstyle}{\global\let\currentrowstyle\relax}
\newcolumntype{W}%
{>{\resetrowstyle\raggedright\arraybackslash\hsize=1.6\hsize}X}
\newcolumntype{Y}%
{>{\currentrowstyle\centering\arraybackslash\hsize=1.05\hsize}X}
\newcolumntype{Z}%
{>{\currentrowstyle\centering\arraybackslash\hsize=0.75\hsize}X}
\newcolumntype{R}%
{>{\currentrowstyle\centering\arraybackslash\hsize=0.85\hsize}X}

\newcommand{\doublecolumn}[1]{\multicolumn{2}{c}{#1}}
\newcommand{\rowstyle}[1]{\gdef\currentrowstyle{#1}#1\ignorespaces}

\renewcommand{\tabularxcolumn}[1]{>{\arraybackslash}m{#1}}

\begin{table}[!tbp]
  \newbegin
  \begin{center}
    \begin{tabularx}{0.95\textwidth}{WYZZR}\toprule[0.12em]
      \rowstyle{\slshape\bfseries}
      Polygon type&Guard type&Upper bound&Lower bound&Reference
      \\\midrule[0.12em]
      &vertex/point&\doublecolumn{$\lbmg$}&\cite{c-ctpg-75,f-spcwt-78}
      \\\cmidrule{2-5}
      linear&edge&$\lfloor\frac{3n}{10}\rfloor$\symbolfootnotemark[2]
      &$\lfloor\frac{n}{4}\rfloor$
      &\cite{s-rrag-92},\cite{o-gnfmg-83}
      \\\cmidrule{2-5}
      &mobile&\doublecolumn{$\lfloor\frac{n}{4}\rfloor$}
      &\cite{o-gnfmg-83}
      \\\midrule
      \multirow{2}{*}{orthogonal}&mobile
      &\doublecolumn{$\lfloor\frac{3n+4}{16}\rfloor$}
      &\cite{a-agpiv-84}
      \\\cmidrule{2-5}
      &edge&\doublecolumn{$\lfloor\frac{3n+4}{16}\rfloor$}
      &\cite{b-egrc-98}
      \\\midrule
      orthogonal with $h$ holes&mobile
      &\doublecolumn{$\lfloor\frac{3n+4h+4}{16}\rfloor$}
      &\cite{ghks-ggprp-96}
      \\\midrule
      \multirow{2}{*}{\pconvex}&vertex
      &\doublecolumn{$\lfloor\frac{2n}{3}\rfloor$}
      &\cite{ktt-gcagv-09}
      \\\cmidrule{2-5}
      &point&$\lfloor\frac{5n}{8}\rfloor$
      &$\lceil\frac{n}{2}\rceil$&\cite{clu-vgac-09},\cite{ktt-gcagv-09}
      \\\midrule
      \multirow{2}{3cm}{monotone \pconvex}&vertex
      &\doublecolumn{$\lfloor\frac{n}{2}\rfloor+1$}
      &\multirow{2}*{\cite{kt-gcagv-08}}
      \\\cmidrule{2-4}
      &point&\doublecolumn{$\lfloor\frac{n}{2}\rfloor$}&
      \\\midrule
      \pconcave&point&\doublecolumn{$2n-4$}&\cite{ktt-gcagv-09}
      \\\midrule[0.1em]
      %
      \multirow{2}{*}{\pconvex}&edge&$\be$\symbolfootnotemark[3]&$\lbeg$
      &\multirow{7}{*}{\textsl{this paper}}
      \\\cmidrule{2-4}
      &mobile&$\bm$&$\lbmg$&
      \\\cmidrule{1-4}
      monotone \pconvex&edge/mobile&\doublecolumn{$\mbd$}&
      \\\cmidrule{1-4}
      monotone locally convex&edge/mobile&\doublecolumn{$\mbd$}&
      \\\bottomrule[0.12em]
    \end{tabularx}
  \end{center}
  \caption{\new{Upper and lower bounds for the number of guards required to
    \guard a polygon with $n$ vertices. We focus on types of polygons
    and types of guards that are relevant to this paper. The upper
    part of the table contains previous results, whereas the lower
    part contains the results in this paper.}}\label{tbl:monitoring}
  \newend
\end{table}
\symbolfootnotetext[2]{\new{Except for $n=3,6$ or $13$, where an extra
  guard may be required.}}
\symbolfootnotetext[3]{\new{Except for $n=4$, where an additional
  guard is required.}}

\newcolumntype{A}%
{>{\resetrowstyle\raggedright\arraybackslash}X}
\newcolumntype{B}%
{>{\currentrowstyle\centering\arraybackslash}X}

\begin{table}[!tbp]
  \newbegin
  \begin{center}
    \begin{tabularx}{0.9\textwidth}{ABBB}\toprule[0.12em]
      \rowstyle{\slshape\bfseries}
      Dominance type&Guard type&Upper \& lower bound&Reference
      \\\midrule[0.12em]
      \multirow{2}*{dominance}&diagonal&$\lfloor\frac{n}{4}\rfloor$
      &\cite{o-gnfmg-83}
      \\\cmidrule{2-4}
      &edge&$\lfloor\frac{3n}{10}\rfloor$\symbolfootnotemark[2]
      &\cite{s-rrag-92}
      \\\midrule[0.1em]
      %
      \multirow{2}*{2-dominance}&diagonal&$\bm$&
      \multirow{2}{*}{\textsl{this paper}}
      \\\cmidrule{2-3}
      &edge&$\be$\symbolfootnotemark[3]&
      \\\bottomrule[0.12em]
    \end{tabularx}
  \end{center}
  \caption{\new{Upper and lower bounds for the number of guards required to
    dominate or 2-dominate the triangulation graph of a polygon with
    $n$ vertices. The upper part of the table refers to previously
    known results, whereas the lower part to the results presented in
    this paper.}}\label{tbl:dominance}
  \newend
\end{table}

Focusing back to the geometric guarding problem, the triangulation
graph $\trg{P}$ of the \pconvex polygon $P$ is a constrained
triangulation graph: based on the geometry of $P$, we require that
certain diagonals of $\trg{P}$ are present; the remaining
non-triangular subpolygons of $\trg{P}$ may be triangulated
arbitrarily. For the edge guarding problem, we prove that any edge
2-dominating set computed for $\trg{P}$ is also an edge \gset
for $P$. Unlike edge guards, a diagonal 2-dominating set computed
for $\trg{P}$ is mapped to a set of mobile guards of $P$, since the
2-dominating set for $\trg{P}$ may contain diagonals of $\trg{P}$ that
are not embeddable as straight-line diagonals of $P$. Using our
results on 2-dominance of triangulation graphs, we then prove that:
(1) we can compute a mobile \gset for $P$ of size at most $\bm$ in
$O(n\log{}n)$ time and $O(n)$ space,
(2) we can compute an edge \gset for $P$ of size at most $\be$ in
$O(n^2)$ time and $O(n)$ space, and 
(3) we can compute an edge \gset for $P$ of size at most $\ubew$ in
$O(n\log{}n)$ time and $O(n)$ space.
Finally, we show that $\lbmg$ mobile or $\lbeg$ edge guards are
sometimes necessary in order to \guard a \pconvex polygon $P$.

In the special case of monotone \pconvex polygons, \ie \pconvex
polygons with the property that there exists a line $L$ such that any
line perpendicular to $L$ intersects the \pconvex polygon at at most
two connected components, the upper and lower bounds on the number of
edge/mobile guards presented above can be further improved. We show
that $\mbd$ edge or mobile guards are always sufficient and sometimes
necessary, while an edge or mobile \gset of that size can be computed
in linear time and space. \new{The same results also hold for monotone
locally convex polygons. Tables \ref{tbl:monitoring} and
\ref{tbl:dominance} summarize the known results relevant to the
problems considered in this paper, as well as our results.}

The rest of the paper is structured as follows.
In Section \ref{sec:trg-diag-guards} we prove our matching upper and
lower bounds on the number of diagonals required in order to
2-dominate a triangulation graph and show how such a 2-dominating set
can be computed in linear time and space.
The next section, Section \ref{sec:trg-edge-guards} deals with the
problem of 2-dominance of triangulation graphs with edge guards.
We first prove our matching upper and lower bounds on the
number of edges required in order to 2-dominate a triangulation
graph. We then prove our relaxed bound and show how the proof is
transformed into a linear time and space algorithm.
In Section \ref{sec:guard-piecewise-convex} we show how to construct
the triangulation graph $\trg{P}$ of a \pconvex polygon
$P$. We describe how a diagonal 2-dominating set of $\trg{P}$ is
mapped to a mobile \gset for $P$. We also show that an edge
2-dominating set for $\trg{P}$ is also an edge \gset for
$P$. Algorithmic considerations are also discussed. We end this
section by providing lower bound constructions for both guarding
problems.
The special case of monotone \pconvex polygons is treated in
Section \ref{sec:monotone}.
Finally in Section \ref{sec:conclusion} we conclude with a discussion
of our results and open problems.


\section{2-dominance of triangulation graphs: diagonal guards}
\label{sec:trg-diag-guards}

\begin{figure}[!t]
  \newbegin
  \begin{center}
    \includegraphics[width=0.99\textwidth]{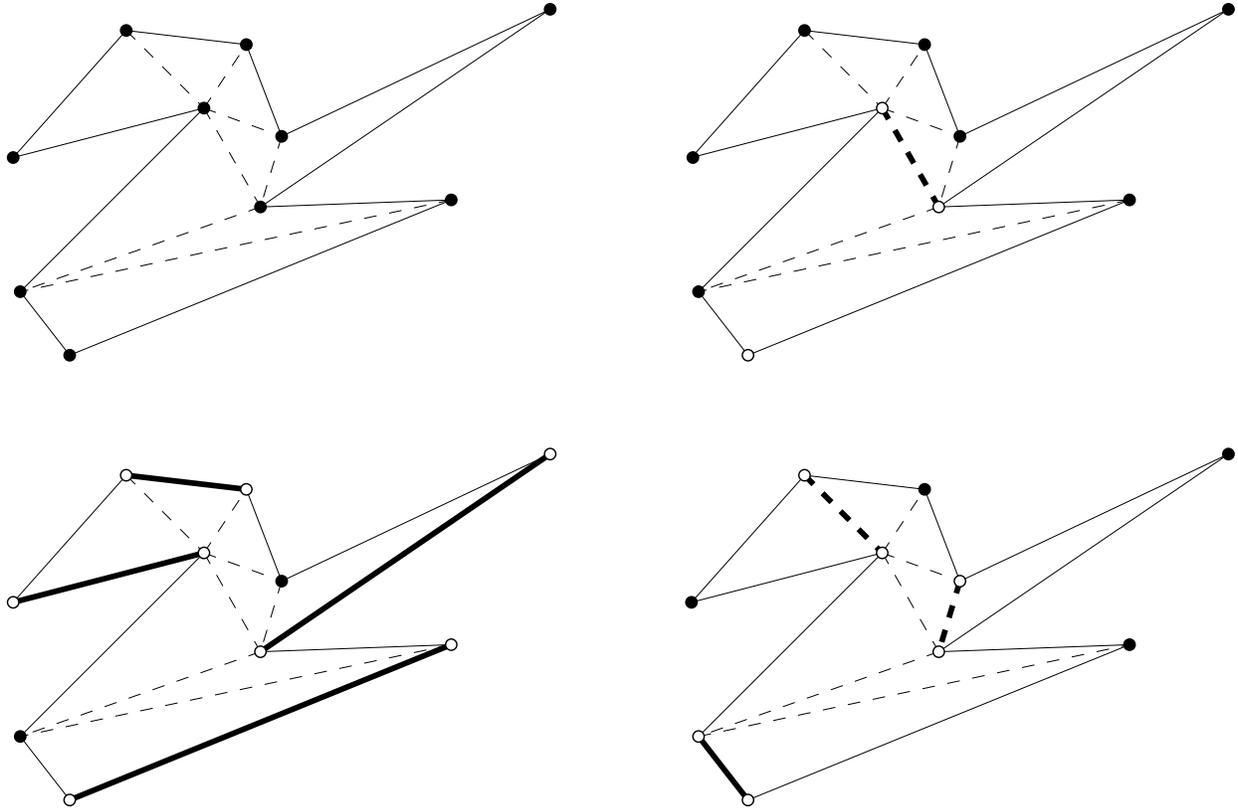}
    \caption{\new{A triangulation graph $T$ with $n=10$ vertices and
        various dominating sets. The diagonals of $T$ are shown with
        dashed lines, whereas the edges of the Hamiltonian cycle in
        $T$ are shown with solid lines. Vertices in a dominating set
        are transparent, whereas edges (\resp diagonals) in
        a dominating set are shown with thick solid (\resp dashed)
        lines. Top left: the triangulation graph $T$. Top right: a
        dominating set of $T$ consisting of a vertex and a
        diagonal. Bottom left: an edge 2-dominating set of $T$. Bottom
        right: a diagonal 2-dominating set of $T$.}}
    \label{fig:trgraph}
  \end{center}
  \newend
\end{figure}

\newbegin
A \emph{triangulation graph} $T$ is a maximal outerplanar graph, \ie a
Hamiltonian planar graph with $n$ vertices and $2n-3$ edges,
all internal faces of which are triangles (see
Fig. \ref{fig:trgraph}(top left)).
The unique Hamiltonian cycle in $T$ is the cycle that bounds the outer
face. The edges that do not belong to the Hamiltonian cycle are called
\emph{diagonals}, whereas the term \emph{edge} is used to refer to the
edges of the Hamiltonian cycle. Given an $n$-vertex linear polygon $P$,
\ie a polygon the edges of which are line segments, its triangulation
graph, denoted by $\trg{P}$, is the planar graph we get when the
polygon has been triangulated.

A \emph{dominating set} $D$ of a triangulation graph $T$ is a
set of vertices, edges or diagonals of $T$ such that at least
one of the vertices of each triangle in $T$ belongs to $D$ (see
Fig. \ref{fig:trgraph}(top right)\footnote{Unless otherwise
  stated, in all figures, edges/diagonals in a dominating/\gset
  are shown as thick solid/dashed lines, while vertices in a
  dominating/\gset are transparent.}).
An \emph{edge (\resp diagonal) dominating set} of $T$ is a
dominating set of $T$ consisting of only edges (\resp edges or
diagonals) of $T$.
A \emph{2-dominating set} $D$ of $T$ is a dominating set of
$T$ that has the property that every triangle in $T$ has
at least two of its vertices in $D$. In a similar manner, an
\emph{edge (\resp diagonal) 2-dominating set} of $T$ is a
2-dominating set of $T$ consisting only of edges (\resp edges or
diagonals) of $T$ (see Fig. \ref{fig:trgraph}(bottom row)).
\newend


\new{In the rest of the paper we shall only refer to triangulation graphs
of polygons.}
Let us, initially, state the following lemma, which is a
direct generalization of Lemmas 1.1 and 3.6 in
\cite{o-agta-87}.

\begin{lemma}\label{lem:diag_existence}
Consider an integer $\lambda\ge{}2$. Let $P$ be a polygon of
$n\ge{}2\lambda$ vertices, and $\trg{P}$ a triangulation graph of
$P$. There exists a diagonal $d$ in $\trg{P}$ that partitions
$\trg{P}$ into two pieces, one of which contains $k$ arcs
corresponding to edges of $P$, where $\lambda\le{}k\le{}2(\lambda-1)$.
\end{lemma}

\begin{proof}
Choose $d$ to be a diagonal of $\trg{P}$ that separates off a
\textsl{minimum} number of polygon edges that is at least $\lambda$.
Let $k\ge\lambda$ be this minimum number, and label the vertices of
$P$ with the labels $0,1,\ldots,n-1$, such that $d$ is $(0,k)$. The
diagonal $d$ supports a triangle whose apex is at vertex $t$,
$0\le{}t\le{}k$. Since $k$ is minimal $t\le{}\lambda-1$ and
$k-t\le{}\lambda-1$. Thus, $\lambda\le{}k\le{}2(\lambda-1)$.
\end{proof}

Before proceeding with the first main result of this section, we state
an intermediate lemma dealing with the diagonal 2-dominance problem
for small values of $n$.

\begin{lemma}\label{lem:guard-combdiag-smallpoly}
Every triangulation graph $\trg{P}$ with $3\le{}n\le{}7$ vertices,
corresponding to a polygon $P$, can be 2-dominated by
$\bm$ diagonal guards.
\end{lemma}

\begin{proof}
Let $v_i$, $1\le{}i\le{}n$ be the vertices of $\trg{P}$, and let $e_i$
be the edge $v_iv_{i+1}$\footnote{Indices are considered to be evaluated
  modulo $n$.}. For each of the \new{five} values for $n$ we are
going to define a diagonal 2-dominating set $D$ of size $\bm$.
\begin{figure}[!t]
  \psfrag{t}[][]{\scriptsize$t$}
  \psfrag{t1}[][]{\scriptsize$t_1$}
  \psfrag{t2}[][]{\scriptsize$t_2$}
  \psfrag{e}[][]{\scriptsize$e$}
  \psfrag{d1}[][]{\scriptsize$d_1$}
  \psfrag{d2}[][]{\scriptsize$d_2$}
  \psfrag{e1}[][]{\scriptsize$e'$}
  \psfrag{e2}[][]{\scriptsize$e''$}
  \begin{center}
    \includegraphics[width=0.85\textwidth]{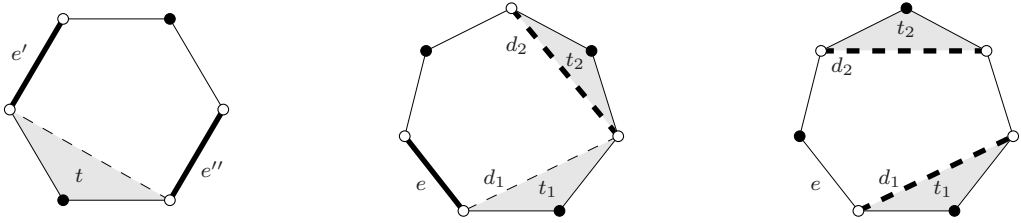}
    \caption{Proof of Lemma \ref{lem:guard-combdiag-smallpoly} for
      $n=6,7$. Left: the case $n=6$. Middle: the case $n=7$ and $d_1$,
      $d_2$ share a vertex. Right: the case $n=7$ and $d_1$, $d_2$ do
      not share a vertex.}
    \label{fig:ears_diags}
  \end{center}
\end{figure}
\begin{mathdescription}
\item[n=3.] Trivial: let $D$ consist of any of the three edges of
  $\trg{P}$.
\item[n=4.] Again trivial: let $D$ consist of the unique diagonal $d$
  of $\trg{P}$.
\item[n=5.] Let $D$ consist of the two diagonals of the pentagon. $D$
  is a 2-dominating set for $\trg{P}$, since the two ears have two of
  their vertices in $D$, whereas the third triangle in $\trg{P}$ has
  all three vertices in $D$.
\item[n=6.] Let $t$ be an ear of $\trg{P}$, and let $e'$ and $e''$ be
  the edges of $P$ incident to $t$ that do not belong to $t$ (see
  Fig. \ref{fig:ears_diags}(left)).
  Set $D=\{e',e''\}$; $D$ is a diagonal 2-dominating set for
  $\trg{P}$, since the triangulation graph $\trg{P}\setminus\{t\}$ has
  all but one of its vertices in $D$, whereas $t$ has two of its
  vertices in $D$.
\item[n=7.] Let $t_1$ and $t_2$ be two ears of $\trg{P}$, and let
  $d_1$ and $d_2$ be the diagonals of $\trg{P}$ supporting these
  ears. The two possible relative positions of $t_1$ and $t_2$ are
  shown in Fig. \ref{fig:ears_diags}: either $d_1$ and $d_2$ share a
  vertex, or $d_1$ and $d_2$ do not share any vertices of $P$.
  In the former case, let $e$ be the edge of $P$ incident to $d_1$
  that is not an edge of $t_1$ or $t_2$. Set $D=\{e,d_2\}$; $D$ is a
  diagonal 2-dominating set for $\trg{P}$, since $t_1$ is 2-dominated
  by vertices of $e$ and $d_2$, $t_2$ is 2-dominated by the two
  vertices of $d_2$, whereas the triangulation graph
  $\trg{P}\setminus\{t_1,t_2\}$ has four of its five vertices in $D$.
  In the latter case, set $D=\{d_1,d_2\}$; $D$ is a diagonal
  2-dominating set for $\trg{P}$, since $t_1$ is 2-dominated by the
  two vertices of $d_1$, $t_2$ is 2-dominated by the two vertices of
  $d_2$, whereas the triangulation graph $\trg{P}\setminus\{t_1,t_2\}$
  has four of its five vertices in $D$.\qedhere
\end{mathdescription}
\end{proof}

Using Lemma \ref{lem:diag_existence} for $\lambda=4$, yields the
following theorem concerning the worst-case number of diagonals that
are sufficient and necessary in order to 2-dominate a
triangulation graph. The inductive proof that follows is not the
simplest possible. The interested reader may find a much simpler
alternative proof in Section \ref{sec:trg-diag-guard-alt} of the
Appendix. The proof in Section \ref{sec:trg-diag-guard-alt}, however,
makes use of edge contractions (to be discussed in detail in Section
\ref{sec:trg-edge-guards}), which make it unsuitable as a basis for a
linear time and space algorithm. On the other hand, the proof
presented below can be implemented in linear time and space, as will
be discussed in Section \ref{sec:compute-diag-dsets}. The proof
below is a detailed, rather technical, case-by-case analysis; we
present it, however, uncondensed, so as to illustrate the details that
pertain to our linear time and space algorithm.

\begin{theorem}\label{thm:guard-combdiag-trgraph}
Every triangulation graph $\trg{P}$ of a polygon $P$ with $n\ge{}3$
vertices can be 2-dominated by $\bm$ diagonal guards. This bound is
tight in the worst-case.
\end{theorem}

\begin{proof}
In Lemma \ref{lem:guard-combdiag-smallpoly}, we have shown the result
for $3\le{}n\le{}7$.
Let us now assume that $n\ge{}8$ and that the theorem holds for all
$n'$ such that $3\le{}n'<n$. By means of Lemma \ref{lem:diag_existence}
with $\lambda=4$, there exists a diagonal $d$ that partitions $\trg{P}$
into two triangulation graphs $T_1$ and $T_2$, where $T_1$ contains
$k$ boundary edges of $\trg{P}$ with $4\le{}k\le{}6$. Let $v_i$,
$0\le{}i\le{}k$, be the $k+1$ vertices of $T_1$, as we encounter them
while traversing $P$ counterclockwise, and let $v_0v_k$ be the common
edge of $T_1$ and $T_2$. For each value of $k$ we are going to define
a diagonal 2-dominating set $D$ for $\trg{P}$ of size $\bm$. In what
follows $d_{ij}$ denotes the diagonal $v_iv_j$, whereas $e_i$ denotes
the edge $v_iv_{i+1}$. Consider each value of $k$ separately.

\begin{mathdescription}
\item[k=4.] In this case $T_2$ contains $n-3$ vertices. By our
  induction hypothesis we can 2-dominate $T_2$ with $f(n-3)=\bm-1$
  diagonal guards. Let $D_2$ be the diagonal 2-dominating set for
  $T_2$. At least one of $v_0$ and $v_4$ is in $D_2$. The cases are
  symmetric, so we can assume without loss of generality that
  $v_0\in{}D_2$. Consider the following cases (see
  Fig. \ref{fig:diagproof2_pentagons}):
  \begin{mathdescription}
  \item[d_{13}\in{}T_1.] Set $D=D_2\cup\{d_{13}\}$.
  \item[d_{24}\in{}T_1.] Set $D=D_2\cup\{d_{24}\}$.
  \item[d_{02},d_{03}\in{}T_1.] Set $D=D_2\cup\{e_2\}$. 
  \end{mathdescription}

  \begin{figure}[!ht]
    \begin{center}
      \psfrag{0}[][]{\scriptsize$v_0$}
      \psfrag{1}[][]{\scriptsize$v_1$}
      \psfrag{2}[][]{\scriptsize$v_2$}
      \psfrag{3}[][]{\scriptsize$v_3$}
      \psfrag{4}[][]{\scriptsize$v_4$}
      \psfrag{d}[][]{\scriptsize$d$}
      \includegraphics[width=\columnwidth]{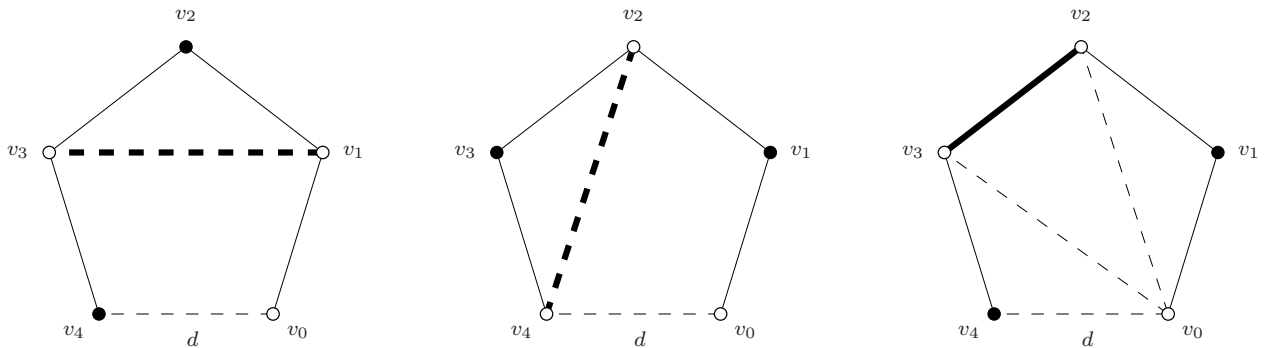}
      \caption{Proof of Theorem \ref{thm:guard-combdiag-trgraph}: the
        case $k=4$. Left: $d_{13}\in{}T_1$. Middle:
        $d_{24}\in{}T_1$. Right: $d_{02},d_{03}\in{}T_1$.}
      \label{fig:diagproof2_pentagons}
    \end{center}
  \end{figure}

\item[k=5.] The presence of diagonals $d_{04}$ and $d_{15}$ would
  violate the minimality of $k$. Let $t$ be the triangle supported by
  $d$ in $T_1$. The apex $v$ of this triangle can either be $v_2$ or
  $v_3$. The two cases are symmetric, so we assume, without loss of
  generality that the apex of $t$ is $v_2$. Consider the triangulation
  graph $T'=T_2\cup\{t\}$. It has $n-3$ vertices, hence, by our
  induction hypothesis, it can be 2-dominated with $f(n-3)=\bm-1$
  diagonal guards. Let $D'$ be the 2-dominating set for $T'$. Consider
  the following cases (see Fig. \ref{fig:diagproof2_hexagons}):
  \begin{mathdescription}
  \item[d_{02}\in{}D_2.] Set $D=D'\cup\{e_3\}$.
  \item[d_{02}\nin{}D_2.] If $d_{25}\in{}D'$, set
    $D=(D'\setminus\{d_{25}\})\cup\{d_{02},e_4\}$. Otherwise, $v_2$
    cannot belong to $D'$ (both edges of $T'$ incident to $v_2$
    do not belong to $D'$). However, the triangle $t$ is 2-dominated in
    $T'$, which implies that both $v_0$ and $v_5$ belong to
    $D'$. Hence, set $D=D'\cup\{e_2\}$.
  \end{mathdescription}

  \begin{figure}[!t]
    \begin{center}
      \psfrag{0}[][]{\scriptsize$v_0$}
      \psfrag{1}[][]{\scriptsize$v_1$}
      \psfrag{2}[][]{\scriptsize$v_2$}
      \psfrag{3}[][]{\scriptsize$v_3$}
      \psfrag{4}[][]{\scriptsize$v_4$}
      \psfrag{5}[][]{\scriptsize$v_5$}
      \psfrag{d}[][]{\scriptsize$d$}
      \psfrag{t}[][]{\scriptsize$t$}
      \includegraphics[width=\columnwidth]{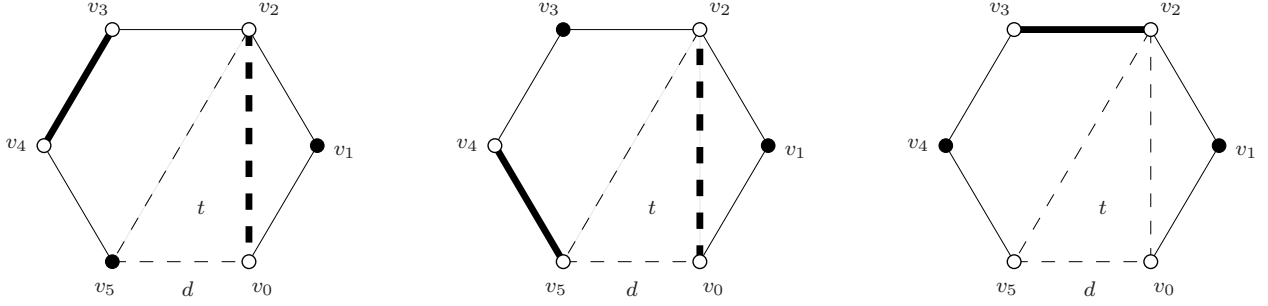}
      \caption{Proof of Theorem \ref{thm:guard-combdiag-trgraph}:
        the case $k=5$. Left: $d_{02}\in{}D'$. Middle:
        $d_{02}\nin{}D'$ and $d_{25}\in{}D'$. Right:
        $d_{02},d_{25}\nin{}D'$.}
      \label{fig:diagproof2_hexagons}
    \end{center}
  \end{figure}

\item[k=6.]  The presence of diagonals $d_{04}$, $d_{05}$, $d_{16}$
  and $d_{26}$ would violate the minimality of $k$. Let $t$ be the
  triangle supported by $d$ in $T_1$. The apex $v$ of this triangle
  must be $v_3$. Let $t'$ be the second triangle in $T_1$ beyond $t$
  supported by the diagonal $d_{03}$, and let $v'$ be its vertex
  opposite to $d_{03}$. Symmetrically, let $t''$ be the second
  triangle in $T_1$ beyond $t$ supported by the diagonal $d_{36}$, and
  let $v''$ be its vertex opposite to $d_{36}$. \new{Consider the
  triangulation graphs $T'=T_2\cup\{t,t'\}$ and
  $T''=T_2\cup\{t,t''\}$. $T'$ and $T''$ have $n-3$ vertices,
  hence, by our induction hypothesis, they can be 2-dominated with
  $f(n-3)=\bm-1$ diagonal guards. Let $D'$ (\resp $D''$) be the
  2-dominating set for $T'$ (\resp $T''$).}

  Let us first consider the case $v'\equiv{}v_2$. Let $d''$ be the
  unique diagonal of the quadrilateral $v_3v_4v_5v_6$. Consider the
  following cases (see Fig. \ref{fig:diagproof2_heptagons}):
  \begin{mathdescription}
  \item[d_{02}\in{}D'.] Set $D=D'\cup\{d''\}$.
  \item[d_{02}\nin{}D'.] We further distinguish between the
    following two cases:
    \begin{mathdescription}
    \item[d_{36}\in{}D'.] If $v_0\in{}D'$, simply set
      $D=(D'\setminus\{d_{36}\})\cup\{e_2,e_5\}$. If $v_0\nin{}D'$,
      the diagonal $d_{03}$ cannot belong to $D'$. Therefore, in order
      for the triangle $t'$ to be 2-dominated by $D'$, we must have
      that $e_2$ in $D'$. Thus, set
      $D=(D'\setminus\{d_{36}\})\cup\{e_0,e_5\}$.
    \item[d_{36}\nin{}D'.] In order for $t'$ to be 2-dominated by
      $D'$ we must have that either $d_{03}\in{}D'$ or
      $e_2\in{}D'$. If $d_{03}\in{}D'$, set
      $D=(D'\setminus\{d_{03}\})\cup\{d_{02},d''\}$; otherwise, set
      $D=(D'\setminus\{e_2\})\cup\{d_{02},d''\}$.
    \end{mathdescription}
  \end{mathdescription}

  \begin{figure}[!tp]
    \begin{center}
      \psfrag{0}[][]{\scriptsize$v_0$}
      \psfrag{1}[][]{\scriptsize$v_1$}
      \psfrag{2}[][]{\scriptsize$v_2$}
      \psfrag{3}[][]{\scriptsize$v_3$}
      \psfrag{4}[][]{\scriptsize$v_4$}
      \psfrag{5}[][]{\scriptsize$v_5$}
      \psfrag{6}[][]{\scriptsize$v_6$}
      \psfrag{d}[][]{\scriptsize$d$}
      \psfrag{dpp}[][]{\scriptsize$d''$}
      \psfrag{t}[][]{\scriptsize$t$}
      \psfrag{tp}[][]{\scriptsize$t'$}
      \includegraphics[width=\columnwidth]{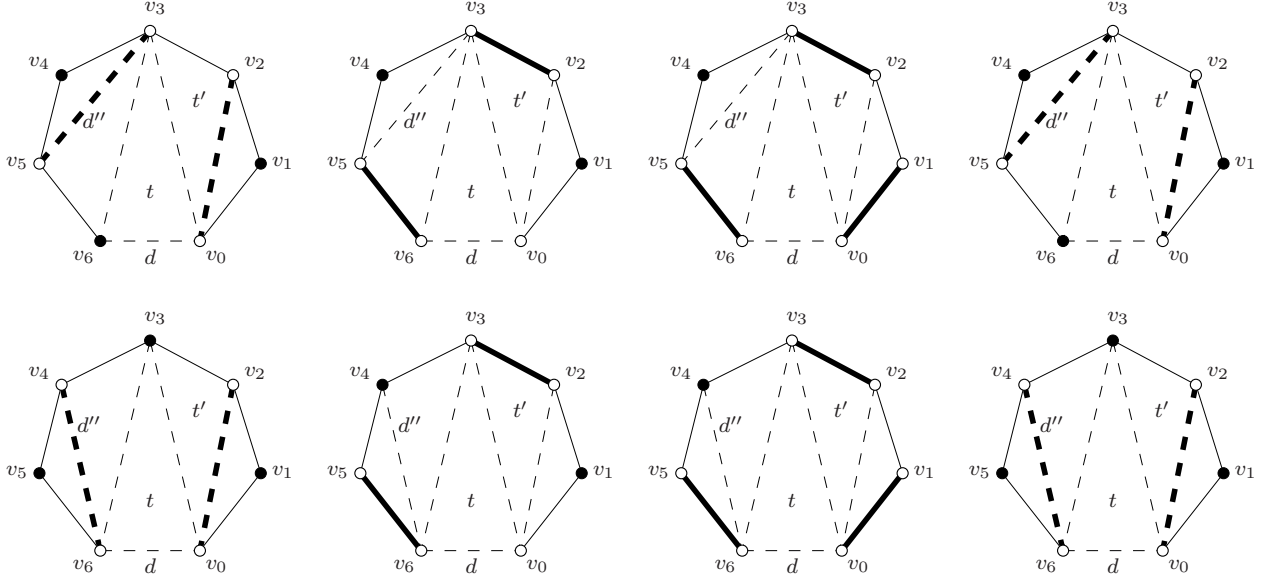}
      \caption{Proof of Theorem \ref{thm:guard-combdiag-trgraph}:
        the case $k=6$ with $v'\equiv{}v_2$. \newbegin
        Top row: $v''\equiv{}v_5$.
        Bottom row: $v''\equiv{}v_4$.
        Left column: $d_{02}\in{}D'$.
        Middle left column: $d_{02}\nin{}D'$ and $d_{36}\in{}D'$ and
        $v_0\in{}D'$.
        Middle right column: $d_{02}\nin{}D'$ and $d_{36}\in{}D'$ 
        and $v_0\nin{}D'$.
        Right column: $d_{02},d_{36}\nin{}D'$.\newend}
      \label{fig:diagproof2_heptagons}
    \end{center}
    \newend
  \end{figure}

  \newbegin
  Let us now consider the case $v'\equiv{}v_1$. We first consider the
  situation $v''\equiv{}v_4$. Consider the following cases (see
  Fig. \ref{fig:diagproof2_heptagons1}):
  \begin{mathdescription}
  \item[d_{46}\in{}D''.] Set $D=D''\cup\{d_{13}\}$.
  \item[d_{46}\nin{}D''.] We further distinguish between the
    following two cases:
    \begin{mathdescription}
    \item[d_{03}\in{}D''.] If $v_6\in{}D''$, simply set
      $D=(D''\setminus\{d_{03}\})\cup\{e_0,e_3\}$. If $v_6\nin{}D''$,
      the diagonal $d_{36}$ cannot belong to $D''$. Therefore, in order
      for the triangle $t''$ to be 2-dominated by $D''$, we must have
      that $e_3$ in $D''$. Thus, set
      $D=(D''\setminus\{d_{36}\})\cup\{e_0,e_5\}$.
    \item[d_{03}\nin{}D''.] In order for $t''$ to be 2-dominated by
      $D''$ we must have that either $d_{36}\in{}D''$ or
      $e_3\in{}D''$. If $d_{36}\in{}D''$, set
      $D=(D''\setminus\{d_{36}\})\cup\{d_{13},d_{46}\}$; otherwise, set
      $D=(D''\setminus\{e_3\})\cup\{d_{13},d_{46}\}$.
    \end{mathdescription}
  \end{mathdescription}
  \newend

  \begin{figure}[!t]
    \newbegin
    \begin{center}
      \psfrag{0}[][]{\scriptsize$v_0$}
      \psfrag{1}[][]{\scriptsize$v_1$}
      \psfrag{2}[][]{\scriptsize$v_2$}
      \psfrag{3}[][]{\scriptsize$v_3$}
      \psfrag{4}[][]{\scriptsize$v_4$}
      \psfrag{5}[][]{\scriptsize$v_5$}
      \psfrag{6}[][]{\scriptsize$v_6$}
      \psfrag{d}[][]{\scriptsize$d$}
      \psfrag{t}[][]{\scriptsize$t$}
      \psfrag{tp}[][]{\scriptsize$t'$}
      \psfrag{tpp}[][]{\scriptsize$t''$}
      \includegraphics[width=\columnwidth]{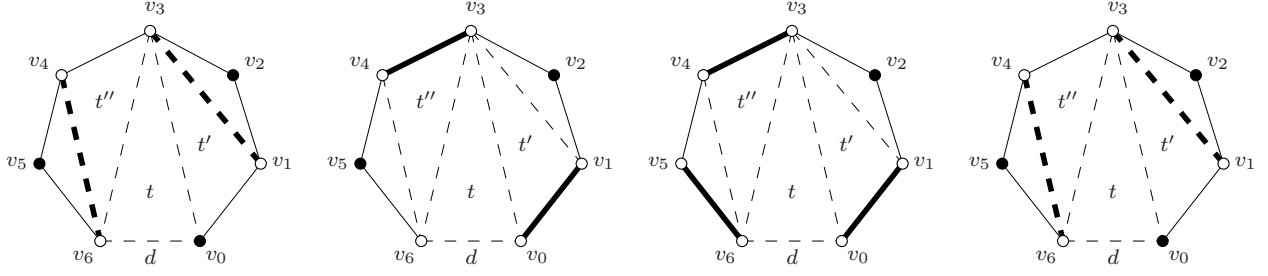}
      \caption{\new{Proof of Theorem \ref{thm:guard-combdiag-trgraph}:
        the case $k=6$ with $v'\equiv{}v_1$ and $v''\equiv{}v_4$.
        Left: $d_{36}\in{}D''$.
        Middle left: $d_{36}\nin{}D''$ and $d_{03}\in{}D''$ and
        $v_6\in{}D''$.
        Middle right: $d_{36}\nin{}D''$ and $d_{03}\in{}D''$ 
        and $v_6\nin{}D''$.
        Right: $d_{03},d_{36}\nin{}D''$.}}
      \label{fig:diagproof2_heptagons1}
    \end{center}
    \newend
  \end{figure}

  \begin{figure}[!t]
    \begin{center}
      \psfrag{0}[][]{\scriptsize$v_0$}
      \psfrag{1}[][]{\scriptsize$v_1$}
      \psfrag{2}[][]{\scriptsize$v_2$}
      \psfrag{3}[][]{\scriptsize$v_3$}
      \psfrag{4}[][]{\scriptsize$v_4$}
      \psfrag{5}[][]{\scriptsize$v_5$}
      \psfrag{6}[][]{\scriptsize$v_6$}
      \psfrag{d}[][]{\scriptsize$d$}
      \psfrag{t}[][]{\scriptsize$t$}
      \psfrag{tp}[][]{\scriptsize$t'$}
      \includegraphics[width=0.7\columnwidth]{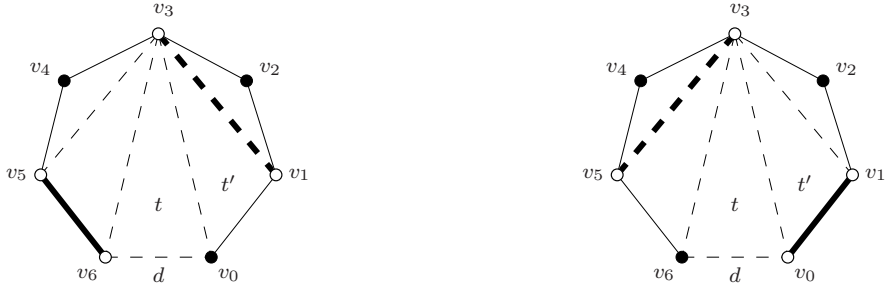}
      \caption{Proof of Theorem \ref{thm:guard-combdiag-trgraph}:
        the case $k=6$ with $v'\equiv{}v_1$ and $v''\equiv{}v_5$.
        Left: $d_{13}\in{}D'$; also $d_{13},d_{03},e_0\nin{}D'$.
        Right: $d_{13}\nin{}D'$ and $d_{03}\in{}D'$; also
        $d_{13},d_{03}\nin{}D'$ and $e_0\in{}D'$.}
      \label{fig:diagproof2_heptagons2}
    \end{center}
  \end{figure}

  The only remaining case is the case where $v'\equiv{}v_1$ and
  $v''\equiv{}v_5$. Consider the following cases (see
  Fig. \ref{fig:diagproof2_heptagons2}):
  \begin{mathdescription}
  \item[d_{13}\in{}D'.] Set $D=D'\cup\{e_5\}$.
  \item[d_{13}\nin{}D'.] We further distinguish between the
    following two cases:
    \begin{mathdescription}
    \item[d_{03}\in{}D'.] Set $D=(D'\setminus\{d_{03}\})\cup\{e_0,d_{35}\}$.
    \item[d_{03}\nin{}D'.] If $e_0\in{}D'$, set
      $D=D'\cup\{d_{35}\}$. Otherwise, \ie if $e_0\nin{}D'$, $v_1$
      cannot be in $D'$. Since the triangle $t'$ is 2-dominated in
      $D'$, both $v_0$ and $v_3$ have to belong to $D'$. Since the
      diagonal $d_{03}$ does not belong to $D'$, the diagonal $d_{36}$
      has to belong to $D'$ in order for $v_3$ to be in $D'$. Thus,
      set $D=(D'\setminus\{d_{36}\})\cup\{d_{13},e_5\}$. 
    \end{mathdescription}
  \end{mathdescription}
\end{mathdescription}

\begin{figure}[!t]
  \psfrag{T1}[][]{$T_1$}
  \psfrag{T2}[][]{$T_2$}
  \psfrag{T3}[][]{$T_3$}
  \psfrag{v0}[][]{\scriptsize$v_0$}
  \psfrag{v1}[][]{\scriptsize$v_1$}
  \psfrag{v2}[][]{\scriptsize$v_2$}
  \psfrag{v3}[][]{\scriptsize$v_3$}
  \psfrag{v4}[][]{\scriptsize$v_4$}
  \psfrag{v5}[][]{\scriptsize$v_5$}
  \psfrag{v3m+1}[][]{\scriptsize$v_{3m+1}$}
  \psfrag{v3m}[][]{\scriptsize$v_{3m}$}
  \psfrag{v3m-1}[][]{\scriptsize$v_{3m-1}$}
  \psfrag{v3m-2}[][]{\scriptsize$v_{3m-2}$}
  \psfrag{v3m-3}[][]{\scriptsize$v_{3m-3}$}
  \psfrag{v3m-4}[][]{\scriptsize$v_{3m-4}$}
  \psfrag{v3m-5}[][]{\scriptsize$v_{3m-5}$}
  \psfrag{v3m-6}[][]{\scriptsize$v_{3m-6}$}
  \hfil%
  \begin{center}
    \includegraphics[width=0.46\columnwidth]{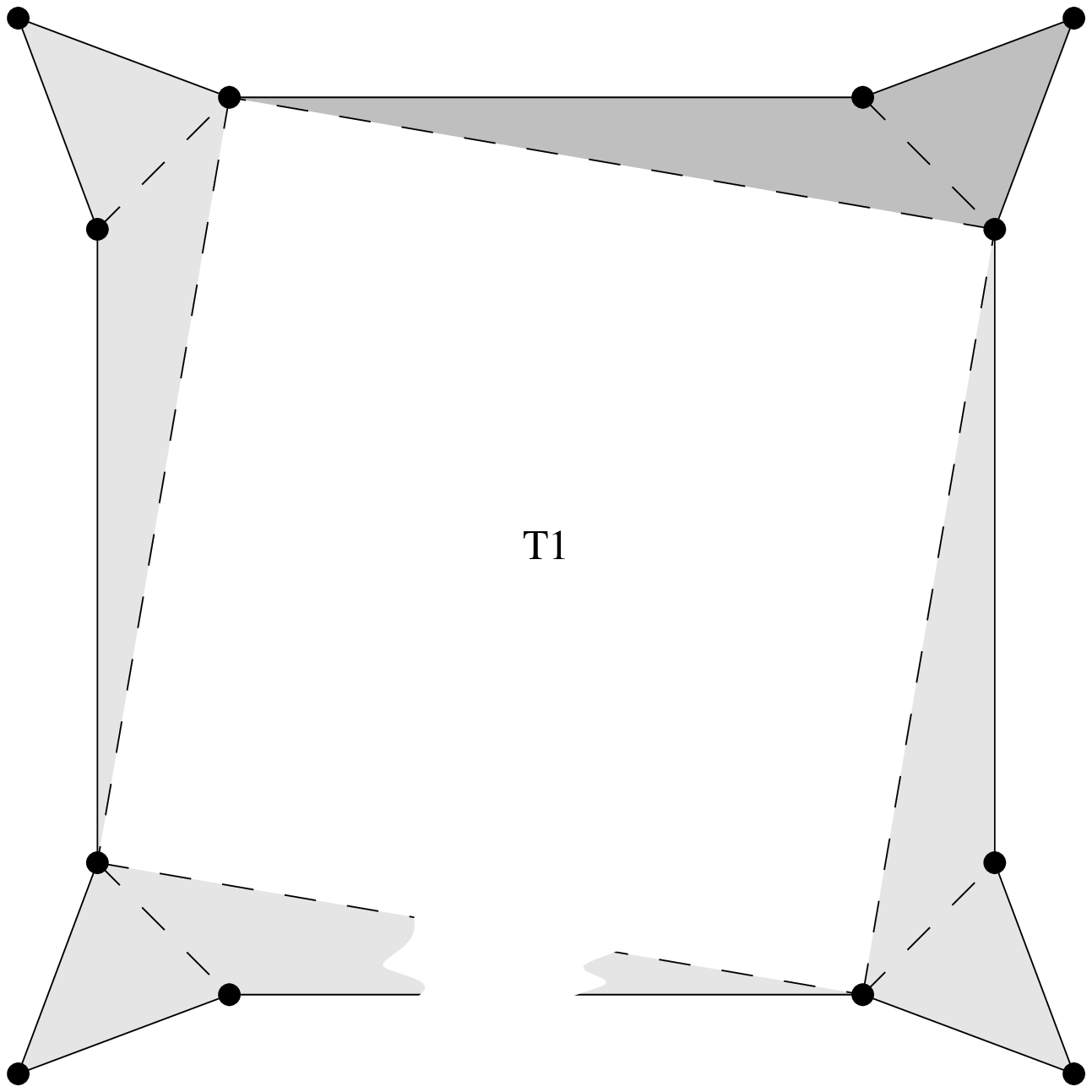}\hfill%
    \includegraphics[width=0.48\columnwidth]{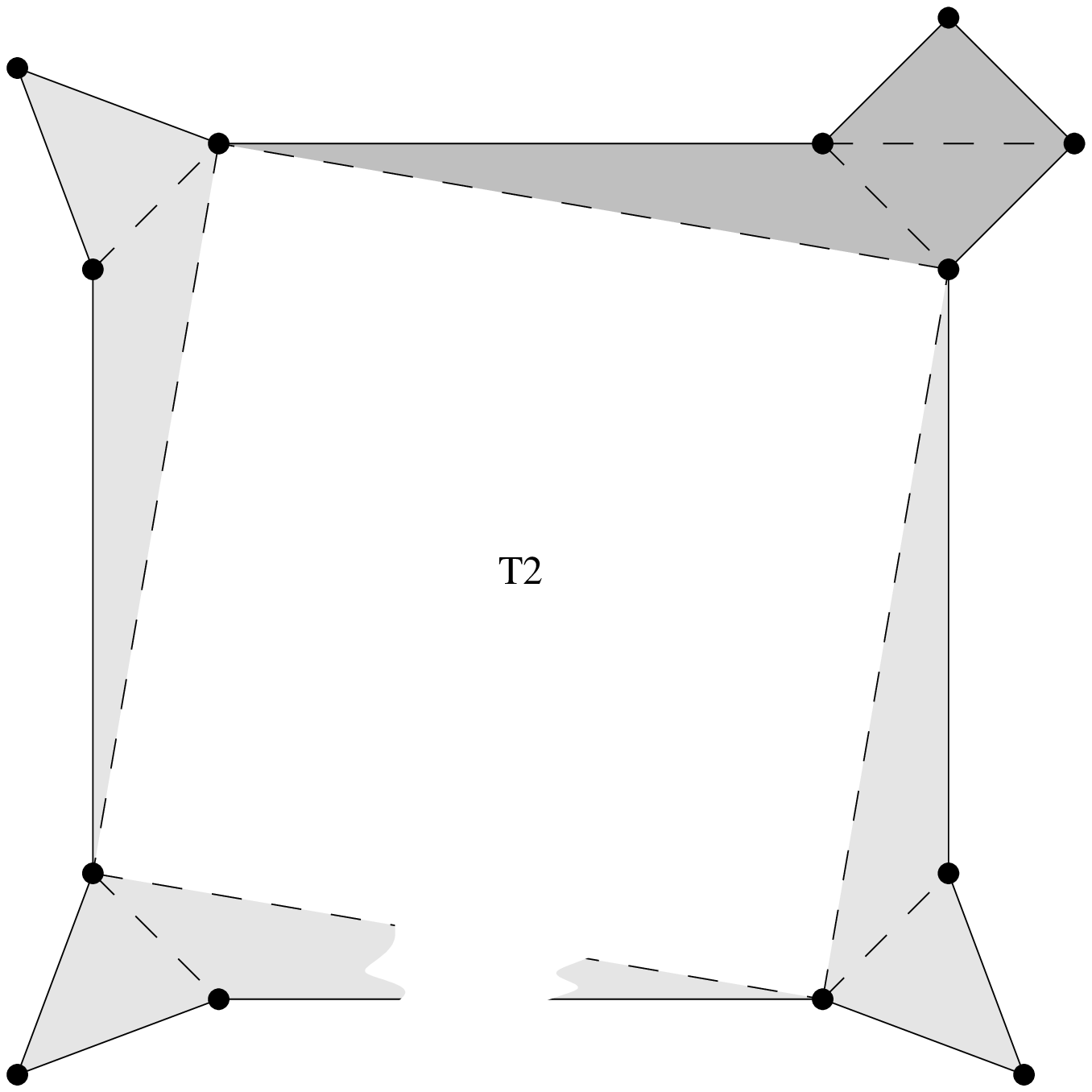}\\[10pt]
    \includegraphics[width=0.5\columnwidth]{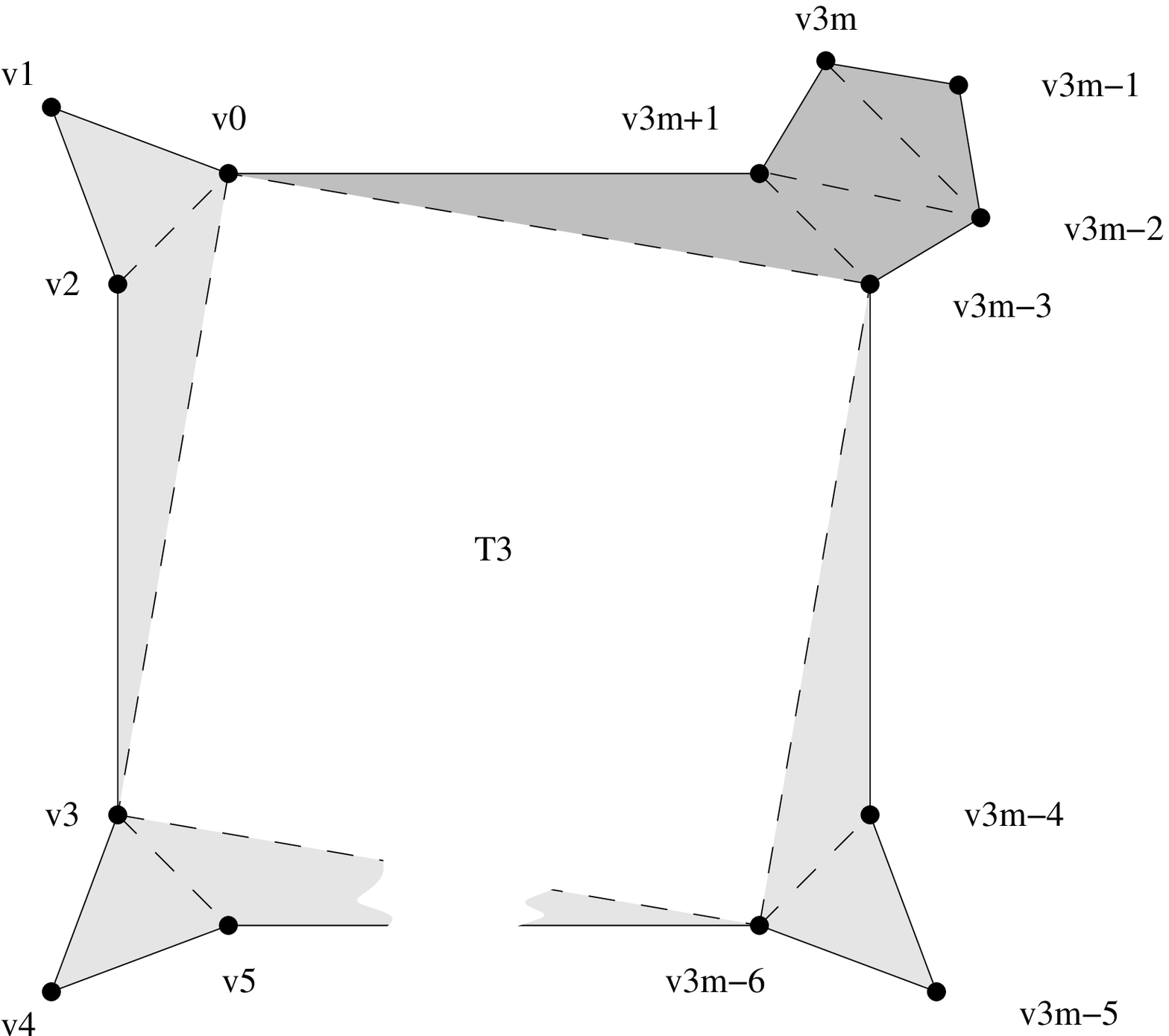}
    \vspace*{-1mm}
  \end{center}
  \caption{Three triangulation graphs $T_i$, $i=1,2,3$, with
    $n=3m+i-1$ vertices, respectively (the central part of the graph
    is triangulated arbitrarily). All three triangulation graphs
    require at least $\bm$ diagonal guards in order to be
    2-dominated.}
  \label{fig:trg-diag-lb}
\end{figure}

Let us now turn our attention to establishing the lower bound.
Consider the triangulation graphs $T_i$, $i=1,2,3$, with $n=3m+i-1$
vertices, shown in Fig. \ref{fig:trg-diag-lb}, and let $D_i$ be the
diagonal 2-dominating set of $T_i$. The central part of $T_i$ is
triangulated arbitrarily. Notice that each subgraph of $T_i$, shown in
either light or dark gray, requires at least one among its edges or
diagonals to be in $D_i$ in order to be 2-dominated. Consider, for
example, the quadrilateral $v_0v_1v_2v_3$ of $T_3$ \new{(the situation
  for all other subgraphs shown in light gray is analogous, whereas the
  subgraphs shown in dark gray have at least as many vertices as those
  shown in light gray, and, thus, could not possibly be 2-dominated
  with fewer diagonal guards with respect to the subgraphs shown in
  light gray)}. Even if both $v_0$
and $v_3$ belong to $D_3$ due to edges or diagonals of the neighboring
shaded subgraphs\new{, or due to diagonals of the central part of $T_3$},
the triangle $v_0v_1v_2$ is not 2-dominated unless either one of the
edges $e_0$, $e_1$, $e_2$, or the diagonal $d_{02}$ belongs to
$D_3$. This observation immediately establishes a lower bound of
$\lfloor\frac{n}{3}\rfloor$.

Let us now assume that $|D_3|=\lfloor\frac{n}{3}\rfloor$. Notice that,
under this assumption, each shaded subgraph in $T_3$ must have
\textsl{exactly one} among its edges or diagonals in $D_3$. Moreover,
none of the diagonals in the central part of $T_3$ (not shown in
Fig. \ref{fig:trg-diag-lb}(bottom)) can belong to $D_3$, since then
the size of $D_3$ would be greater than $\lfloor\frac{n}{3}\rfloor$.
Consider the triangulated hexagon
$H:=v_0v_{3m-3}v_{3m-2}v_{3m-1}v_{3m}v_{3m+1}$. In order for $H$ to be
2-dominated with exactly one of its edges or diagonals, both $v_0$ and
$v_{3m-3}$ have to be in $D_3$ due to edges or diagonals in the
neighboring shaded subgraphs, while the unique edge or diagonal of $H$
in $D_3$ must be the diagonal $d_{3m-2,3m}$. Since we require that
$v_{3m-3}$ must belong to $D_3$ via an edge or diagonal of the
quadrilateral $v_{3m-6}v_{3m-5}v_{3m-4}v_{3m-3}$, and at the same time
we require that exactly one of the edges or diagonals of
$v_{3m-6}v_{3m-5}v_{3m-4}v_{3m-3}$ to be in $D_3$, the edge $e_{3m-4}$ must
belong to $D_3$ and $v_{3m-6}$ must be in $D_3$ due to an edge or
diagonal in the quadrilateral $v_{3m-9}v_{3m-8}v_{3m-7}v_{3m-6}$. 
Cascading this argument, we conclude that, since $v_3$ must belong to
$D_3$ due to an edge or diagonal of the quadrilateral $v_0v_1v_2v_3$,
and at the same time exactly one of the edges or diagonals of
$v_0v_1v_2v_3$ must be in $D_3$, the edge $e_2$ must belong to $D_3$
and $v_0$ must belong to $D_3$ due to an edge or diagonal in the
hexagon $H$. But this yields a contradiction, since the unique
edge or diagonal of $H$ in $D_3$ is $d_{3m-2,3m}$, which is not
incident to $v_0$. Hence $T_3$ requires $\bm$ diagonal guards in
order to be \guard[ed].
\end{proof}


\subsection{Computing diagonal 2-dominating sets}
\label{sec:compute-diag-dsets}

The proof of Theorem \ref{thm:guard-combdiag-trgraph} can almost
immediately be transformed into an $O(n)$ time and space algorithm. The
triangulation graph $\trg{P}$ of $P$ is assumed to be represented via
a half-edge representation. Half-edges and vertices in our
representation are assumed to have additional flags for indicating
whether a half-edge is a boundary edge of the polygon, or whether a
half-edge or a vertex of $\trg{P}$ is marked as being in the diagonal
2-dominating set of $\trg{P}$. Under these assumptions, adding or
removing a half-edge or a vertex from the sought-for 2-dominating set,
querying a half-edge or a vertex for membership in the 2-dominating
set, as well as forming the triangulation graph for the recursive
calls, all take $O(1)$ time.

\begin{figure}[!b]
  \begin{center}
    \psfrag{u}[][]{}
    \psfrag{d}[][]{\scriptsize$d$}
    \includegraphics[width=\columnwidth]{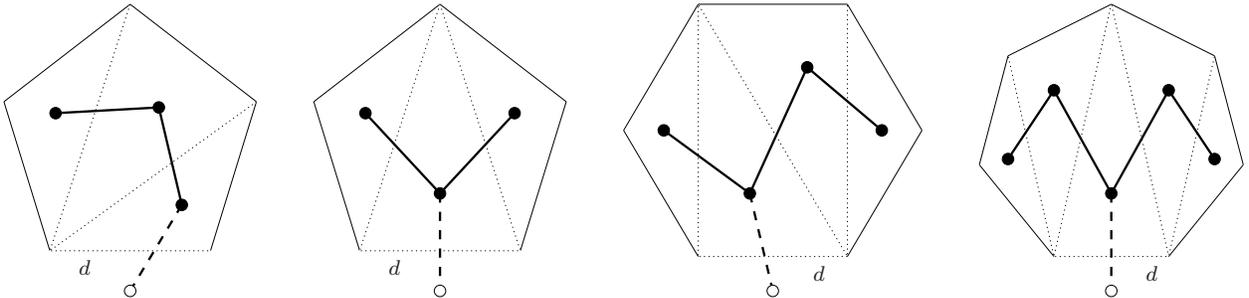}
  \end{center}
  \caption{The four possible configurations for the dual trees
    $\Delta_1$ for $4\le{}k\le{}6$, shown as thick solid lines. The
    diagonal $d$ separates $T_1$ from $T_2$. The triangulations shown
    are indicative: all other triangulations yield
    isomorphic trees.}
  \label{fig:dualtrees_diag}
\end{figure}

Consider a diagonal $d$ that separates $\trg{P}$ into two
triangulation graphs $T_1$ and $T_2$, where $T_1$ contains $k=4,5$
or $6$ edges of $P$; recall from the proof of Lemma
\ref{lem:diag_existence} (for $\lambda=4$) that the value of $k$ is
minimal. Let $\Delta$ be the dual tree of $\trg{P}$, $\Delta_1$ the
dual tree of $T_1$ and $\Delta_1'=\Delta_1\cup\{d'\}$, where $d'$ is
the dual edge of $d$ in $\Delta$.
$\Delta_1$ consists of a subtree of $\Delta$ with 2, 3 or 4
edges of $\Delta$, connected with the rest of $\Delta$ via a degree-2
or a degree-3 node (see Fig. \ref{fig:dualtrees_diag}). Moreover, for
$n\ge{}13$, the subtrees $\Delta_1'$ corresponding to different
diagonals $d$ of $\trg{P}$ must be edge disjoint (otherwise the number
of vertices of $P$ would be less than 13).

Having made these observations we can now describe the algorithm for
computing the diagonal 2-dominating set $D$ for
$\trg{P}$. We first describe the initialization steps:
\begin{enumerate}
\item
  Initialize $D$ to be empty.
\item
  Create a queue $Q$, and initialize it to be empty. $Q$ will consist
  of diagonals of $\trg{P}$.
\item
  For each diagonal $d$ of $\trg{P}$ determine whether it separates off
  \new{$k$ edges of $P$ in $\trg{P}$, with $4\le{}k\le{}6$ and $k$ being
  minimal}. In other words, determine if the dual edge $d'$ of $d$ in
  $\Delta$ is adjacent to subtrees of the form shown in
  Fig. \ref{fig:dualtrees_diag}. If so, put $d$ in $Q$.
\end{enumerate}

The recursive part of the algorithm is as follows:
\begin{enumerate}
\item If the number of vertices of $\trg{P}$ is less than 13, find a
  diagonal 2-dominating set $D$ and return.
\item
  If $Q$ is not empty:
  \begin{enumerate}
  \item
    Pop a diagonal $d$ out of $Q$.
  \item
    If $T_2$ has less than 13 vertices, empty the queue $Q$ and find a
    2-dominating set $D_2$ for $T_2$. Based on $D_2$, and according to
    the cases in the proof of Theorem
    \ref{thm:guard-combdiag-trgraph}, compute $D$ and return.
  \item
    \new{Using the cases in the proof of Theorem
    \ref{thm:guard-combdiag-trgraph}, determine the triangulation
    graph $\hat{T}$} for which we are supposed to find the 2-dominating
    set recursively, and let $\hat{\Delta}$ be the dual tree of
    $\hat{T}$. Let $V$ be the set of vertices in
    $\hat{\Delta}\cap{}\Delta_1'$. For any $v\in{}V$
    determine if $v$ is a leaf-node to a subtree of $\hat{\Delta}$
    like the subtrees in Fig. \ref{fig:dualtrees_diag}. If so, add the
    \new{appropriate} diagonal to $Q$. \new{Neither one of the trees
      $\hat{\Delta}$ and $\hat{\Delta}\cap{}\Delta_1'$, nor the set
      $V$ are computed explicitly; the set $V$ is, in fact, evaluated
      using the cases in the proof of Theorem
      \ref{thm:guard-combdiag-trgraph} without computing
      $\hat{\Delta}\cap{}\Delta_1'$.}
  \item
    Recursively, find a diagonal 2-dominating $\hat{D}$ for $\hat{T}$,
    using $Q$ as the queue.
  \item
    Construct from $\hat{D}$ a diagonal 2-dominating set $D$ for
    $\trg{P}$ and return.
  \end{enumerate}
\end{enumerate}

The initialization part of our algorithm takes linear time, since
Step 2 of the initialization takes constant time per diagonal. 
Let $T(n)$ be the time spent for the recursive part of our algorithm.
Step 1 of the recursive part obviously takes constant time. Step 2
of the recursive part takes $T(n-3)+O(1)$ time. Let us be more
precise. Popping a diagonal from $Q$ takes $O(1)$ time. Step 2(b)
takes $O(1)$ \new{time} since we need to solve our problem for a constant value
of $n$. Determining the case for $d$ takes $O(1)$ time. $V$ has
constant size \new{and can be computed in constant time}, while checking for
new diagonals to be added to the queue $Q$, as well as adding them to
$Q$  also takes $O(1)$ time. Therefore, Step 2(c) costs $O(1)$
time. Step 2(d) is the recursive call, so it takes $T(n-3)$
time. Clearly, Step 2(e) takes $O(1)$ time, since constructing $D$ is
a matter \new{of} updating some flags.

From the analysis above we conclude that the cost $T(n)$ for the
recursive part of our algorithm satisfies the recursive relation

\[
  T(n) = \begin{cases}
    T(n-3)+O(1),& n\ge{}13\\
    O(1),&3\le{}n\le{}12
    \end{cases}
\]
which yields $T(n)=O(n)$. Since initialization takes linear
time, and our space requirements are obviously linear in the size of
$P$ (we do not duplicate parts of $\trg{P}$ for the recursive calls,
but rather set appropriately the boundary flags for some half-edges),
we arrive at the following theorem.

\begin{theorem}\label{thm:trg-diag-timespace}
Given the triangulation graph $\trg{P}$ of a polygon $P$ with
$n\ge{}3$ vertices, we can compute a diagonal 2-dominating set for
$\trg{P}$ of size at most $\bm$ in $O(n)$ time and space.
\end{theorem}


\section{2-dominance of triangulation graphs: edge guards}
\label{sec:trg-edge-guards}

Let $\trg{P}$ be a triangulation graph of a polygon $P$, and let $u$ and
$v$ be two nodes of $\trg{P}$ connected via an edge $e$. The
\emph{contraction} of $e$ is a transformation that removes the nodes
$u$ and $v$ and replaces them with a new node $x$, that is adjacent to
every node that $u$ and $v$ was adjacent to.
The contraction transformation can be used to prove the following
lemma, which is the analogue of Lemma 3.2 in \cite{o-agta-87} in the
context of 2-dominance.

\begin{lemma}\label{lem:contraction-gset}
Suppose $f(n)$ diagonal (\resp edge) guards are always sufficient to
2-dominate an $n$-node triangulation graph. If $\trg{P}$ is an
arbitrary triangulation graph of a polygon $P$, $v$ any vertex of $P$
and $e$ any of the two incident edges of $v$, then $\trg{P}$ can be
2-dominated with $f(n-1)$ diagonal (\resp edge) guards, plus a vertex
guard at $v$. Moreover, $e$, if specified, does not belong to the
2-dominating set of $\trg{P}$.
\end{lemma}

\begin{proof}
Let $u$ be the chosen vertex at which the guard is to be placed. If
the edge $e$ is specified, let $v$ be the node adjacent to $u$ across $e$;
otherwise, let $e$ be any of two the edges of $P$ incident to $u$, and
$v$ the node adjacent to $u$ across $e$. Let $t_e$ be the triangle of
$\trg{P}$ adjacent to $e$ and let $w$ be the third vertex of $t_e$,
besides $u$ and $v$.
Edge contract $\trg{P}$ across $e$, producing the triangulation graph
$\trg{P}'$ of $n-1$ nodes. Since $\trg{P}'$ is a triangulation graph
of a polygon (cf. Lemma 3.1 in \cite{o-agta-87}), it can be 2-dominated
by $f(n-1)$ diagonal (\resp edge) guards.

Let $x$ be the node of $\trg{P}'$ that replaced $u$ and $v$, and let
$D'$ be the 2-dominating set of $\trg{P}'$ consisting of $f(n-1)$
diagonal (\resp edge) guards. Suppose that no guard is placed at $x$,
that is $x$ is not an endpoint of a edge or diagonal (\resp edge) in
$D'$. Then $D=D'\cup\{u\}$ is a dominating set for $\trg{P}$, since
the guard at $u$ dominates $t_e$, whereas the remaining triangles of
$\trg{P}$ are dominated by edges or diagonals (resp. edges) in
$D'$. Moreover, every triangle in $\trg{P}$, except the triangles
adjacent to $u$ or $v$, has two of its vertices in $D'$, and thus in
$D$. Since $x$ is not in $D'$, all the vertices of $\trg{P}'$ adjacent
to $x$ have to be in $D'$. Hence, all triangles adjacent to $u$ or
$v$, except $t_e$ have two of their vertices in $D'$ and thus in
$D$. Finally, $t_e$ has also two vertices in $D$, namely $u$ and
$w$. Thus, $D$ is a 2-dominating set for $\trg{P}$.

Suppose now that a guard is used at $x$ in $D'$. If $xw$ is an edge or
diagonal guard in $D'$, assign $xw$ to $vw$. Every other edge or
diagonal guard $g$ in $D'$ incident to $x$, if any, becomes an edge or
diagonal guard in $D$, incident to either $u$ or $v$, depending on
whether $g$ is incident to $u$ or $v$ in $\trg{P}$. As in the previous
case, every triangle in $\trg{P}$ is dominated and has at least
two of its vertices in $D$. More precisely, every triangle in $\trg{P}'$ not
containing $x$ has two of its vertices in $D'$ and thus in $D$. Every
triangle $t'$ in $\trg{P}'$ containing $x$ is now a triangle in $\trg{P}$
containing either $u$ or $v$ or both (this is the case for $t_e$). 
Therefore every triangle in $\trg{P}$, except $t_e$, that contains $u$
or $v$ has one vertex in $D'$ plus either $u$ or $v$. Clearly, $t_e$
has both $u$ and $v$ in $D$.
\end{proof}

Before proceeding with the first main result of this section, let us
state and prove an intermediate lemma concerning edge 2-dominating
sets for small values of $n$.

\begin{lemma}\label{lem:guard-combedge-smallpoly}
Every triangulation graph $\trg{P}$ with $3\le{}n\le{}9$ vertices,
corresponding to a polygon $P$, can be 2-dominated by
$\be$ edge guards, except for $n=4$, where one additional guard is
required.
\end{lemma}

\begin{proof}
Let $v_i$, $1\le{}i\le{}n$ be the vertices of $\trg{P}$, and let $e_i$
be the edge $v_iv_{i+1}$. For each value of $n$ we are going to define
an edge 2-dominating set $D$ of size $\be$.
\begin{mathdescription}
\item[n\in\{3,4,5,7\}.] Set $D$ to be the set of edges of $P$ with odd
  index.
\item[n=6.] See proof of Lemma \ref{lem:guard-combdiag-smallpoly}.
\item[n=8.] Let $t_1$ and $t_2$ be two ears of $\trg{P}$ and
  consider their relative positions as shown in Fig. 
  \ref{fig:ears_edges_n=8}. In each case define the set $D$ as shown
  in Fig. \ref{fig:ears_edges_n=8}.
  \begin{figure}[!hpt]
    \psfrag{t1}[][]{\scriptsize$t_1$}
    \psfrag{t2}[][]{\scriptsize$t_2$}
    \begin{center}
      \includegraphics[width=0.7\columnwidth]{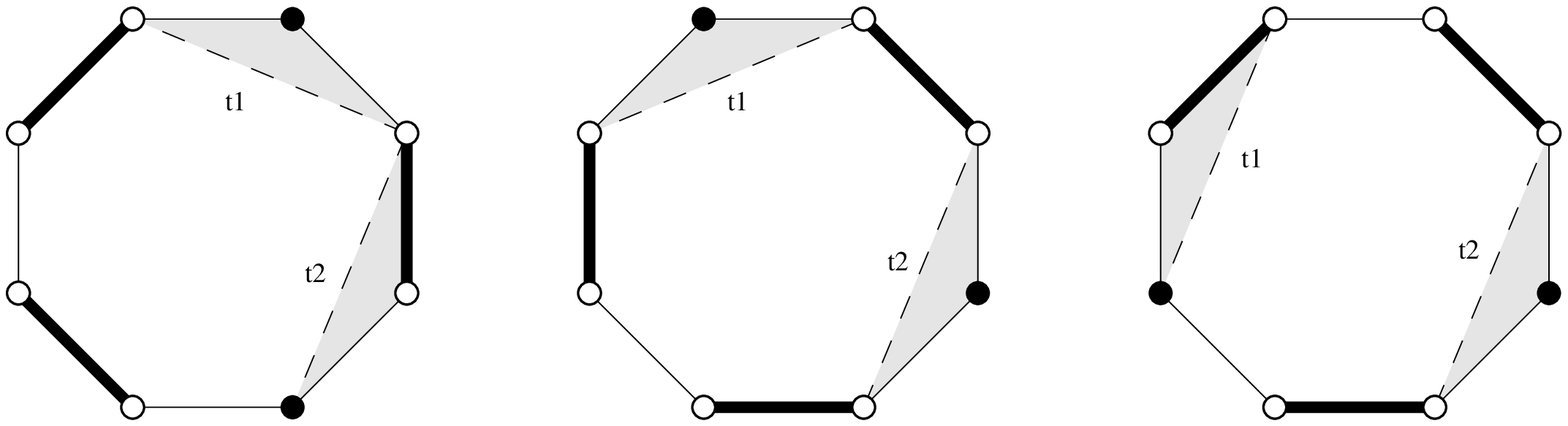}
      \caption{Proof of Lemma \ref{lem:guard-combedge-smallpoly} for
        $n=8$. \new{The shaded triangles $t_1$ and $t_2$ are two ears of
          $\trg{P}$. The subfigures correspond to the three possible
          relative positions of $t_1$ and $t_2$ in $\trg{P}$.}}
      \label{fig:ears_edges_n=8}
    \end{center}
    \begin{center}
      \psfrag{t}[][]{\scriptsize$t$}
      \psfrag{tp}[][]{\scriptsize$t'$}
      \psfrag{tpp}[][]{\scriptsize$t''$}
      \psfrag{0}[][]{\scriptsize$v_0$}
      \psfrag{1}[][]{\scriptsize$v_1$}
      \psfrag{2}[][]{\scriptsize$v_2$}
      \psfrag{3}[][]{\scriptsize$v_3$}
      \psfrag{4}[][]{\scriptsize$v_4$}
      \psfrag{5}[][]{\scriptsize$v_5$}
      \psfrag{6}[][]{\scriptsize$v_6$}
      \psfrag{7}[][]{\scriptsize$v_7$}
      \psfrag{8}[][]{\scriptsize$v_8$}
      \includegraphics[width=0.87\columnwidth]{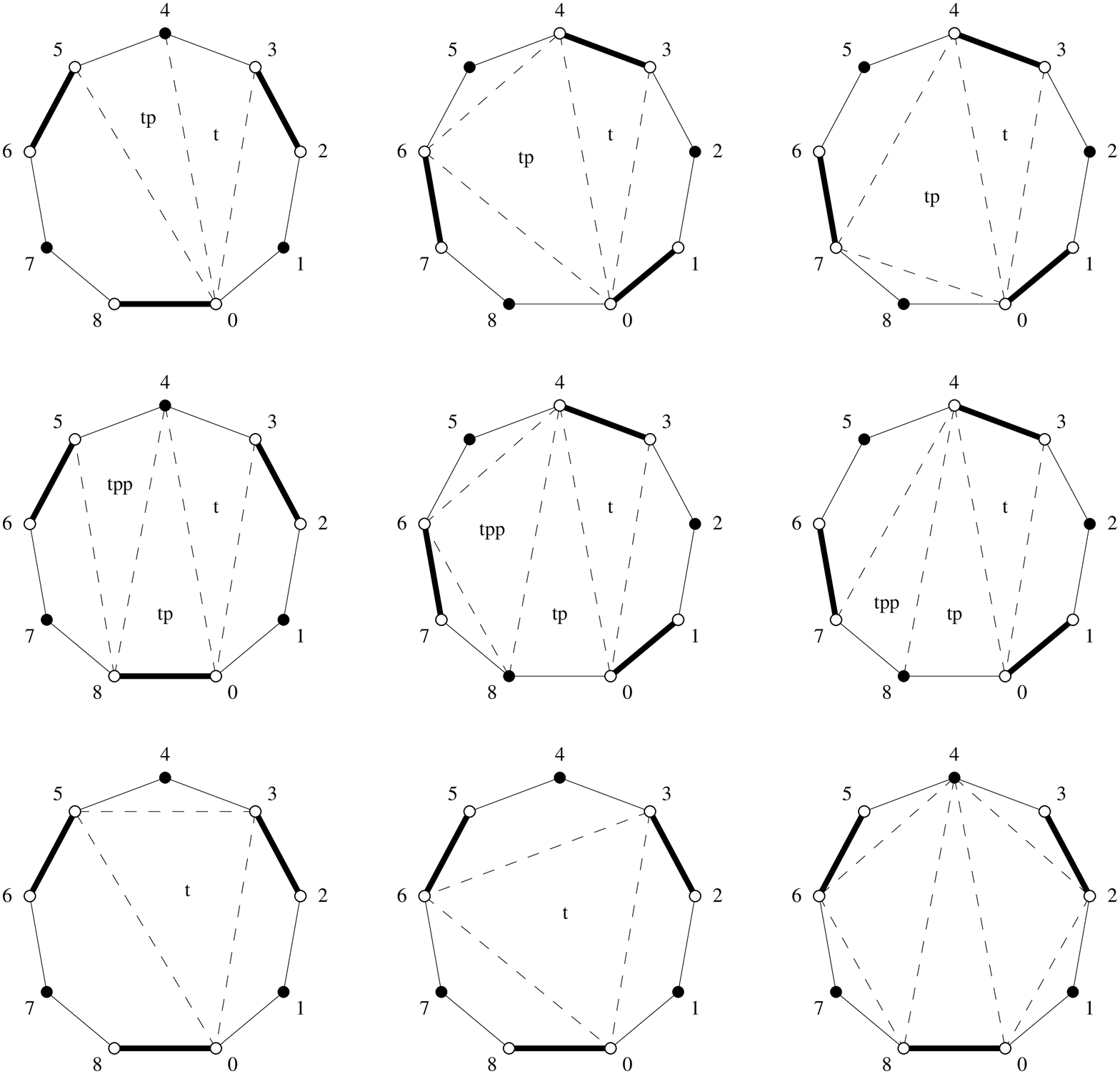}
      \caption{Proof of Lemma \ref{lem:guard-combedge-smallpoly} for
        $n=9$. 
        \new{Top row (left to right):
          $k=3$, $v\equiv{}v_4$ and $v'\equiv{}v_5$;
          $k=3$, $v\equiv{}v_4$ and $v'\equiv{}v_6$;
          $k=3$, $v\equiv{}v_4$ and $v'\equiv{}v_7$.
          Middle row (left to right):
          $k=3$, $v\equiv{}v_4$, $v'\equiv{}v_8$ and $v''\equiv{}v_5$;
          $k=3$, $v\equiv{}v_4$, $v'\equiv{}v_8$ and $v''\equiv{}v_6$;
          $k=3$, $v\equiv{}v_4$, $v'\equiv{}v_8$ and $v''\equiv{}v_7$.
          Bottom row (left to right):
          $k=3$ and $v\equiv{}v_5$;
          $k=3$ and $v\equiv{}v_6$; 
          $k=4$.}}
      \label{fig:ears_edges_n=9}
    \end{center}
  \end{figure}
\item[n=9.] Since $n\ge{}6$, by means of Lemma
  \ref{lem:diag_existence} with $\lambda=3$, there exists diagonal $d$
  that partitions $\trg{P}$ into two triangulation graphs $T_1$ and
  $T_2$, where $T_1$ contains $k$ boundary edges of $\trg{P}$,
  $3\le{}k\le{}4$. Let $d\equiv{}d_{0k}$ be the common edge of $T_1$ and
  $T_2$, where $d_{ij}$ denotes the diagonal $v_iv_j$. Consider each
  of the two values of $k$ separately (see also
  Fig. \ref{fig:ears_edges_n=9}):
  \begin{mathdescription}
  \item[k=3.] Let $t$ be the triangle adjacent to the diagonal
    $d_{03}$ in $T_2$ and let $v$ be its apex. The cases
    $v\equiv{}v_4$, $v\equiv{}v_8$ and $v\equiv{}v_5$, $v\equiv{}v_7$
    are symmetric, so we only need to consider the cases
    $v\in\{v_4,v_5,v_6\}$:
    \begin{mathdescription}
    \item[v\equiv{}v_4.] Let $t'$ be the triangle incident to $d_{04}$
      in the hexagon $v_0v_4v_5v_6v_7v_8$, and let $v'$ be its
      apex. Consider the subcases:
      \begin{mathdescription}
      \item[v'\equiv{}v_5.] Set $D=\{e_2,e_5,e_8\}$.
      \item[v'\in\{v_6,v_7\}.] Set $D=\{e_0,e_3,e_6\}$.
      \item[v'\equiv{}v_8.] Let $t''\ne{}t'$ be the triangle supported
        by $d_{48}$ and let $v''$ be its apex. If $v''\equiv{}v_5$,
        set $D=\{e_2,e_5,e_8\}$. Otherwise, if
        $v''\in\{v_6,v_7\}$, set $D=\{e_0,e_3,e_6\}$.
      \end{mathdescription}
    \item[v\in\{v_5,v_6\}.] Set $D=\{e_2,e_5,e_8\}$.
    \end{mathdescription}
  \item[k=4.] By the minimality of $k$, the apex of the triangle
    supported by $d_{04}$ in $T_1$ must be $v_2$. Again, by the
    minimality of $k$, the diagonals $d_{47}$, $d_{58}$ and $d_{06}$
    cannot exist. This implies that either $d_{48}$ or $d_{05}$ must
    belong to $\trg{P}$. The two cases are symmetric, so we can
    assume, without loss of generality, that $d_{48}\in{}\trg{P}$. 
    Again, by the minimality of $k$, the diagonals $d_{46}$ and
    $d_{68}$ must be in $\trg{P}$. In this case set
    $D=\{e_2,e_5,e_8\}$. \qedhere
  \end{mathdescription}
\end{mathdescription}
\end{proof}

\newbegin
In the next two theorems we state and prove the first two main results
of this section concerning worst-case upper and lower bounds on the
number of edge guards required in order to 2-dominate a triangulation
graph.
\newend

\begin{theorem}\label{thm:guard-combedge-trgraph}
Let $P$ be a polygon with $n\ge{}3$ vertices and $\trg{P}$ its
triangulation graph. $\be$ edge guards are always sufficient
\new{in order to 2-dominate $\trg{P}$, except for $n=4$, where one
additional guard is required.}
\end{theorem}

\begin{proof}
In Lemma \ref{lem:guard-combedge-smallpoly}, we have shown the result
for $3\le{}n\le{}9$.
Let us now assume that $n\ge{}10$ and that the theorem holds for all
$n'$ such that $5\le{}n'<n$. By means of Lemma
\ref{lem:diag_existence} with $\lambda=5$, there exists diagonal $d$
that partitions $\trg{P}$ into two triangulation graphs $T_1$ and
$T_2$, where $T_1$ contains $k$ boundary edges of $\trg{P}$,
$5\le{}k\le{}8$. Let $v_0,\ldots,v_k$ be the $k+1$ vertices 
of $T_1$, as we encounter them while traversing $P$
counterclockwise, and let $v_0v_k$ be the common edge of $T_1$
and $T_2$. For each value of $k$ we are going to define an edge
2-dominating set $D$ for $\trg{P}$ of size $\be$. In what follows
$d_{ij}$ denotes the diagonal $v_iv_j$, whereas $e_i$ denotes the edge
$v_iv_{i+1}$. Consider each of the four values of $k$ separately:
\begin{mathdescription}
\item[k=5.] Let $t$ be the triangle supported by $d$ in $T_1$, and let
  $v$ be the apex of this triangle. $|T_2|=n-4$, and by Lemma
  \ref{lem:contraction-gset} there exists a 2-dominating set $D_0$
  (\resp $D_5$) for $T_2$, consisting of $f(n-5)$ edge guards plus
  $v_0$ (\resp $v_5$), such that $d\nin{}D_0$ (\resp $d\nin{}D_5$). 
  If $v\in\{v_3,v_4\}$, set $D=D_0\cup\{e_0,e_3\}$. If $v\in\{v_1,v_2\}$, 
  set $D=D_5\cup\{e_1,e_4\}$ (see Fig. \ref{fig:hexagons}).
  \begin{figure}[!t]
    \begin{center}
      \psfrag{0}[][]{\scriptsize$v_0$}
      \psfrag{1}[][]{\scriptsize$v_1$}
      \psfrag{2}[][]{\scriptsize$v_2$}
      \psfrag{3}[][]{\scriptsize$v_3$}
      \psfrag{4}[][]{\scriptsize$v_4$}
      \psfrag{5}[][]{\scriptsize$v_5$}
      \psfrag{d}[][]{\scriptsize$d$}
      \psfrag{t}[][]{\scriptsize$t$}
      \includegraphics[width=\columnwidth]{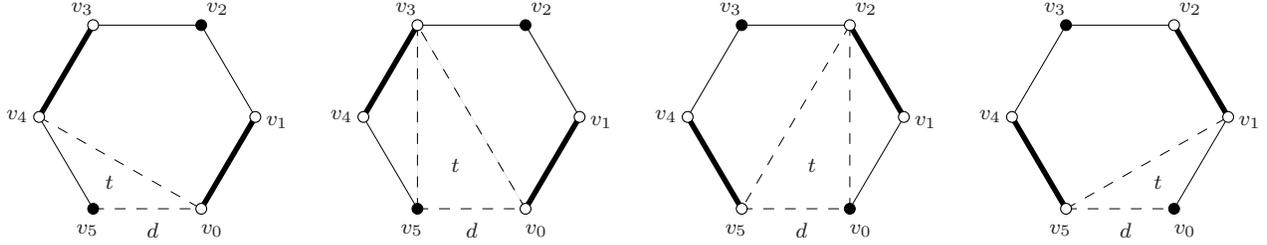}
      \caption{Proof of Theorem \ref{thm:guard-combedge-trgraph}. The case
        $k=5$. Left two: $v\in\{v_3,v_4\}$. Right two:  $v\in\{v_1,v_2\}$.}
      \label{fig:hexagons}
    \end{center}
  \end{figure}
\item[k=6.] The presence of diagonals $d_{05}$ or $d_{16}$
  would violate the minimality of $k$. Let $t$ be the triangle supported
  by $d$ in $T_1$. The apex $v$ of this triangle should be $v_2$, $v_3$
  or $v_4$. The cases $v\equiv{}v_2$ and $v\equiv{}v_4$ are symmetric,
  so we only consider the cases $v\equiv{}v_2$ and $v\equiv{}v_3$.
  Since $T_2$ has $n-5$ vertices, by our induction hypothesis we have
  that $T_2$ can be dominated with $f(n-5)=\be-2$ edge guards. Let
  $D_2$ be the edge 2-dominating set for $T_2$. Consider the following
  cases (see also Fig. \ref{fig:heptagons}):
  \begin{mathdescription}
  \item[d_{06}\in{}D_2.]
    Set $D=(D_2\setminus\{d_{06}\})\cup\{e_0,e_2,e_5\}$.
  \item[d_{06}\nin{}D_2.] Since $D_2$ is a 2-dominating set for $T_2$,
    either $v_0$ or $v_6$ belongs to $D_2$. If $v_0\in{}D_2$, set
    $D=D_2\cup\{e_2,e_4\}$. Otherwise, $v_6\in{}D_2$, in which case
    set $D=D_2\cup\{e_1,e_3\}$.
  \end{mathdescription}
  \begin{figure}[!t]
    \begin{center}
      \psfrag{T1}[][]{\scriptsize$t$}
      \psfrag{T2}[][]{\scriptsize$t$}
      \psfrag{0}[][]{\scriptsize$v_0$}
      \psfrag{1}[][]{\scriptsize$v_1$}
      \psfrag{2}[][]{\scriptsize$v_2$}
      \psfrag{3}[][]{\scriptsize$v_3$}
      \psfrag{4}[][]{\scriptsize$v_4$}
      \psfrag{5}[][]{\scriptsize$v_5$}
      \psfrag{6}[][]{\scriptsize$v_6$}
      \psfrag{d}[][]{\scriptsize$d$}
      \includegraphics[width=\columnwidth]{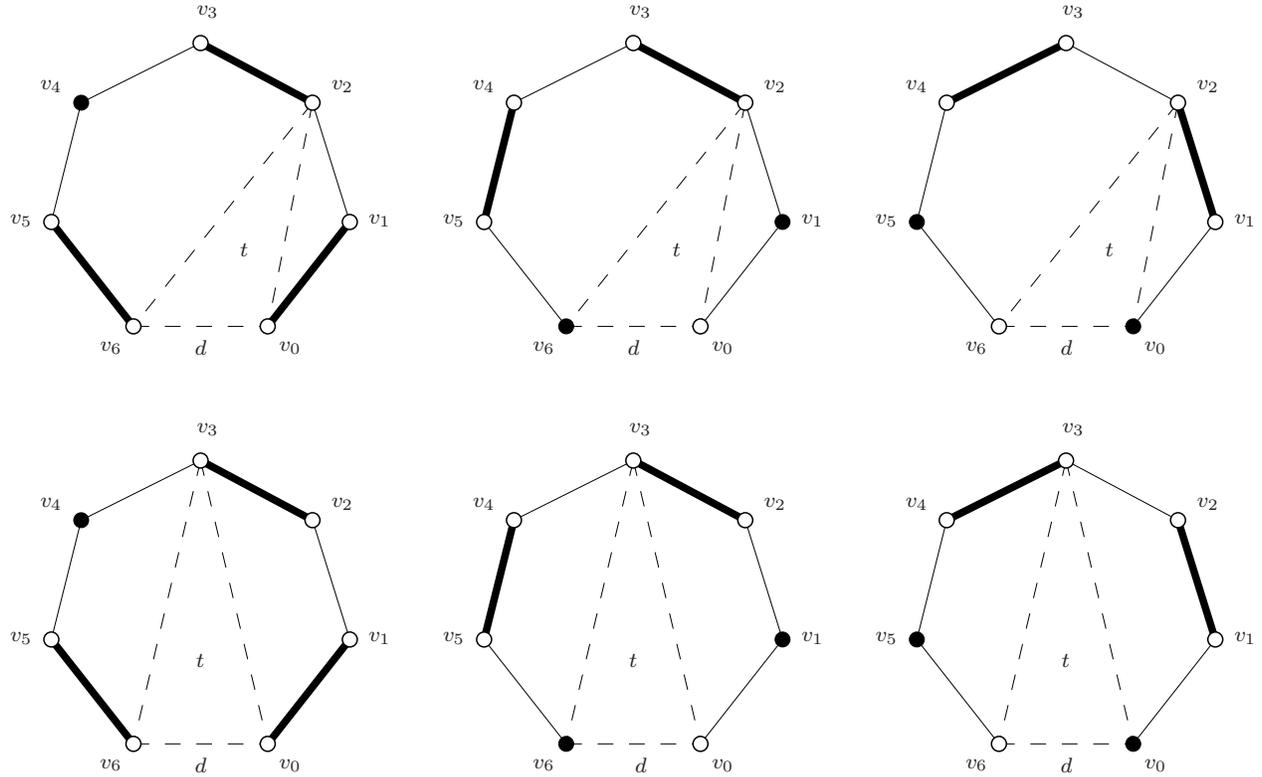}
      \caption{Proof of Theorem \ref{thm:guard-combedge-trgraph}. The case
        $k=6$. Top row: the apex of $t$ is $v_2$. Bottom row: the apex of
        $t$ is $v_3$. Left column: $d_{06}\in{}D_2$. Middle column:
        $d_{06}\nin{}D_2,v_0\in{}D_2$. Right column:
        $d_{06}\nin{}D_2,v_6\in{}D_2$.}
      \label{fig:heptagons}
    \end{center}
  \end{figure}
  \begin{figure}[!t]
    \begin{center}
      \psfrag{d}[][]{\scriptsize$d$}
      \psfrag{T2}[][]{\scriptsize$T_2$}
      \psfrag{t}[][]{\scriptsize$t$}
      \psfrag{0}[][]{\scriptsize$v_0$}
      \psfrag{1}[][]{\scriptsize$v_1$}
      \psfrag{2}[][]{\scriptsize$v_2$}
      \psfrag{3}[][]{\scriptsize$v_3$}
      \psfrag{4}[][]{\scriptsize$v_4$}
      \psfrag{5}[][]{\scriptsize$v_5$}
      \psfrag{6}[][]{\scriptsize$v_6$}
      \psfrag{7}[][]{\scriptsize$v_7$}
      \includegraphics[width=0.7\columnwidth]{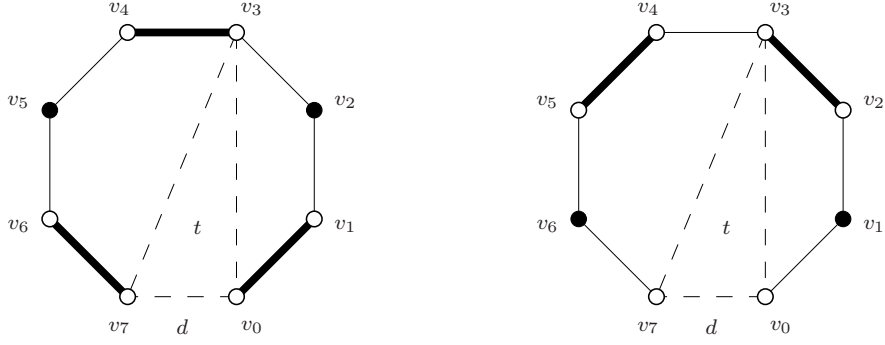}
      \caption{Proof of Theorem \ref{thm:guard-combedge-trgraph}. The case
        $k=7$. Left: $|D'\cap\{d_{03},d_{37}\}|\ge{}1$. Right:
        $d_{03},d_{37}\nin{}D'$.}
      \label{fig:case_k7}
    \end{center}
  \end{figure}
\item[k=7.] The presence of diagonals $d_{06}$,
  $d_{05}$, $d_{17}$ or $d_{27}$ would violate the minimality of
  $k$. Let $t$ be the triangle supported by $d$ in $T_1$. The apex $v$
  of this triangle is either $v_3$ or $v_4$. The two cases are
  symmetric, so we can assume without loss of generality that the apex
  of $t$ is $v_3$ (see Fig. \ref{fig:case_k7}).
  Consider the triangulation graph $T'=T_2\cup\{t\}$. It
  has $n-5$ vertices and, by our induction hypothesis, it can be
  2-dominated with $f(n-5)=\be-2$ edge guards. Let $D'$ be the
  2-dominating set of $T'$. Consider the following two cases:
  \begin{mathdescription}
  \item[|D'\cap\{d_{03},d_{37}\}|\ge{}1.]
    Set $D=(D'\setminus\{d_{03},d_{37}\})\cup\{e_0,$ $e_3,e_6\}$.
  \item[d_{03},d_{37}\nin{}D'.]
    In this case $v_3$ cannot be in $D'$, since either $d_{03}$ or
    $d_{37}$ would have to be in $D'$. This implies that both $v_0$
    and $v_7$ have to be in $D'$ (2-dominance of $t$). Set
    $D=D'\cup\{e_2,e_4\}$.
  \end{mathdescription}
  \begin{figure}[!tp]
    \begin{center}
      \psfrag{d}[][]{\scriptsize$d$}
      \psfrag{Tp1}[][]{\scriptsize$T'$}
      \psfrag{Tp2}[][]{\scriptsize$T''$}
      \psfrag{T2}[][]{\scriptsize$T_2$}
      \psfrag{t}[][]{\scriptsize$t$}
      \psfrag{tp}[][]{\scriptsize$t'$}
      \psfrag{0}[][]{\scriptsize$v_0$}
      \psfrag{1}[][]{\scriptsize$v_1$}
      \psfrag{2}[][]{\scriptsize$v_2$}
      \psfrag{3}[][]{\scriptsize$v_3$}
      \psfrag{4}[][]{\scriptsize$v_4$}
      \psfrag{5}[][]{\scriptsize$v_5$}
      \psfrag{6}[][]{\scriptsize$v_6$}
      \psfrag{7}[][]{\scriptsize$v_7$}
      \psfrag{8}[][]{\scriptsize$v_8$}
      \includegraphics[width=\columnwidth]{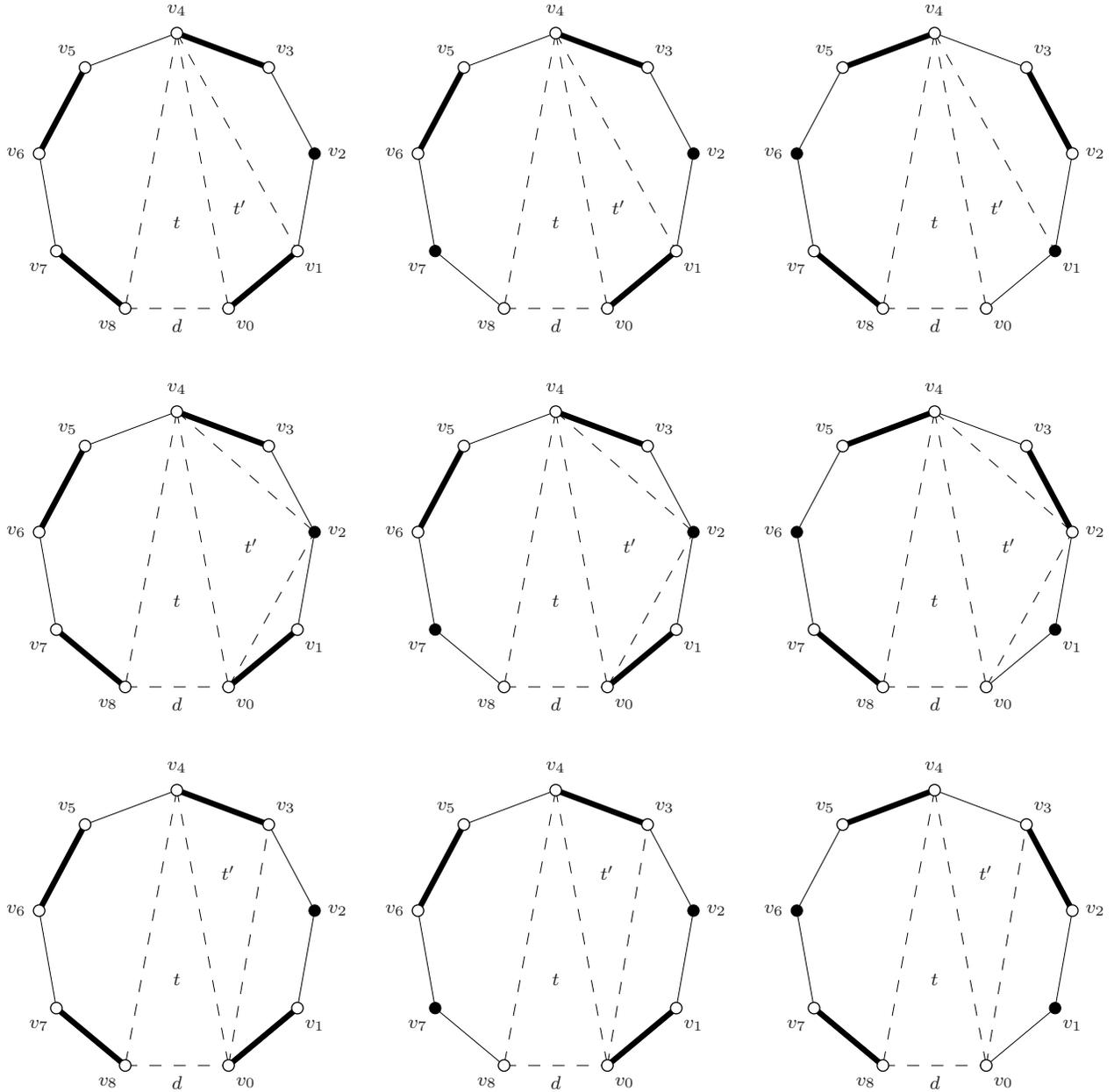}
      \caption{Proof of Theorem \ref{thm:guard-combedge-trgraph}. The case
        $k=8$. 
        Rows (top to bottom): $v'\equiv{}v_1$; $v'\equiv{}v_2$;
        $v'\equiv{}v_3$.
        Top row (left to right): $d_{14},d_{48}\in{}D'$;
        $d_{14}\in{}D',d_{48}\nin{}D'$, $v_8\in{}D'$, \new{and also
          $d_{14},d_{48}\nin{}D'$}; 
        $d_{14}\in{}D',d_{48}\nin{}D'$, $v_0\in{}D'$, and also
        $d_{14}\nin{}D',d_{48}\in{}D'$.
        Middle row (left to right): $|\{d_{02},d_{24},d_{48}\}\cap{}D'|\ge{}2$;
        $d_{02}\in{}D',d_{24},d_{48}\nin{}D'$, and also 
        $d_{24}\in{}D',d_{02},d_{48}\nin{}D'$, $v_8\in{}D'$;
        $d_{24}\in{}D',d_{02},d_{48}\nin{}D'$, $v_0\in{}D'$, and also
        $d_{48}\in{}D',d_{02},d_{24}\nin{}D'$.
        Bottom row (left to right): $d_{03},d_{48}\in{}D'$, and also
        $d_{03}\in{}D',d_{48}\nin{}D'$, $e_3\in{}D'$, as well as
        $d_{03}\nin{}D',d_{48}\in{}D'$, $e_3\in{}D'$;
        $d_{03}\in{}D',d_{48}\nin{}D'$, $e_3\nin{}D'$, and also
        $d_{03},d_{48}\nin{}D'$, $v_8\in{}D'$;
        $d_{03}\nin{}D',d_{48}\in{}D'$, $e_3\nin{}D'$, and also
        $d_{03},d_{48}\nin{}D'$, $v_0\in{}D'$.}
      \label{fig:case_k8}
    \end{center}
  \end{figure}
\item[k=8.] The presence of diagonals $d_{07}$, $d_{06}$, $d_{05}$,
  $d_{18}$, $d_{28}$ or $d_{38}$ would violate the minimality of
  $k$. Thus, the apex of the triangle $t$ in $T_1$ that 
  is supported by $d$ is $v_4$. Let $t'\ne{}t$ be the triangle
  \new{incident to $d_{04}$}, and let $v'$ be its vertex opposite
  $d_{04}$. Clearly, $v'\in\{v_1,v_2,v_3\}$. Consider the
  triangulation graph $T'=T_2\cup\{t,t'\}$. It has $n-5$ vertices and,
  by our induction hypothesis, it can be 2-dominated with
  $f(n-5)=\be-2$ edge guards. Let $D'$ be the 2-dominating set of
  $T'$. Consider the following cases (see also
  Fig. \ref{fig:case_k8}):
  \begin{mathdescription}
  \item[v'\equiv{}v_1.] Consider the following subcases:
    \begin{mathdescription}
    \item[d_{14},d_{48}\in{}D'.] Set
      $D=(D'\setminus\{d_{14},d_{48}\})\cup\{e_0,e_3,$ $e_5,e_7\}$.
    \item[d_{14}\in{}D',d_{48}\nin{}D'.] If $v_8\in{}D'$, set
      $D=(D'\setminus\{d_{14}\})$ $\cup\{e_0,e_3,\new{e_5}\}$. Otherwise,
      $v_0\in{}D'$ (2-dominance of $t$), in which case
      set $D=(D'\setminus\{d_{14}\})\cup\{e_2,e_4,e_7\}$. 
    \item[d_{14}\nin{}D',d_{48}\in{}D'.] In this case either $v_0$ or
      $v_1$ belongs to $D'$ (2-dominance of $t'$). Since
      $d_{14}\nin{}D'$, we must have that either $v_0\in{}D'$ or
      $\new{e_0}\in{}D'$, which implies, in either case, that
      $v_0\in{}D'$. Hence, set
      $D=(D'\setminus\{d_{48}\})\cup\{e_2,e_4,e_7\}$.
    \item[d_{14},d_{48}\nin{}D'.] In this case $v_4\nin{}D'$, which
      implies that $v_0,v_1,v_8\in{}D'$. But then
      $e_0\in{}D'$. Therefore, set $D=D'\cup\{e_3,e_5\}$.
    \end{mathdescription}
  \item[v'\equiv{}v_2.] Notice that in this case it is not possible
    that $d_{02},d_{24},d_{48}\nin{}D'$, since then $v_2,v_4\nin{}D'$,
    which contradicts the 2-dominance of $t'$ by $D'$ in
    $T'$. Consider the remaining subcases:
    \begin{mathdescription}
    \item[|\{d_{02},d_{24},d_{48}\}\cap{}D'|\ge{}2.] Set
      $D=(D'\setminus\{d_{02},d_{24},$ $d_{48}\})\cup\{e_0,e_3,$ $e_5,e_7\}$.
    \item[d_{02}\in{}D',d_{24},d_{48}\nin{}D'.] Then $v_4\nin{}D'$,
      which implies that $v_8\in{}D'$ (2-dominance of $t$). Set
      $D=(D'\setminus\{d_{02}\})\cup\{e_0,e_3,e_5\}$.
    \item[d_{24}\in{}D',d_{02},d_{48}\nin{}D'.] $v_0$ or $v_8$
      belongs to $D'$ (2-dominance of $t$). If $v_0\in{}D'$, set
      $D=(D'\setminus\{d_{24}\})\cup\{e_2,e_4,e_7\}$. Otherwise, if
      $v_8\in{}D'$, set $D=(D'\setminus\{d_{24}\})\cup\{e_0,e_3,e_5\}$.
    \item[d_{48}\in{}D',d_{02},d_{24}\nin{}D'.] Then $v_2\nin{}D'$,
      which implies that $v_0\in{}D'$ (2-dominance of $t'$). Set
      $D=(D'\setminus\{d_{48}\})\cup\{e_2,e_4,e_7\}$.
    \end{mathdescription}
  \item[v'\equiv{}v_4.] Consider the following subcases:
    \begin{mathdescription}
    \item[d_{03},d_{48}\in{}D'.]
      Set $D=(D'\setminus\{d_{03},d_{48}\})\cup\{e_0,e_3,$ $e_5,e_7\}$.
    \item[d_{03}\in{}D',d_{48}\nin{}D'.] If $e_3\in{}D'$, set
      $D=(D'\setminus\{d_{03}\})$ $\cup\{e_0,e_5,e_7\}$. Otherwise,
      $v_4\nin{}D'$, \ie both $v_0$ and $v_8$ belong to $D'$. Set
      $D=(D'\setminus\{d_{03}\})\cup\{e_0,e_3,e_5\}$.
    \item[d_{03}\nin{}D',d_{48}\in{}D'.] If $e_3\in{}D'$, set
      $D=(D'\setminus\{d_{48}\})$ $\cup\{e_0,e_5,e_7\}$. Otherwise,
      $v_3\nin{}D'$, \ie $v_0$ belongs to $D'$ (2-dominance of
      $t'$). Set $D=(D'\setminus\{d_{48}\})\cup\{e_2,e_4,e_7\}$.
    \item[d_{03},d_{48}\nin{}D'.]
      Since $d_{03},d_{48}\nin{}D'$, $t'$ can be 2-dominated in
      $D'$ only if $e_3\in{}D'$. Now, if $v_8\in{}D'$, set
      $D=D'\cup\{e_0,e_5\}$; otherwise, \ie if $v_8\nin{}D'$,
      $v_0$ has to be in $D'$, in which case set
      $D=(D'\setminus\{e_3\})\cup\{e_2,e_4,e_7\}$.\qedhere
    \end{mathdescription}
  \end{mathdescription}
\end{mathdescription}
\end{proof}

\newbegin

\begin{theorem}\label{thm:guard-combedge-trgraph-lb}
There exists a family of triangulation graphs with $n\ge{}3$ vertices
any edge 2-dominating set of which has cardinality at least $\be$,
except for $n=4$, where any edge 2-dominating set has cardinality at
least $2$.
\end{theorem}

\begin{proof}
Our claim is trivial for $n\in\{3,4\}$. We are first going to prove
the lower bound for all $n=5m+k$, where $m\ge{}1$ and
$k\in\{0,1,3,4\}$. The case $n=5m+2$, for $m\ge{}1$, is a bit more
complicated and is dealt with separately.

Consider the triangulation graphs $\Gamma_{5m}$, $\Gamma_{5m+1}$,
$\Gamma_{5m+3}$ and $\Gamma_{5m+4}$, $m\ge{}1$, shown in
Fig. \ref{fig:trg-edge-lb}. The central part of these graphs is
triangulated arbitrarily. $\Gamma_{5m+i}$, $i=0,1,3,4$, consists of
$n=5m+i$ vertices, and requires a minimum of two edge guards per
hexagon shown in light gray (this is true even if the two vertices of
these hexagons that also belong to the neighboring shaded polygons are
in the 2-dominating set due to edges of these polygons). Moreover,
$\Gamma_{5m}$ and $\Gamma_{5m+1}$ require two 
more edge guards for the hexagon and heptagon, respectively, shown in
dark gray, whereas $\Gamma_{5m+3}$ and $\Gamma_{5m+4}$ require 
three more edge guards for the enneagon and decagon shown in dark
gray (this is true even if the two vertices of these polygons that
also belong to the neighboring shaded polygons are in the 2-dominating
set due to edges of these polygons). Hence,  $\Gamma_{5m}$,
$\Gamma_{5m+1}$, $\Gamma_{5m+3}$ and $\Gamma_{5m+4}$ require $\be$
edge guards in order to be 2-dominated.

\begin{figure}[!tp]
  \newbegin
  \psfrag{T1}[][]{$\Gamma_{5m}$}
  \psfrag{T2}[][]{$\Gamma_{5m+1}$}
  \psfrag{T4}[][]{$\Gamma_{5m+3}$}
  \psfrag{T5}[][]{$\Gamma_{5m+4}$}
  \psfrag{v0}[][]{\scriptsize$v_0$}
  \psfrag{v1}[][]{\scriptsize$v_1$}
  \psfrag{v2}[][]{\scriptsize$v_2$}
  \psfrag{v3}[][]{\scriptsize$v_3$}
  \psfrag{v4}[][]{\scriptsize$v_4$}
  \psfrag{v5}[][]{\scriptsize$v_5$}
  \psfrag{v6}[][]{\scriptsize$v_6$}
  \psfrag{v7}[][]{\scriptsize$v_7$}
  \psfrag{v8}[][]{\scriptsize$v_8$}
  \psfrag{v9}[][]{\scriptsize$v_9$}
  \psfrag{v5m+3}[][]{\scriptsize$v_{5m+3}$}
  \psfrag{v5m+2}[][]{\scriptsize$v_{5m+2}$}
  \psfrag{v5m+1}[][]{\scriptsize$v_{5m+1}$}
  \psfrag{v5m}[][]{\scriptsize$v_{5m}$}
  \psfrag{v5m-1}[][]{\scriptsize$v_{5m-1}$}
  \psfrag{v5m-2}[][]{\scriptsize$v_{5m-2}$}
  \psfrag{v5m-3}[][]{\scriptsize$v_{5m-3}$}
  \psfrag{v5m-4}[][]{\scriptsize$v_{5m-4}$}
  \psfrag{v5m-5}[][]{\scriptsize$v_{5m-5}$}
  \psfrag{v5m-6}[][]{\scriptsize$v_{5m-6}$}
  \psfrag{v5m-7}[][]{\scriptsize$v_{5m-7}$}
  \psfrag{v5m-8}[][]{\scriptsize$v_{5m-8}$}
  \psfrag{v5m-9}[][]{\scriptsize$v_{5m-9}$}
  \psfrag{v5m-10}[][]{\scriptsize$v_{5m-10}$}
  \begin{center}
    \includegraphics[width=0.45\columnwidth]{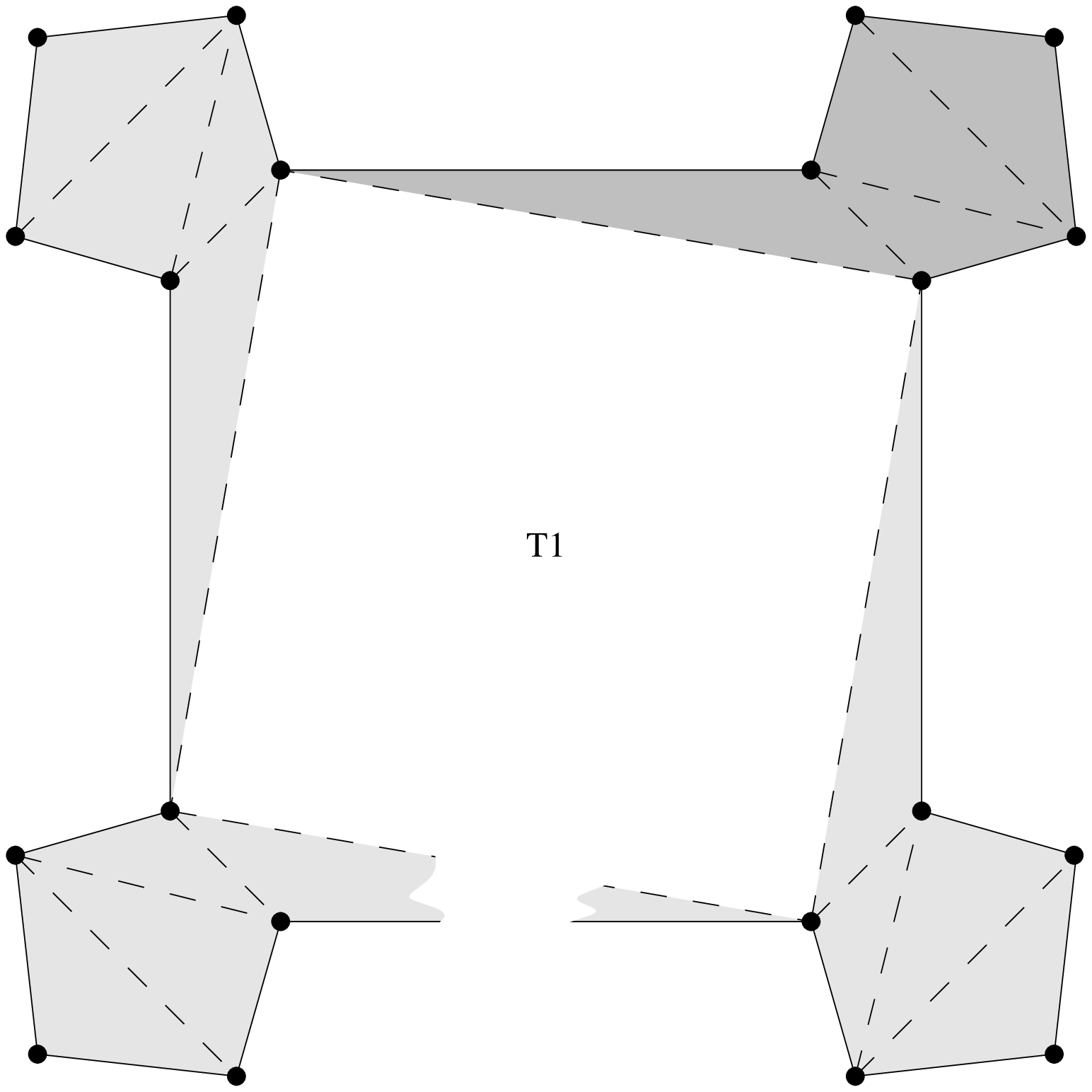}\hfil%
    \includegraphics[width=0.45\columnwidth]{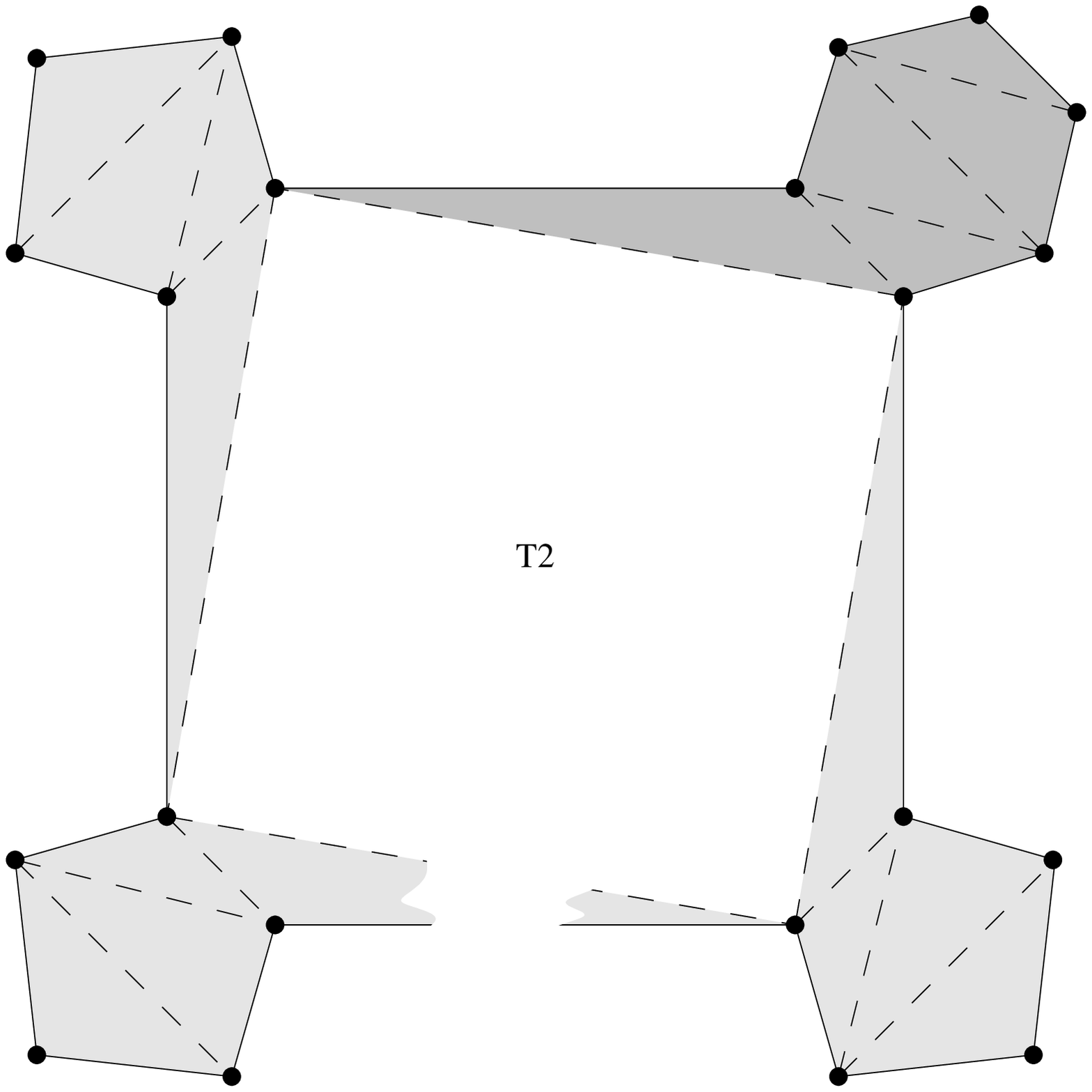}\\[5mm]
    \includegraphics[width=0.45\columnwidth]{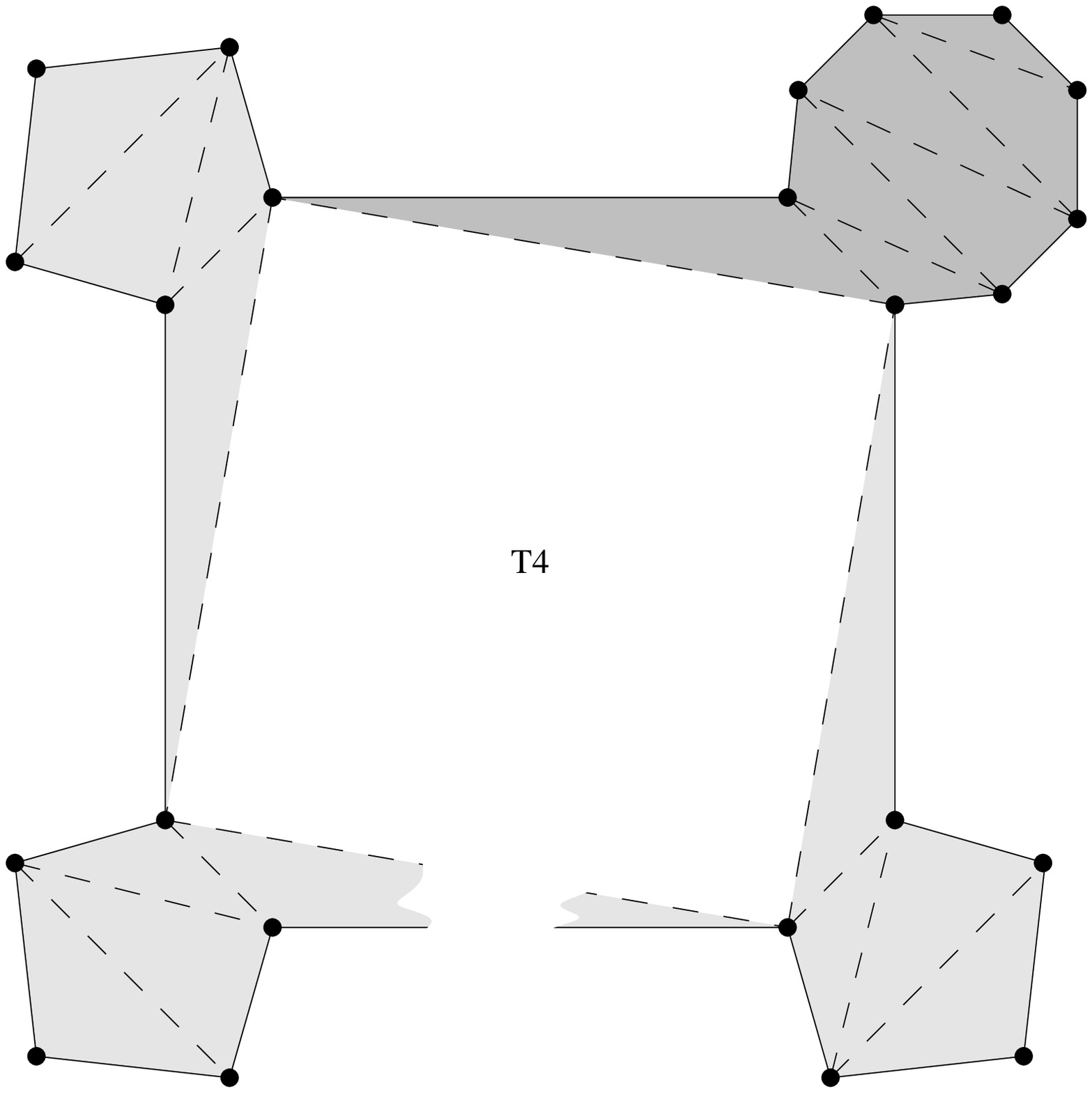}\hfil%
    \includegraphics[width=0.45\columnwidth]{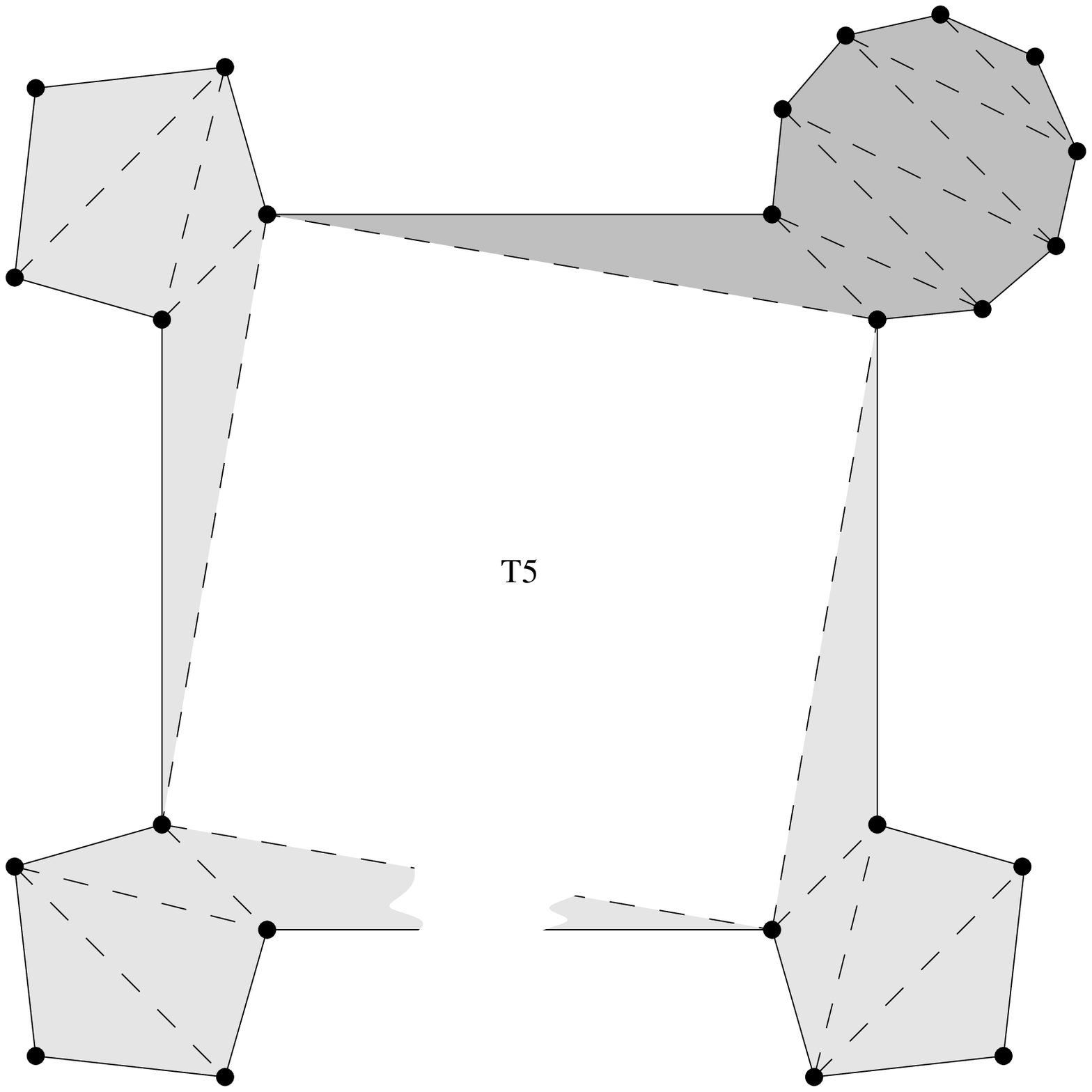}
    \caption{\new{The triangulation graphs $\Gamma_{5m+i}$, $i=0,1,3,4$,
      with $n=5m+i$ vertices, respectively (the central parts of the
      graphs are triangulated arbitrarily). All four triangulation
      graphs require at least $\be$ edge guards in order to be
      2-dominated.}}
    \label{fig:trg-edge-lb}
  \end{center}
  \newend
\end{figure}

\newcommand{\figtextsize}[1]{\footnotesize{#1}}
\begin{figure}[!tb]
  \newbegin
  \begin{center}
    \psfrag{G7}[][]{$\Gamma_7$}
    \psfrag{G12}[][]{$\Gamma_{12}$}
    \psfrag{G17}[][]{$\Gamma_{17}$}
    \psfrag{G22}[][]{$\Gamma_{22}$}
    \psfrag{e0}[][]{\figtextsize{$e_0$}}
    \psfrag{e6}[][]{\figtextsize{$e_6$}}
    \psfrag{1}[][]{\figtextsize{$v_0$}}
    \psfrag{2}[][]{\figtextsize{$v_1$}}
    \psfrag{3}[][]{\figtextsize{$v_2$}}
    \psfrag{4}[][]{\figtextsize{$v_3$}}
    \psfrag{5}[][]{\figtextsize{$v_4$}}
    \psfrag{6}[][]{\figtextsize{$v_5$}}
    \psfrag{7}[][]{\figtextsize{$v_6$}}
    \psfrag{8}[][]{\figtextsize{$v_7$}}
    \psfrag{9}[][]{\figtextsize{$v_8$}}
    \psfrag{10}[][]{\figtextsize{$v_9$}}
    \psfrag{11}[][]{\figtextsize{$v_{10}$}}
    \psfrag{12}[][]{\figtextsize{$v_{11}$}}
    \psfrag{13}[][]{\figtextsize{$v_{12}$}}
    \psfrag{14}[][]{\figtextsize{$v_{13}$}}
    \psfrag{15}[][]{\figtextsize{$v_{14}$}}
    \psfrag{16}[][]{\figtextsize{$v_{15}$}}
    \psfrag{17}[][]{\figtextsize{$v_{16}$}}
    \psfrag{18}[][]{\figtextsize{$v_{17}$}}
    \psfrag{19}[][]{\figtextsize{$v_{18}$}}
    \psfrag{20}[][]{\figtextsize{$v_{19}$}}
    \psfrag{21}[][]{\figtextsize{$v_{20}$}}
    \psfrag{22}[][]{\figtextsize{$v_{21}$}}
    \includegraphics[width=0.44\columnwidth]{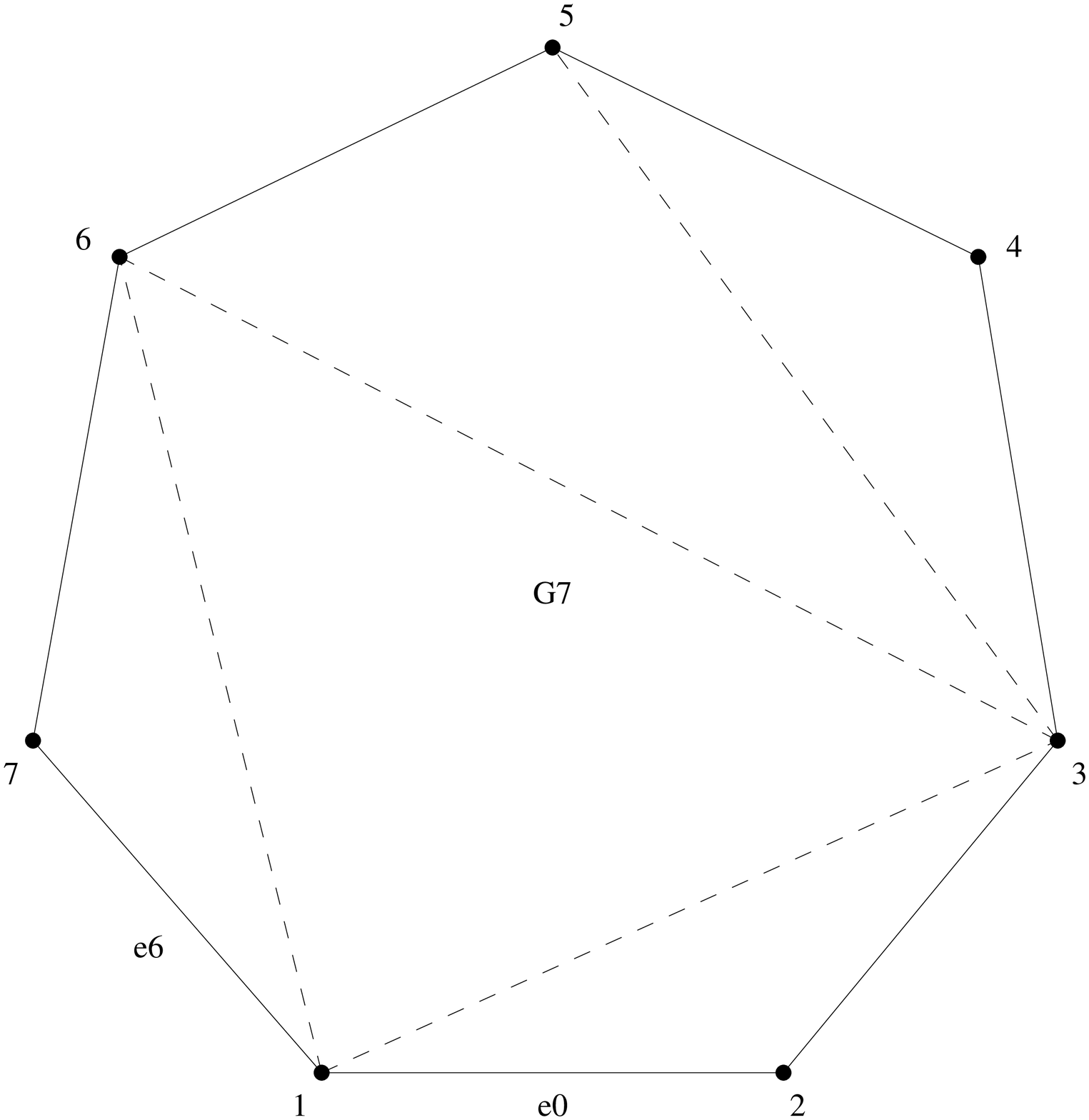}\hfill%
    \includegraphics[width=0.45\columnwidth]{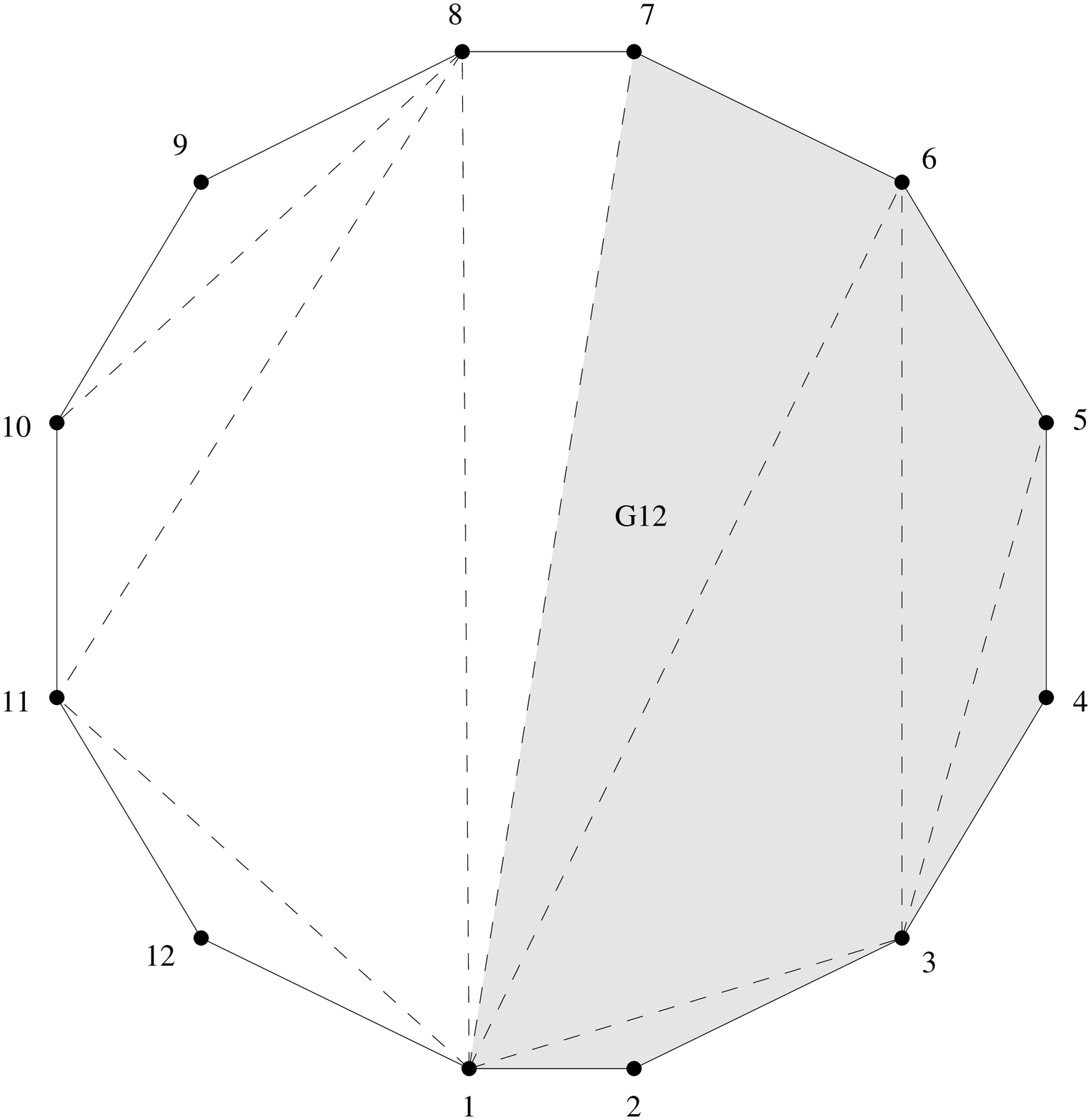}\\[15pt]%
    \includegraphics[width=0.44\columnwidth]{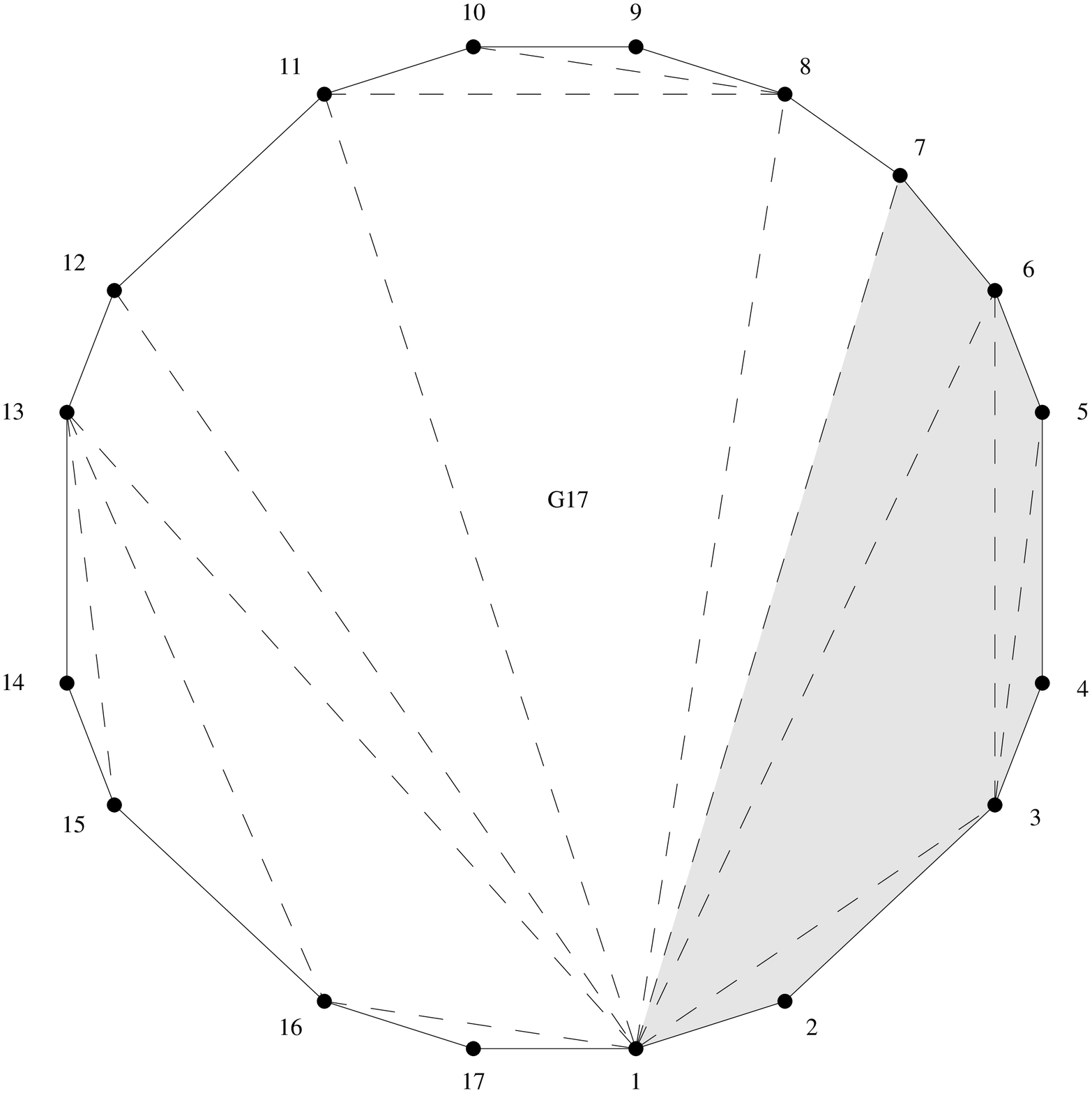}\hfill%
    \includegraphics[width=0.45\columnwidth]{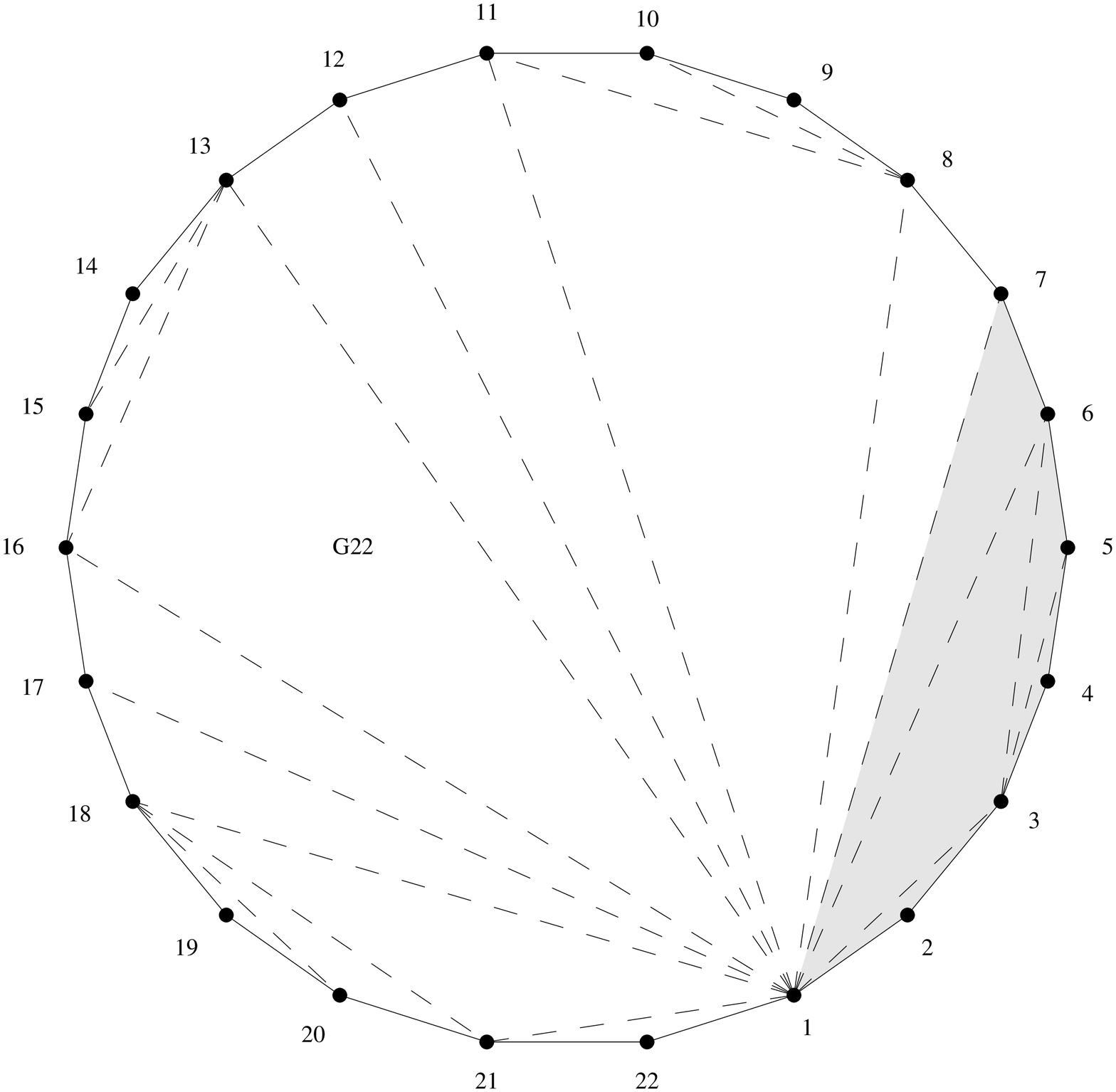}\hfill%
  \end{center}
  \caption{\new{The triangulation graphs $\Gamma_7$, $\Gamma_{12}$,
    $\Gamma_{17}$ and $\Gamma_{22}$, with $n=7,12,17$ and $22$
    vertices, respectively. Each of these graphs requires $\be$ edge
    guards in order to be 2-dominated. The shaded part of the graph
    $\Gamma_n$, $n=12,17,22$, corresponds to the graph $\Gamma_7$
    that is glued to $\Gamma_{n-5}$ in order to construct $\Gamma_n$.}}
  \label{fig:lbe}
  \newend
\end{figure}

To prove the lower bound for all remaining $n\ge{}7$, we are going to
inductively construct a family of triangulation graphs
$\Gamma_{5m+2}$, $m\ge{}1$, as follows.
The triangulation graph $\Gamma_7$ is shown in Fig. \ref{fig:lbe}(top
left).
$\Gamma_{12}$ is constructed by \emph{gluing} two copies $\Gamma_7'$
and $\Gamma_7''$ of $\Gamma_7$ along the edge $e_0$ of $\Gamma_7'$ and
the edge $e_6$ of $\Gamma_7''$, such that the vertex $v_0$ (\resp
$v_1$) of $\Gamma_7'$ is identified with the vertex $v_0$ (\resp
$v_6$) of $\Gamma_7''$ (see Fig. \ref{fig:lbe}(top right)). In
$\Gamma_{12}$, $v_0$ is the vertex that used to be $v_0$ in both
$\Gamma_7'$ and $\Gamma_7''$, while all other vertices are numbered in
the counterclockwise sense.
$\Gamma_{5m+7}$, $m\ge{}2$, is constructed by gluing $\Gamma_{5m+2}$
with $\Gamma_7$ along the edge $e_0$ of $\Gamma_{5m+2}$ and the edge
$e_6$ of $\Gamma_7$, such that the vertex $v_0$ (\resp $v_1$) of
$\Gamma_{5m+2}$ is identified with the vertex $v_0$ (\resp $v_6$) of
$\Gamma_7$ (see Fig. \ref{fig:lbe}(bottom row) for $\Gamma_{17}$ and
$\Gamma_{22}$). In $\Gamma_{5m+7}$, $v_0$ is the vertex that used to
be $v_0$ in both $\Gamma_{5m+2}$ and $\Gamma_7$, while all other
vertices are numbered in the counterclockwise sense.

We are now ready to proceed with our proof of the lower bound for the
triangulation graphs $\Gamma_{5m+2}$, $m\ge{}1$. More precisely, we
will show, by induction on $m$, that every edge 2-dominating set of the
triangulation graph $\Gamma_{5m+2}$ has size at least $2m+1$.
We start by the base case, \ie $m=1$. $\Gamma_7$ cannot be 2-dominated
by less than three edges, since then we would be able to find an edge
$e$ of $\Gamma_7$ such that its two endpoints are not in the edge
2-dominating set of $\Gamma_7$, and thus the triangle of $\Gamma_7$
incident to $e$ would not be 2-dominated. Let us now assume that our
claim holds true for some $m\ge{}1$, \ie every edge 2-dominating set
of $\Gamma_{5m+2}$ has size at least $2m+1$.

Consider the triangulation graph $\Gamma_{5m+7}$. Let $D$ be
an edge 2-dominating set for $\Gamma_{5m+7}$, and let us assume
that $|D|<2(m+1)+1$, \ie $|D|\le{}2m+2$.
Let $T_1$ and $T_2$ be the triangulation graphs that we get by
\emph{cutting} $\Gamma_{5m+7}$ along the diagonal $d_{06}$, with
$T_2$ being the one containing the vertex $v_1$ (see
Fig. \ref{fig:lbe2}(left)), and, moreover, let $T_3$ and $T_4$ be the
triangulation graphs that we get by cutting $\Gamma_{5m+7}$ along
the diagonal $d_{0,5m+1}$, with $T_4$ being the one containing the
vertex $v_1$ (see Fig. \ref{fig:lbe2}(right)). Notice that 
$T_1$ and $T_4$ (\resp $T_2$ and $T_3$) are isomorphic to
$\Gamma_{5m+2}$ (\resp $\Gamma_7$). Let $D_1$ (\resp $D_2$) be the
subset of $D$ containing the edges of $D$ in $T_1$ (\resp $T_2$), and
define $D_3$ and $D_4$ analogously. Finally, notice that the sets
$D_1'=D_1\cup\{d_{06}\}$ and $D_4'=D_4\cup\{d_{0,5m+1}\}$ are
edge 2-dominating sets of $T_1$ and $T_4$, respectively.
It is easy to verify that $|D_2|\ge{}2$ (\resp $|D_3|\ge{}2$), since
otherwise we would be able to find an edge in $\{e_1,e_2,e_3,e_4\}$
(\resp $\{e_{5m+2},e_{5m+3},e_{5m+4},e_{5m+5}\}$) such that its two
endpoints are not endpoints of edges in $D$; notice that this is true
even if both $v_0$ and $v_6$ (\resp $v_0$ and $v_{5m+1}$) belong to
$D$ due to edges in $D_1$ (\resp $D_4$).
Consider the following cases:

\begin{figure}[t]
  \newbegin
  \begin{center}
    \psfrag{T1}[][]{$T_1$}
    \psfrag{T2}[][]{$T_2$}
    \psfrag{T3}[][]{$T_3$}
    \psfrag{T4}[][]{$T_4$}
    \psfrag{1}[][]{\figtextsize{$v_0$}}
    \psfrag{2}[][]{\figtextsize{$v_1$}}
    \psfrag{3}[][]{\figtextsize{$v_2$}}
    \psfrag{4}[][]{\figtextsize{$v_3$}}
    \psfrag{5}[][]{\figtextsize{$v_4$}}
    \psfrag{6}[][]{\figtextsize{$v_5$}}
    \psfrag{7}[][]{\figtextsize{$v_6$}}
    \psfrag{8}[][]{\figtextsize{$v_7$}}
    \psfrag{9}[][]{\figtextsize{$v_8$}}
    \psfrag{10}[][]{\figtextsize{$v_9$}}
    \psfrag{11}[][]{\figtextsize{$v_{10}$}}
    \psfrag{12}[][]{\figtextsize{$v_{5m+1}$}}
    \psfrag{13}[][]{\figtextsize{$v_{5m+2}$}}
    \psfrag{14}[][]{\figtextsize{$v_{5m+3}$}}
    \psfrag{15}[][]{\figtextsize{$v_{5m+4}$}}
    \psfrag{16}[][]{\figtextsize{$v_{5m+5}$}}
    \psfrag{17}[][]{\figtextsize{$v_{5m+6}$}}
    \includegraphics[width=0.45\textwidth]{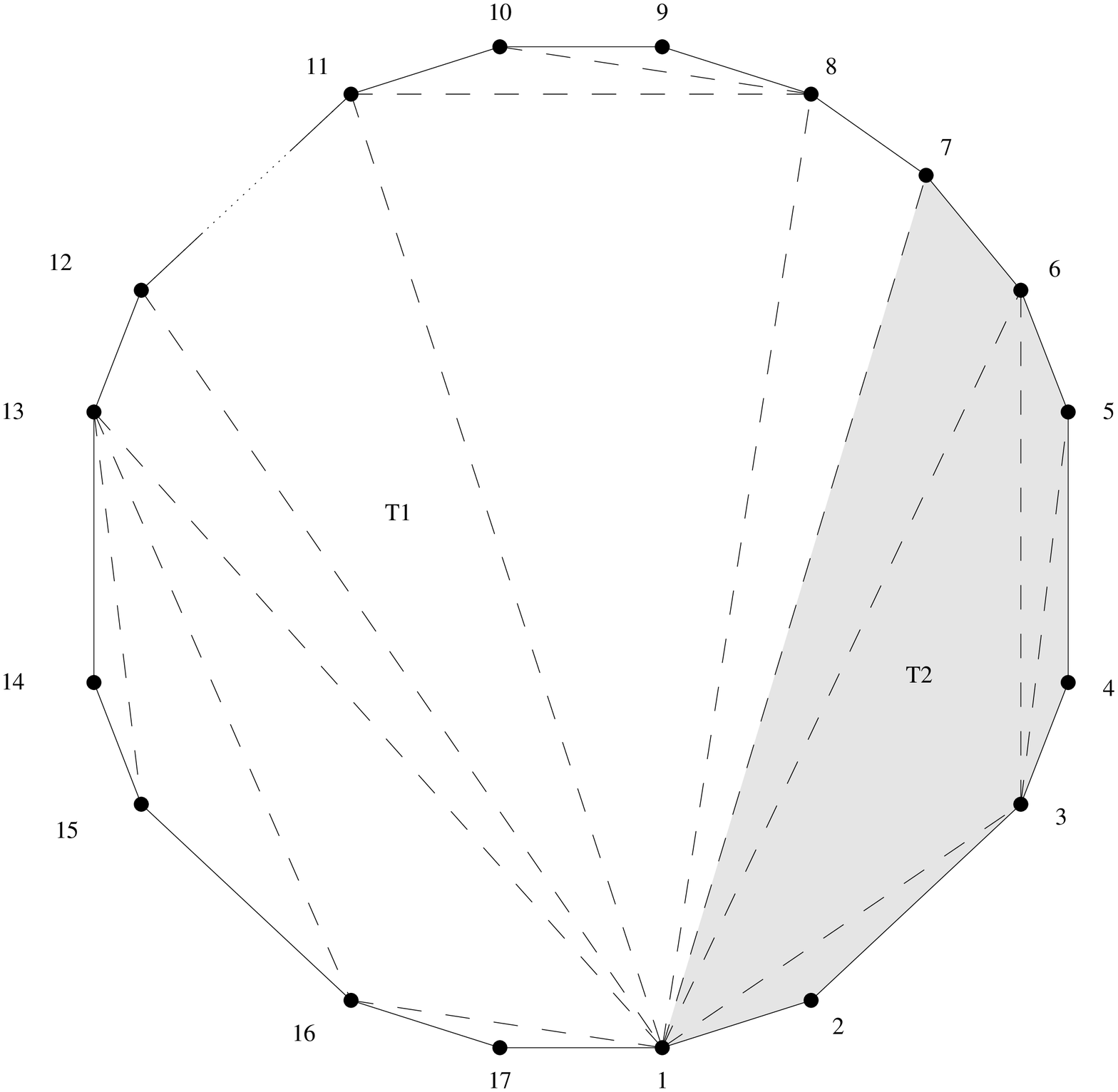}\hfill%
    \includegraphics[width=0.45\textwidth]{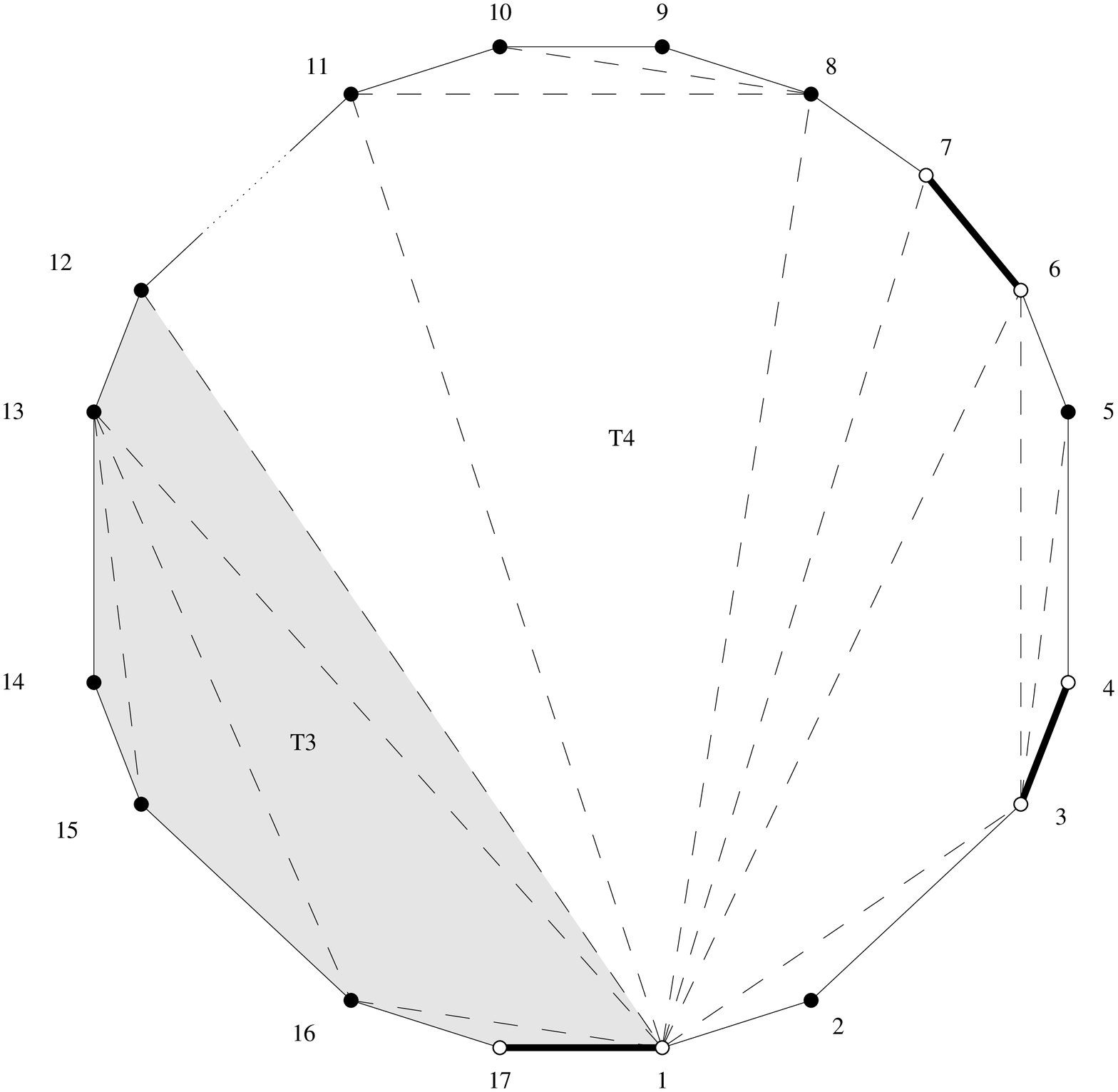}
  \end{center}
  \caption{\new{The triangulation graph $\Gamma_{5m+7}$. The shaded
    subgraphs of $\Gamma_{5m+7}$ are the triangulation graphs
    $T_2$ and $T_3$, while the non-shaded subgraphs of
    $\Gamma_{5m+7}$ are the triangulation graph $T_1$ and $T_4$,
    respectively. In the right subfigure we also depict some of the
    edges in $D$ when $D_2=\{e_2,e_5\}$.}}
  \label{fig:lbe2}
  \newend
\end{figure}

\begin{mathdescription}
\item[|D_2|\ge{}3.] In this case we have
  $|D_1|=|D|-|D_2|\le{}(2m+2)-3=2m-1$, which further implies that
  $|D_1'|=|D_1|+1\le{}2m$. This contradicts our inductive assumption,
  since $D_1'$ is an edge 2-dominating set of $T_1$, and thus of
  $\Gamma_{5m+2}$.
\item[|D_2|=2.] In this case $|D_1|=|D|-|D_2|\le(2m+2)-2=2m<2m+1$.
  Observe that $D_2$ can only be one of the following
  four subsets of $\{e_0,e_1,\ldots,e_5\}$ of size two:
  $\{e_1,e_3\}$, $\{e_1,e_4\}$, $\{e_2,e_4\}$ and $\{e_2,e_5\}$.
  All other subsets of size two of $\{e_0,e_1,\ldots,e_5\}$, except
  $\{e_0,e_3\}$, are such that there exists an edge
  in $\{e_1,e_2,e_3,e_4\}$ with the property that its two endpoints
  are not endpoints of edges in $D$. Lastly, if $D_2$ was equal to
  $\{e_0,e_3\}$, the triangle $v_0v_2v_5$ would not be 2-dominated by
  $D$. Consider the following subcases:
  \begin{mathdescription}
  \item[D_2\in\{\{e_1,e_3\},\{e_1,e_4\},\{e_2,e_4\}\}.]
    Refer to Fig. \ref{fig:lbe2}(left). Notice that none of the
    vertices of edges in $D_2$ is a vertex of a triangle in $T_1$, \ie
    the vertices of edges in $D_2$ do not contribute to the
    2-domination of triangles in $T_1$. This further implies that the
    triangles in $T_1$ are essentially 2-dominated by the edges in
    $D_1$, which suggests the existence of an edge 2-dominating set
    for $\Gamma_{5m+2}$ of size $|D_1|=2m<2m+1$, a contradiction with
    respect to our inductive hypothesis.
  \item[D_2=\{e_2,e_5\}.]
    Refer to Fig. \ref{fig:lbe2}(right). In order for the triangle
    $v_0v_1v_2$ to be 2-dominated we must have that $e_{5m+6}\in{}D_1$,
    and, more importantly, that $e_{5m+6}\in{}D_3$.
    Recall that $|D_3|\ge{}2$; we argue that in this case
    $|D_3|\ge{}3$. To verify that, suppose that $|D_3|=2$. Then the
    unique edge in $D_3\setminus\{e_{5m+6}\}$ cannot be one of
    $e_{5m+1}$, $e_{5m+2}$, $e_{5m+4}$ or $e_{5m+5}$, since then we
    would be able to find an edge in
    $\{e_{5m+2},e_{5m+3},e_{5m+4},e_{5m+5}\}$, such that its two
    endpoints are not endpoints of edges in $D$; moreover,
    if the unique edge in $D_3\setminus\{e_{5m+6}\}$ is $e_{5m+3}$,
    the triangle $v_0v_{5m+2}v_{5m+5}$ is not 2-dominated by $D$.
    Since $|D_3|\ge{}3$, we get that the size of $D_4$ has to be
    $|D_4|=|D|-|D_3|\le{}(2m+2)-3=2m-1$, which gives that
    $|D_4'|=|D_4|+1\le{}2m$. As for the case $|D_2|\ge{}3$ above, the
    bound on the size of $|D_4'|$ contradicts our inductive assumption,
    since $D_4'$ is an edge 2-dominating set of $T_4$, and thus of
    $\Gamma_{5m+2}$.\qedhere
  \end{mathdescription}
\end{mathdescription}
\end{proof}
\newend


\subsection{Computing edge 2-dominating sets in linear time}
\label{sec:compute-edge-dsets}

Unlike the case of diagonal 2-dominating sets, the proof of Theorem
\ref{thm:guard-combedge-trgraph} uses edge contractions, which yields
an $O(n^2)$ time and $O(n)$ space algorithm. A linear time and space
algorithm is, however, feasible by relaxing the requirement on the
size of the edge 2-dominating set. More precisely, we prove in this
subsection that we can 2-dominate a triangulation graph with $\ubew$
edge guards. Although this result is weaker with respect to the result
of Theorem \ref{thm:guard-combedge-trgraph}, the proof technique is
analogous to the technique in the proof of Theorem
\ref{thm:guard-combdiag-trgraph}, \ie it does not use edge
contractions. Consequently, in analogy to the considerations of
Section \ref{sec:compute-diag-dsets}, we can devise a linear time and
space algorithm for computing an edge 2-dominating set of size at most
$\ubew$.

\begin{theorem}\label{thm:guard-combedge-trgraph-weak}
Every triangulation graph $\trg{P}$ of a polygon $P$ with $n\ge{}3$
vertices can be 2-dominated by $\ubew$ edge guards, except for $n=4$,
where one additional guard is required.
\end{theorem}

\begin{proof}
By Theorem \ref{thm:guard-combedge-trgraph}, and since
$\be\new{=}\ubew$ for all $3\le{}n\le{}11$, we conclude that our
theorem holds true for all $n$, with $3\le{}n\le{}11$.

Let us now assume that $n\ge{}12$ and that the theorem holds for all
$n'$ such that $5\le{}n'<n$. By Lemma \ref{lem:diag_existence} with
$\lambda=6$, there exists a diagonal $d$ that partitions $\trg{P}$
into two triangulation graphs $T_1$ and $T_2$, where $T_1$ contains
$k$ boundary edges of $\trg{P}$ with $6\le{}k\le{}10$. Let $v_i$,
$0\le{}i\le{}k$, be the $k+1$ vertices of $T_1$, as we encounter them
while traversing $P$ counterclockwise, and let $v_0v_k$ be the common
edge of $T_1$ and $T_2$. For each value of $k$ we are going to define
an edge 2-dominating set $D$ for $\trg{P}$ of size $\ubew$. In what
follows $d_{ij}$ denotes the diagonal $v_iv_j$, whereas $e_i$ denotes
the edge $v_iv_{i+1}$. Consider each value of $k$ separately. 

  \begin{figure}[!t]
    \begin{center}
      \psfrag{0}[][]{\scriptsize$v_0$}
      \psfrag{1}[][]{\scriptsize$v_1$}
      \psfrag{2}[][]{\scriptsize$v_2$}
      \psfrag{3}[][]{\scriptsize$v_3$}
      \psfrag{4}[][]{\scriptsize$v_4$}
      \psfrag{5}[][]{\scriptsize$v_5$}
      \psfrag{6}[][]{\scriptsize$v_6$}
      \psfrag{d}[][]{\scriptsize$d$}
      \psfrag{t}[][]{\scriptsize$t$}
      \psfrag{tp}[][]{\scriptsize$t'$}
      \includegraphics[width=0.9\textwidth]{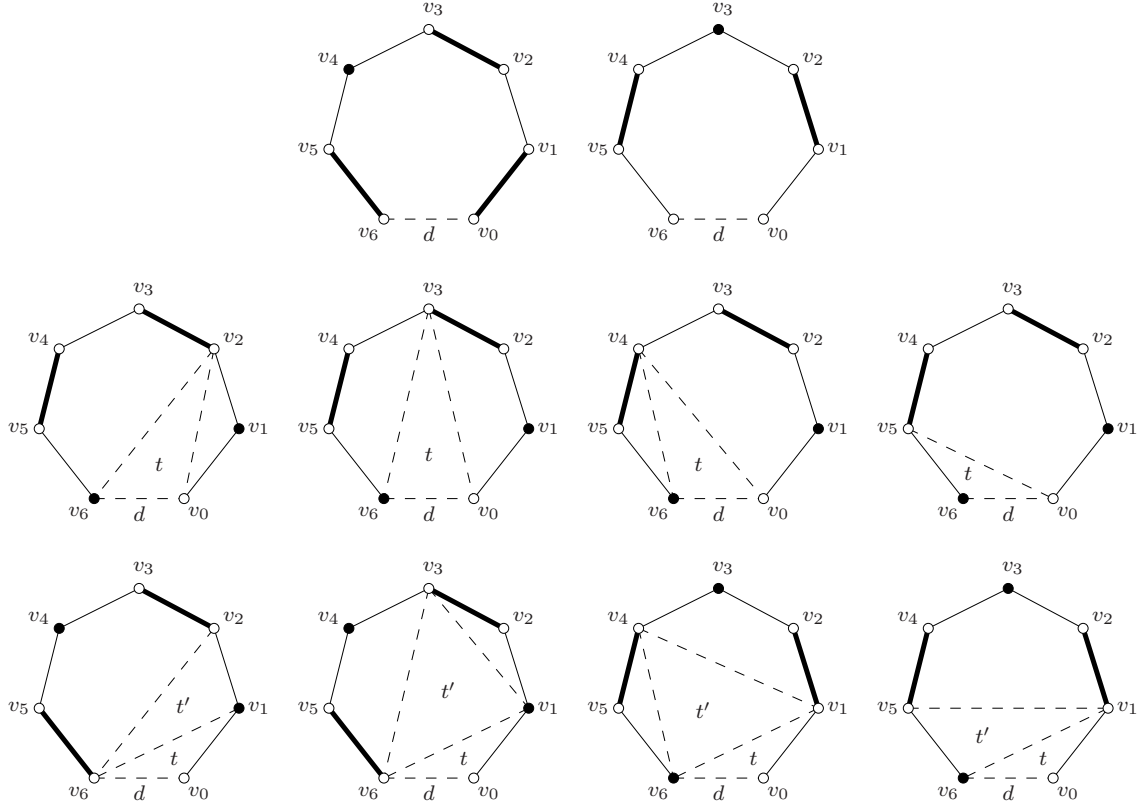}
    \end{center}
    \caption{Proof of Theorem \ref{thm:guard-combedge-trgraph-weak}: 
      the case $k=6$. Top row: the case $d_{06}\in{}D_2$ (left) and
      the case $d_{06}\nin{}D_2$, $v_0,v_6\in{}D_2$ (right). 
      Middle and bottom rows: the case \new{$d_{06}\nin{}D_2$,}
      $v_0\in{}D_2,v_6\nin{}D_2$.
      Middle row: the case $v\in\{v_2,v_3,v_4,v_5\}$; from left to
      right: $v\equiv{}v_2$, $v\equiv{}v_3$, $v\equiv{}v_4$,
      $v\equiv{}v_5$. Bottom row: the case $v\equiv{}v_1$; from left
      to right: $v'\equiv{}v_2$, $v'\equiv{}v_3$, $v'\equiv{}v_4$,
      $v'\equiv{}v_5$.}
    \label{fig:edgeproofw_k6}
  \end{figure}
  \begin{figure}[!t]
    \begin{center}
      \psfrag{0}[][]{\scriptsize$v_0$}
      \psfrag{1}[][]{\scriptsize$v_1$}
      \psfrag{2}[][]{\scriptsize$v_2$}
      \psfrag{3}[][]{\scriptsize$v_3$}
      \psfrag{4}[][]{\scriptsize$v_4$}
      \psfrag{5}[][]{\scriptsize$v_5$}
      \psfrag{6}[][]{\scriptsize$v_6$}
      \psfrag{7}[][]{\scriptsize$v_7$}
      \psfrag{d}[][]{\scriptsize$d$}
      \psfrag{t}[][]{\scriptsize$t$}
      \psfrag{tp}[][]{\scriptsize$t'$}
      \includegraphics[width=0.9\textwidth]{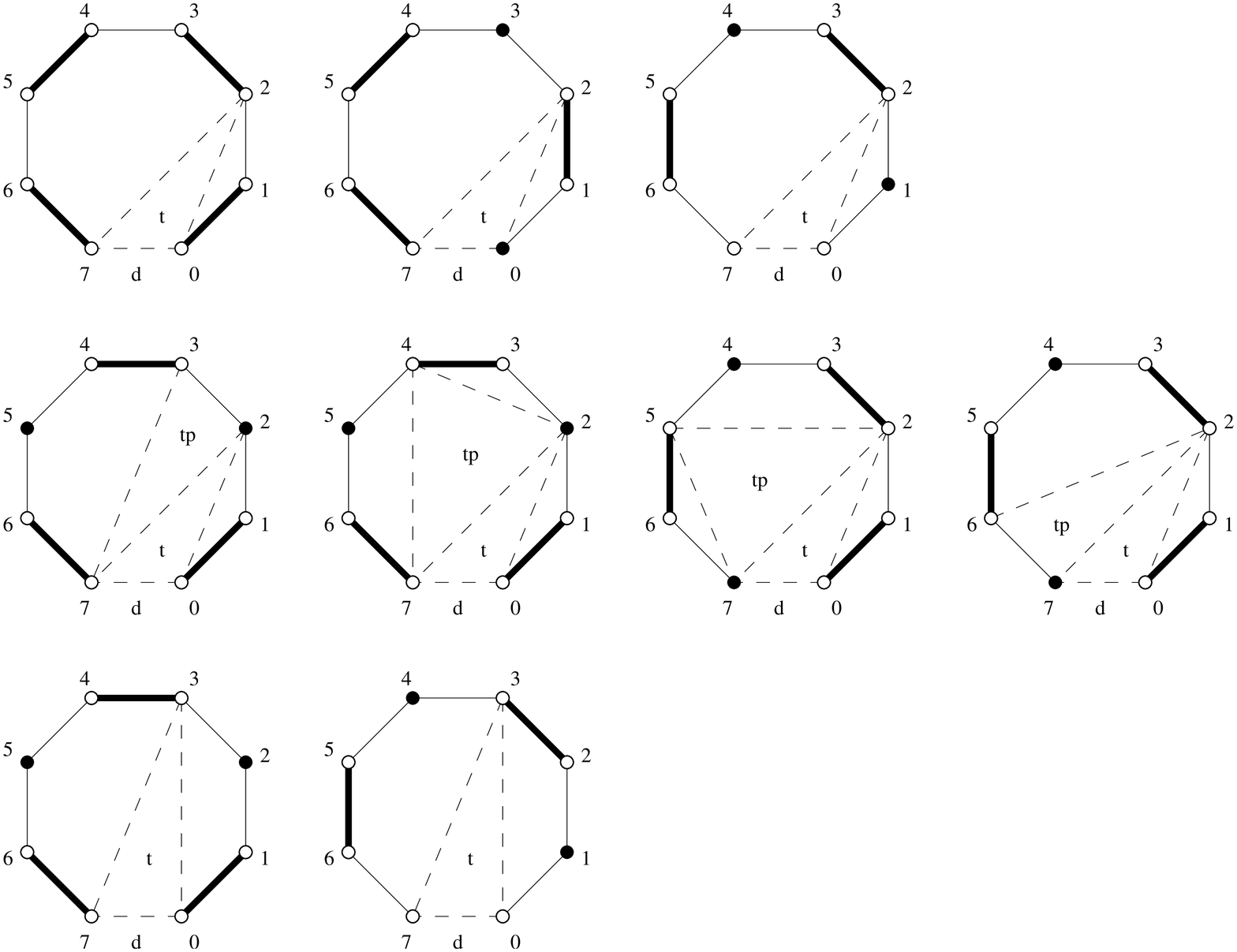}
    \end{center}
    \caption{Proof of Theorem \ref{thm:guard-combedge-trgraph-weak}: 
      the case $k=7$. Top and middle rows: the case $v\equiv{}v_2$.
      Bottom row: the case $v\equiv{}v_3$.
      Top row (from left to right): the case $d_{02},d_{27}\in{}D'$;
      the case $d_{02}\nin{}D',d_{27}\in{}D'$; the case
      $d_{02},d_{27}\nin{}D'$.
      Middle row: the case $d_{02}\in{}D',d_{27}\nin{}D'$; from left
      to right: $v'\equiv{}v_3$, $v'\equiv{}v_4$, $v'\equiv{}v_5$,
      $v'\equiv{}v_6$.
      Bottom row (from left to right): the case $d_{02}\in{}D'$ or
      $d_{27}\in{}D'$; the case $d_{02},d_{27}\nin{}D'$.}
    \label{fig:edgeproofw_k7}
  \end{figure}

\begin{mathdescription}
\item[k=6.] In this case $T_2$ contains $n-5$ vertices. By our
  induction hypothesis we can dominate $T_2$ with
  $f(n-5)\le\ubew-2$ edge guards. Let $D_2$ be the edge
  2-dominating set for $T_2$. Consider the following cases: (see
  Fig. \ref{fig:edgeproofw_k6}):
  \begin{mathdescription}
  \item[d_{06}\in{}D_2.] Set
    $D=(D_2\setminus\{d_{06}\})\cup\{e_0,e_2,e_5\}$.
  \item[d_{06}\nin{}D_2.] Since $T_2$ is 2-dominated by $D_2$, at
    least one of the vertices $v_0$ and $v_6$ belongs to $D_2$. We
    distinguish between the following subcases:
    \begin{mathdescription}
    \item[v_0,v_6\in{}D_2.] Set $D=D_2\cup\{e_1,e_4\}$.
    \item[v_0\in{}D_2,v_6\nin{}D_2.] Let $t$ be the triangle
      supported by $d$ in $T_1$ and let $v$ be its vertex opposite to
      $d$. If $v\in\{v_2,v_3,v_4,v_5\}$, set $D=D_2\cup\{e_2,e_4\}$.
      If $v\equiv{}v_1$, let $t'$ be the second triangle supported by
      $d_{16}$ beyond the triangle $t$, and let $v'$ be its vertex
      opposite to $d_{16}$. If $v'\in\{v_2,v_3\}$, set
      $D=D_2\cup\{e_2,e_5\}$. Otherwise, \ie if $v'\in\{v_4,v_5\}$,
      set $D=D_2\cup\{e_1,e_4\}$.
    \item[v_0\nin{}D_2,v_6\in{}D_2.] This case is symmetric to the
      previous one. Let $t$ be the triangle supported by $d$ in $T_1$
      and let $v$ be its vertex opposite to $d$. If
      $v\in\{v_1,v_2,v_3,v_4\}$, set $D=D_2\cup\{e_1,e_3\}$. 
      If $v\equiv{}v_5$, let $t'$ be the second triangle supported by
      $d_{05}$ beyond the triangle $t$, and let $v'$ be its vertex
      opposite to $d_{05}$. If $v'\in\{v_1,v_2\}$, set
      $D=D_2\cup\{e_1,e_4\}$. Otherwise, \ie if $v'\in\{v_3,v_4\}$,
      set $D=D_2\cup\{e_0,e_3\}$.
    \end{mathdescription}
  \end{mathdescription}
\item[k=7.] The presence of diagonals $d_{06}$ or $d_{17}$ would
  violate the minimality of $k$. Let $t$ be the triangle supported by
  $d$ in $T_1$ and let $v$ its vertex opposite to $d$. Consider the
  triangulation graph $T'=T_2\cup\{t\}$. It has $n-5$ vertices, hence,
  by our induction hypothesis, it can be 2-dominated with
  $f(n-5)\le\ubew-2$ edge guards. Let $D'$ be the 2-dominating set
  for $T'$. Clearly, $v'\in\{v_2,v_3,v_4,v_5\}$; furthermore notice
  that the cases $v\equiv{}v_2$ and $v\equiv{}v_5$, and $v\equiv{}v_3$
  and $v\equiv{}v_4$ are symmetric. We, therefore, consider only the
  cases $v\equiv{}v_2$ and $v\equiv{}v_3$ (see Fig. \ref{fig:edgeproofw_k7}):
  \begin{mathdescription}
  \item[v\equiv{}v_2.] We distinguish between the following subcases:
    \begin{mathdescription}
    \item[d_{02},d_{27}\in{}D'.] Set
      $D=(D'\setminus\{d_{02},d_{27}\})\cup\{e_0,e_2,e_4,e_6\}$.
    \item[d_{02}\in{}D',d_{27}\nin{}D'.] Let $t'\ne{}t$ be the
      triangle supported by $d_{27}$, and let $v'$ be its vertex
      opposite to $d_{27}$. If $v'\in\{v_3,v_4\}$, set
      $D=(D'\setminus\{d_{02}\})\cup\{e_0,e_3,e_6\}$. Otherwise, if
      $v'\in\{v_5,v_6\}$, set $D=(D'\setminus\{d_{02}\})\cup\{e_0,e_2,e_5\}$. 
    \item[d_{02}\nin{}D',d_{27}\in{}D'.] Set
      $D=(D'\setminus\{d_{27}\})\cup\{e_1,e_4,e_6\}$.
    \item[d_{02},d_{27}\nin{}D'.] In this case $v_2$ cannot belong
      to $D'$. Hence in order for $t$ to be 2-dominated we must have
      that $v_0,v_7\in{}D'$. Hence, set $D=D'\cup\{e_2,e_5\}$.
    \end{mathdescription}
  \item[v\equiv{}v_3.] Consider the following subcases:
    \begin{mathdescription}
    \item[d_{02}\text{\em{} or }d_{27}\in{}D'.]
      Set $D=(D_2\setminus\{d_{02},d_{27}\})\cup\{e_0,e_3,e_6\}$.
    \item[d_{02},d_{27}\nin{}D'.] In this case $v_3$ cannot belong
      to $D'$. Hence in order for $t$ to be 2-dominated we must have
      that $v_0,v_7\in{}D'$. Hence, set $D=D'\cup\{e_2,e_5\}$.
    \end{mathdescription}
\item[k=8.] The presence of diagonals $d_{07}$, $d_{06}$, $d_{18}$ or
  $d_{28}$ would violate the minimality of $k$. Let $t$ be the
  triangle supported by $d$ in $T_1$ and let $v$ its vertex opposite
  to $d$. In this case $T_2$ contains $n-7$ vertices, hence, it can be
  2-dominated with $f(n-7)=\ubew-3$ edge guards. Let $D_2$ be the
  2-dominating set for $T_2$. Clearly, $v'\in\{v_3,v_4,v_5\}$;
  furthermore notice that the cases $v\equiv{}v_3$ and $v\equiv{}v_5$
  are symmetric. We, therefore, consider only the cases $v\equiv{}v_3$
  and $v\equiv{}v_4$. In fact, both cases can be treated
  jointly. Consider the following subcases (see
  Fig. \ref{fig:edgeproofw_k8}):
  \begin{mathdescription}
  \item[d_{08}\in{}D_2.] Set
    $D=(D_2\setminus\{d_{08}\})\cup\{e_0,e_3,e_5,e_7\}$.
  \item[d_{08}\nin{}D_2.] Then either $v_0$ or $v_8$ belongs to $D_2$.
    \begin{mathdescription}
    \item[v_0\in{}D_2.] Set $D=D_2\cup\{e_2,e_4,e_7\}$.
    \item[v_8\in{}D_2.] Set $D=D_2\cup\{e_0,e_3,e_5\}$.
    \end{mathdescription}
  \end{mathdescription}
  \begin{figure}[!hb]
    \begin{center}
      \psfrag{0}[][]{\scriptsize$v_0$}
      \psfrag{1}[][]{\scriptsize$v_1$}
      \psfrag{2}[][]{\scriptsize$v_2$}
      \psfrag{3}[][]{\scriptsize$v_3$}
      \psfrag{4}[][]{\scriptsize$v_4$}
      \psfrag{5}[][]{\scriptsize$v_5$}
      \psfrag{6}[][]{\scriptsize$v_6$}
      \psfrag{7}[][]{\scriptsize$v_7$}
      \psfrag{8}[][]{\scriptsize$v_8$}
      \psfrag{d}[][]{\scriptsize$d$}
      \psfrag{t}[][]{\scriptsize$t$}
      \includegraphics[width=0.8\textwidth]{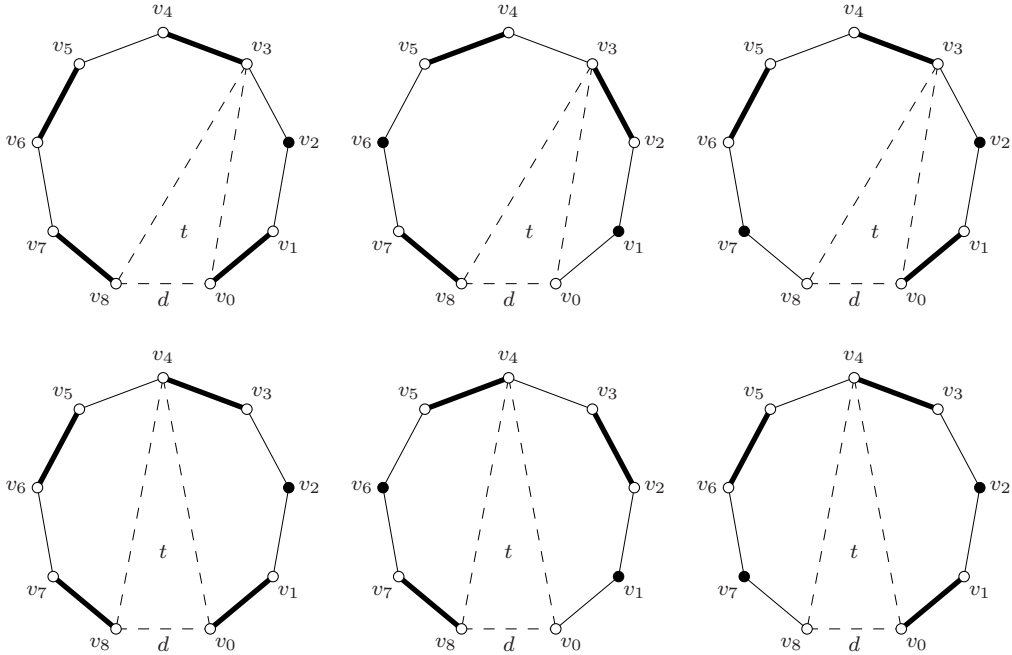}
    \end{center}
    \caption{Proof of Theorem \ref{thm:guard-combedge-trgraph-weak}: 
      the case $k=8$. Top row: $v\equiv{}v_3$. Bottom row:
      $v\equiv{}v_4$. Left column: the case $d_{08}\in{}D_2$. Middle
      column: the case $d_{08}\nin{}D_2$ and $v_0\in{}D_2$. Right
      column: the case $d_{08}\nin{}D_2$ and $v_8\in{}D_2$.}
    \label{fig:edgeproofw_k8}
  \end{figure}
\item[k=9.] The presence of diagonals $d_{08}$, $d_{07}$, $d_{06}$,
  $d_{19}$, $d_{29}$ or $d_{39}$ would violate the minimality of
  $k$. Let $t$ be the triangle supported by $d$ in $T_1$ and let $v$
  its vertex opposite to $d$. Consider the triangulation graph
  $T'=T_2\cup\{t\}$, and let $D'$ be its edge 2-dominating set. $T'$
  has $n-7$ vertices, hence, by our induction hypothesis, $D'$
  consists of $f(n-7)=\ubew-3$ edge guards. Clearly, $v'\in\{v_4,v_5\}$.
  The two cases are symmetric, so we only need to consider the case
  $v\equiv{}v_4$. Consider the following subcases (see
  Fig. \ref{fig:edgeproofw_k9}):
  \begin{mathdescription}
  \item[d_{04}\text{\em{} or }d_{49}\in{}D'.]
    Set $D=(D_2\setminus\{d_{04},d_{49}\})\cup\{e_0,e_3,e_5,e_8\}$.
  \item[d_{04},d_{49}\nin{}D'.] In this case $v_4$ cannot belong
    to $D'$. Hence in order for $t$ to be 2-dominated we must have
    that $v_0,v_9\in{}D'$. Hence, set $D=D'\cup\{e_2,e_4,e_6\}$.
  \end{mathdescription}
  \begin{figure}[!ht]
    \begin{center}
      \psfrag{0}[][]{\scriptsize$v_0$}
      \psfrag{1}[][]{\scriptsize$v_1$}
      \psfrag{2}[][]{\scriptsize$v_2$}
      \psfrag{3}[][]{\scriptsize$v_3$}
      \psfrag{4}[][]{\scriptsize$v_4$}
      \psfrag{5}[][]{\scriptsize$v_5$}
      \psfrag{6}[][]{\scriptsize$v_6$}
      \psfrag{7}[][]{\scriptsize$v_7$}
      \psfrag{8}[][]{\scriptsize$v_8$}
      \psfrag{9}[][]{\scriptsize$v_9$}
      \psfrag{d}[][]{\scriptsize$d$}
      \psfrag{t}[][]{\scriptsize$t$}
      \includegraphics[width=0.6\textwidth]{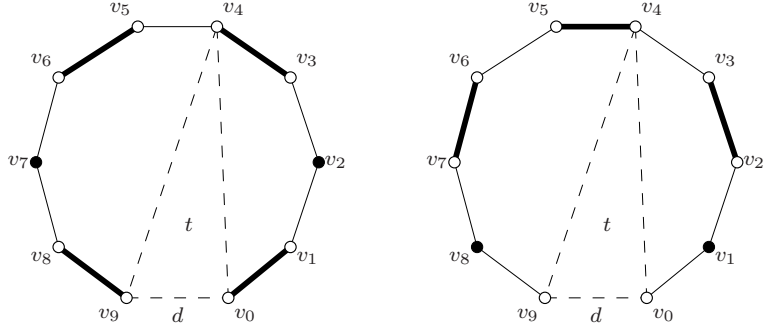}
    \end{center}
    \caption{Proof of Theorem \ref{thm:guard-combedge-trgraph-weak}: 
      the case $k=9$. Left: the case $d_{04}$ or
      $d_{49}\in{}D'$. Right: the case: $d_{04},d_{49}\nin{}D'$.}
    \label{fig:edgeproofw_k9}
  \end{figure}
  \end{mathdescription}
  \begin{figure}[!t]
    \begin{center}
      \psfrag{0}[][]{\scriptsize$v_0$}
      \psfrag{1}[][]{\scriptsize$v_1$}
      \psfrag{2}[][]{\scriptsize$v_2$}
      \psfrag{3}[][]{\scriptsize$v_3$}
      \psfrag{4}[][]{\scriptsize$v_4$}
      \psfrag{5}[][]{\scriptsize$v_5$}
      \psfrag{6}[][]{\scriptsize$v_6$}
      \psfrag{7}[][]{\scriptsize$v_7$}
      \psfrag{8}[][]{\scriptsize$v_8$}
      \psfrag{9}[][]{\scriptsize$v_9$}
      \psfrag{10}[][]{\scriptsize$v_{10}$}
      \psfrag{d}[][]{\scriptsize$d$}
      \psfrag{t}[][]{\scriptsize$t$}
      \psfrag{tp}[][]{\scriptsize$t'$}
      \psfrag{tpp}[][]{\scriptsize$t''$}
      \psfrag{dp}[][]{\scriptsize$d'$}
      \psfrag{dpp}[][]{\scriptsize$d''$}
      \includegraphics[width=0.9\textwidth]{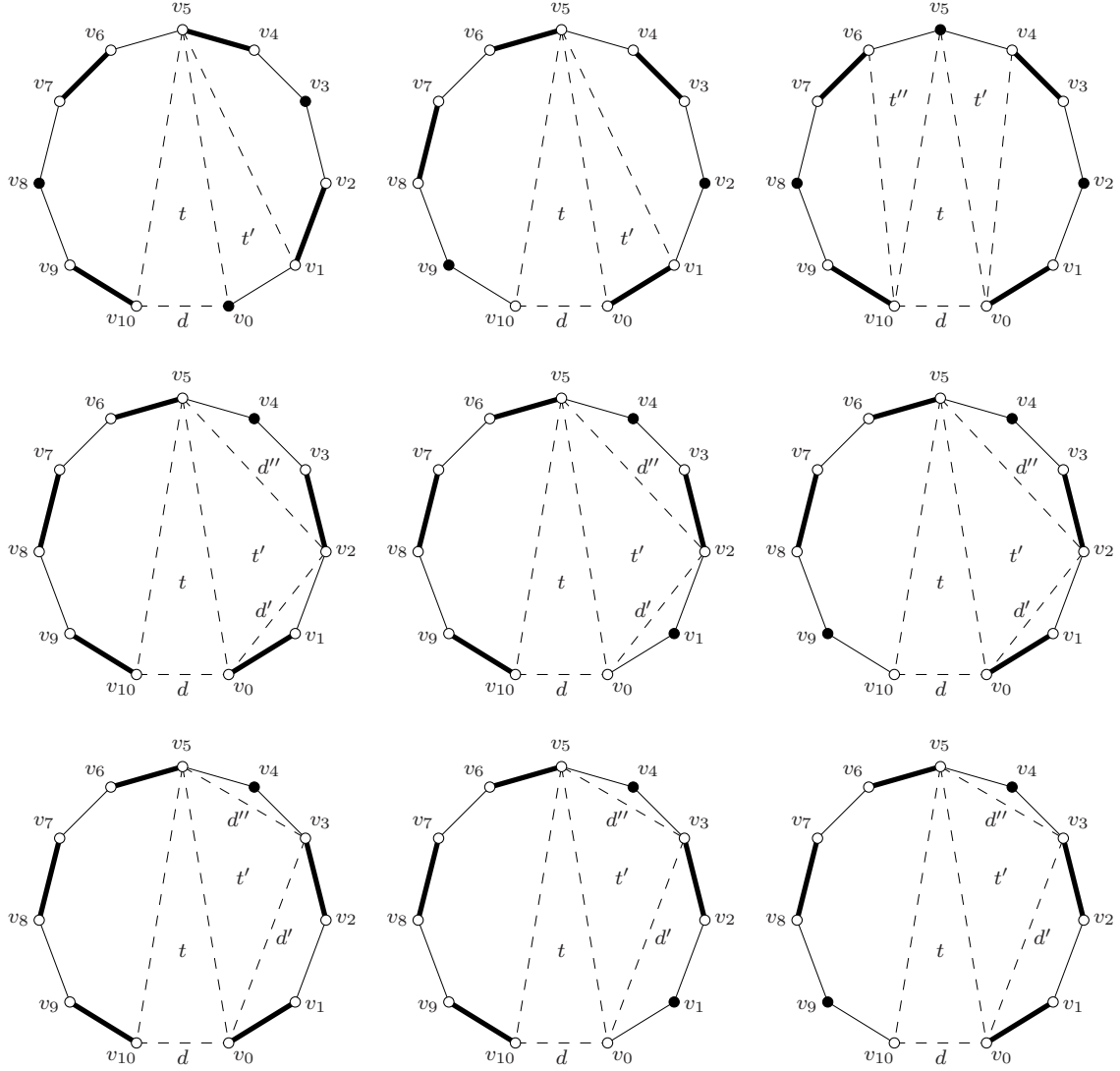}
    \end{center}
    \caption{Proof of Theorem \ref{thm:guard-combedge-trgraph-weak}:
      the case $k=10$. Top row (from left to right): the case
      $v'\equiv{}v_1$ and $d_{15}$ or $d_{5,10}\in{}D'$; the case
      $v'\equiv{}v_1$ and $d_{15},d_{5,10}\nin{}D'$; the case
      $v'\equiv{}v_4$. Middle and bottom rows: the cases
      $v'\equiv{}v_2$ and $v'\equiv{}v_3$, respectively. From left
      to right (middle and bottom rows): the case
      \new{$D'\cap\{d',d'',d_{5,10}\}|\ge{}2$; the case
      $|D'\cap\{d',d'',d_{5,10}\}|=1$ and $v_0\in{}D'\setminus\{d'\}$;
      the case $|D'\cap\{d',d'',d_{5,10}\}|=1$ and
      $v_0\nin{}D'\setminus\{d'\}$.}}
    \label{fig:edgeproofw_k10}
  \end{figure}
\item[k=10.] The presence of diagonals $d_{09}$, $d_{08}$, $d_{07}$,
  $d_{06}$, $d_{1,10}$, $d_{2,10}$ $d_{3,10}$ or $d_{4,10}$ would
  violate the minimality of $k$. Let $t$ be the triangle supported by
  $d$ in $T_1$. Clearly, the vertex of $t$ opposite to $d$ is
  $v_5$. Let $t'\ne{}t$ be the triangle in $T_1$ supported by
  $d_{05}$, and let $v'$ be its vertex opposite to $d_{05}$. Consider
  the triangulation graph $T'=T_2\cup\{t,t'\}$, and let $D'$ be its
  edge 2-dominating set. $T'$ has $n-7$ vertices, hence, by our
  induction hypothesis, $D'$ contains $f(n-7)=\ubew-3$ edge
  guards. Clearly, $v'\in\{v_1,v_2,v_3,v_4\}$. Consider each of the
  \new{following} three cases for $v'$ (see
  Fig. \ref{fig:edgeproofw_k10}):
  \begin{mathdescription}
  \item[v'\equiv{}v_1.] We distinguish between the following subcases:
     \begin{mathdescription}
    \item[d_{15}\text{\em{} or }d_{5,10}\in{}D'.] Set
      $D=(D'\setminus\{d_{15},d_{5,10}\})\cup\{e_1,e_4,e_6,e_9\}$.
    \item[d_{15},d_{5,10}\nin{}D'.] In this case $v_5$ cannot
      belong to $D'$. Hence in order for $t$ and $t'$ to be
      2-dominated we must have that $v_0,v_1,v_{10}\in{}D'$. Since
      $d_{15}\nin{}D'$, we must have that $e_0\in{}D'$, in order
      for $v_1$ to be in $D'$. Hence, given that $e_0,v_{10}\in{}D'$,
      set $D=D'\cup\{e_3,e_5,e_7\}$.
     \end{mathdescription}
  \item[v'\in\{v_2,v_3\}.] \new{Let $d'$ be the diagonal $v_0v'$ and $d''$
    the diagonal $v'v_5$. Notice that at least one of $d'$, $d''$ and
    $d_{5,10}$ must belong to $D'$, since otherwise both}
    $v'$ and $v_5$ would not belong to $D'$ (both their
    incident edges in $T'$ would not belong to $D'$), which implies
    that the triangle $t'$ would not be 2-dominated by
    $D'$. Given this fact, we distinguish between the following cases:
    \begin{mathdescription}\newbegin
    \item[|D'\cap\{d',d'',d_{5,10}\}|\ge{}2,] \ie at least two
      among $d'$, $d''$ and $d_{5,10}$ belong to $D'$. Set
      $D=(D'\setminus\{d',d'',d_{5,10}\})\cup\{e_0,e_2,e_5,e_7,e_9\}$.
    \item[|D'\cap\{d',d'',d_{5,10}\}|=1,] \ie exactly one among
      $d'$, $d''$ and $d_{5,10}$ belongs to $D'$. Consider the
      two cases:
      \begin{mathdescription}
      \item[v_0\in{}D'\setminus\{d'\}.] Set
        $D=(D'\setminus\{d',d'',d_{5,10}\})\cup\{e_2,e_5,e_7,e_9\}$.
      \item[v_0\nin{}D'\setminus\{d'\}.] In order for $t$ to be
        2-dominated by $D'$, we must have that $v_{10}$ in
        $D'$. Hence, set
        $D=(D'\setminus\{d',d'',d_{5,10}\})\cup\{e_0,e_2,e_5,e_7\}$.
      \end{mathdescription}\newend
    \end{mathdescription}
  \item[v'\equiv{}v_4.] Let $t''\ne{}t$ be the triangle in $T_1$
    supported by $d_{5,10}$, and let $v''$ be its vertex opposite
    $d_{5,10}$. If $v''\not\equiv{}v_6$, we have a configuration
    that is symmetric to one of the cases $v'\equiv{}v_1$,
    $v'\equiv{}v_2$ or $v'\equiv{}v_3$, treated above. Hence, we
    only need to consider the case $v''\equiv{}v_6$. We distinguish
    between the following cases:
    \begin{mathdescription}
    \item[d_{04}\text{\em{} or }d_{5,10}\in{}D'.] Set
      $D=(D'\setminus\{d_{04},d_{5,10}\})\cup\{e_0,e_3,e_6,e_9\}$.
    \item[d_{04},d_{5,10}\nin{}D'.] In order for $t'$ to be
      2-dominated by $D'$, either $v_4$ or $v_5$ has to belong to
      $D'$. Since both $d_{04}$ and $d_{5,10}$ do not belong to $D'$,
      we conclude that $e_4$ must belong to $D'$. Hence, set
      $D=(D'\setminus\{e_4\})\cup\{e_0,e_3,e_6,e_9\}$.\qedhere
    \end{mathdescription}
  \end{mathdescription}
\end{mathdescription}
\end{proof}

\begin{figure}[!t]
  \begin{center}
    \psfrag{u}[][]{}
    \psfrag{d}[][]{\tiny$d$}
    \includegraphics[width=\columnwidth]{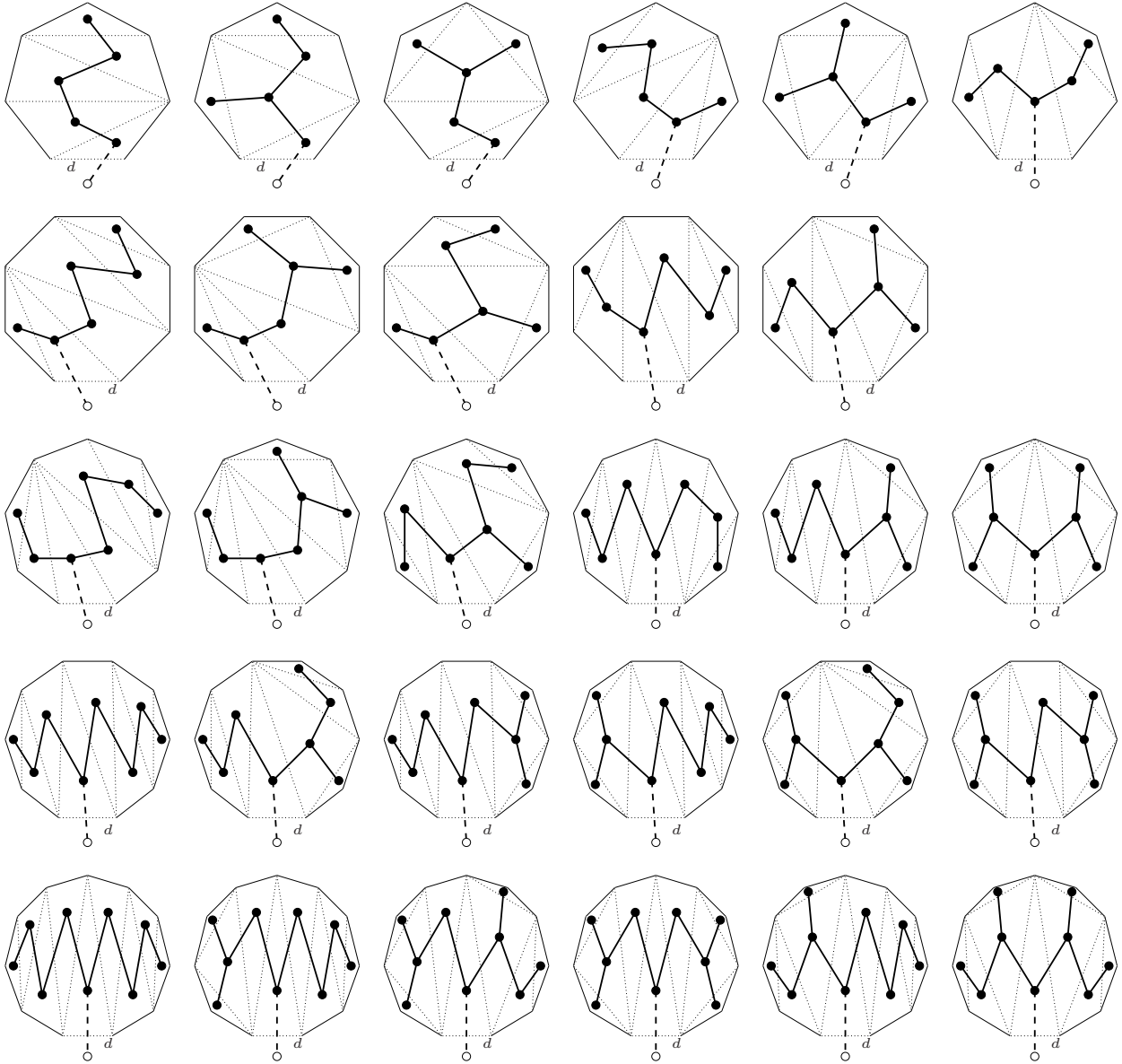}
  \end{center}
  \caption{The 29 possible configurations for the dual trees
    $\Delta_1$ for $6\le{}k\le{}10$, shown as thick solid lines. The
    diagonal $d$ separates $T_1$ from $T_2$. The triangulations shown
    are indicative: all other triangulations yield
    isomorphic trees.}
  \label{fig:dualtrees_edge}
\end{figure}

In a manner analogous to the case of diagonal 2-dominating sets, the
proof of Theorem \ref{thm:guard-combedge-trgraph-weak} can almost
immediately be transformed into an $O(n)$ time and space
algorithm. The algorithm is, in fact, almost identical to the
algorithm presented in Section \ref{sec:compute-diag-dsets} for
computing diagonal 2-dominating sets. The differences, which by
no means alter the spirit of the algorithm, are related to how the
proof of Theorem \ref{thm:guard-combedge-trgraph-weak} is
incorporated. More precisely, the values of $k$ are $6, 7, 8, 9$ and
$10$, instead of $4,5$ and $6$, whereas the dual trees $\Delta$ are
those in Fig. \ref{fig:dualtrees_edge}, instead of those in
Fig. \ref{fig:dualtrees_diag}. Finally, the cut-off value for the
recursion is 21 (instead of 13): for $n\ge{}21$, the subtrees
$\Delta_1'$ corresponding to different diagonals $d$ of $\trg{P}$ must
be edge disjoint (otherwise the number of vertices of $P$ would be
less than 21).

The analysis of the edge 2-dominance linear time algorithm, sketched
above, is entirely analogous to the analysis of the algorithm for
computing diagonal 2-dominating sets. Initialization takes linear time
and space, whereas the recursive part of the algorithm requires linear
space, \new{and} its time requirements satisfy the recursive relation

\[
  T(n) \le \begin{cases}
    T(n-5)+O(1),& n\ge{}21\\
    O(1),&3\le{}n\le{}20
    \end{cases}
\]
which, clearly, yields $T(n)=O(n)$. Hence, we arrive at the following
theorem.

\begin{theorem}\label{thm:trg-edge-timespace-weak}
Given the triangulation graph $\trg{P}$ of a polygon $P$ with
$n\ge{}3$ vertices, we can compute an edge 2-dominating set for
$\trg{P}$ of size at most $\ubew$ (except for $n=4$, where one
additional edge guard is required) in $O(n)$ time and space.
\end{theorem}


\section{Piecewise-convex polygons}
\label{sec:guard-piecewise-convex}

\begin{figure}[!b]
\begin{center}
  \includegraphics[width=0.35\columnwidth]{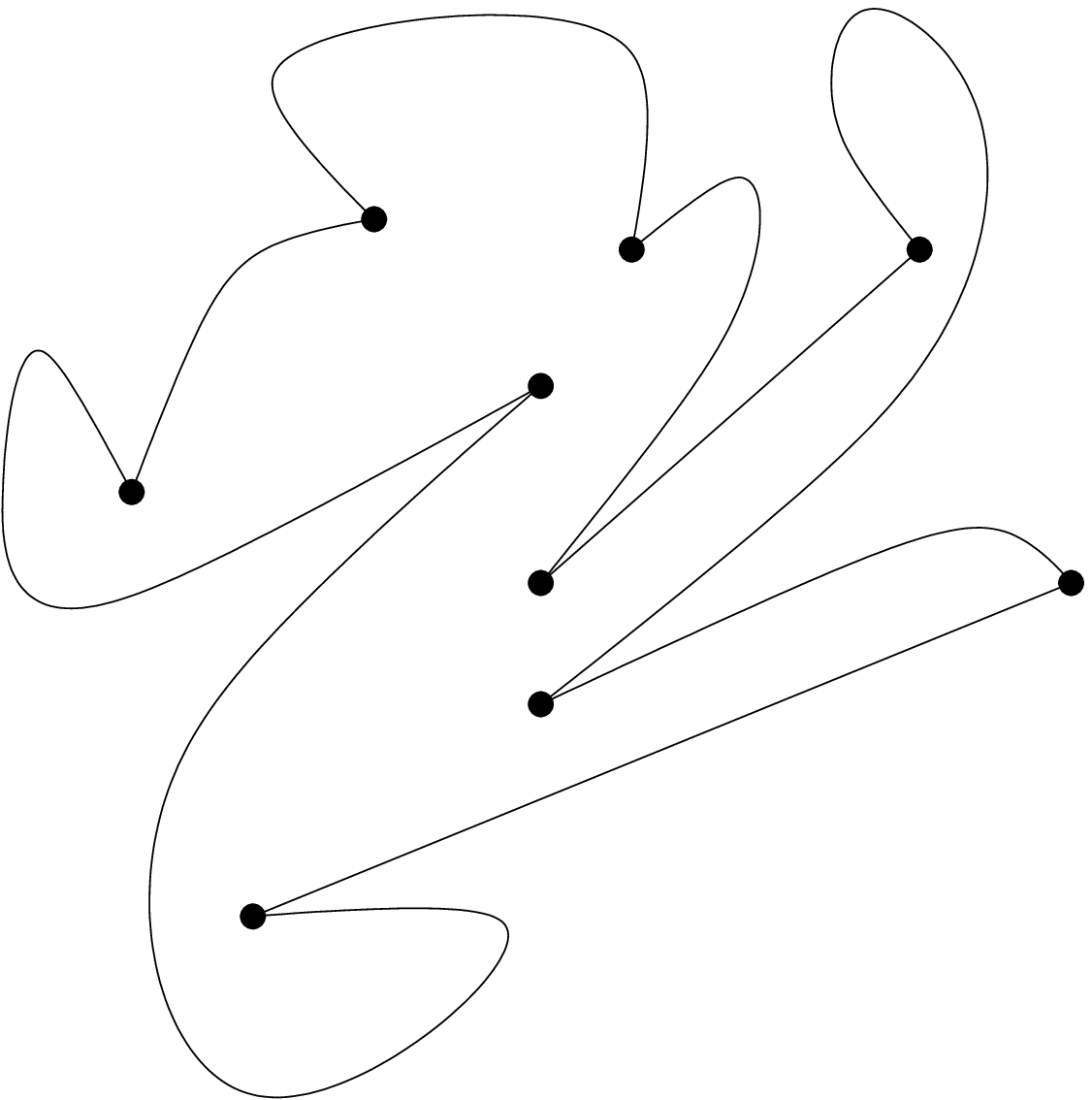}\hfill%
  \includegraphics[width=0.5\columnwidth]{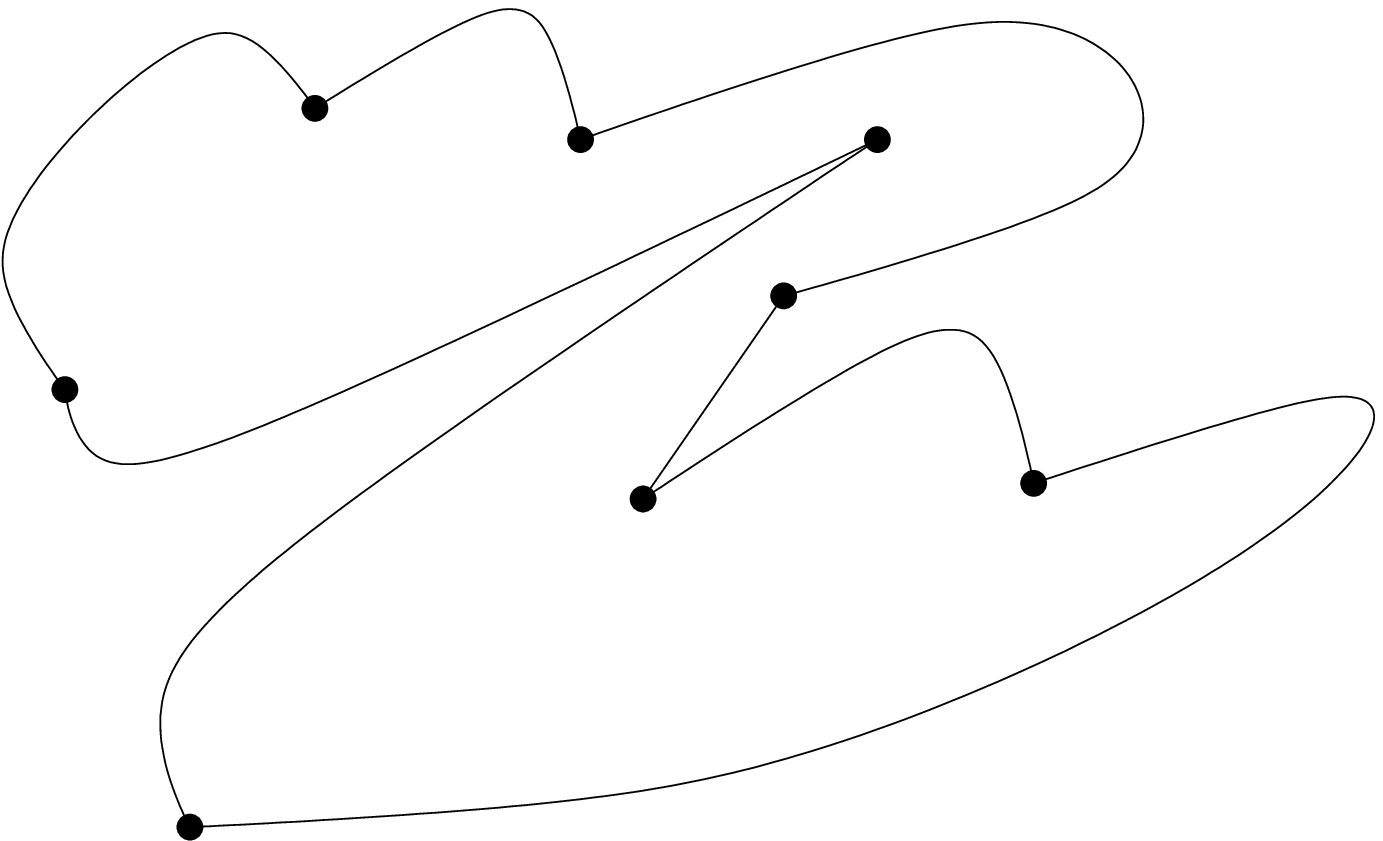}
  \caption{Left: A locally convex polygon. Right: A \pconvex polygon.}
  \label{fig:poly-types}
\end{center}
\end{figure}

Let $v_1,\ldots,v_n$, $n\ge{}2$, be a sequence of points and 
$a_1,\ldots,a_n$ a set of curvilinear arcs, such that $a_i$ has as
endpoints the points $v_i$ and $v_{i+1}$. We will assume that
the arcs $a_i$ and $a_j$, $i\ne j$, do not intersect, except when
$j=i-1$ or $j=i+1$, in which case they intersect only at the points
$v_i$ and $v_{i+1}$, respectively. We define a \emph{curvilinear polygon} $P$
to be the closed region of the plane delimited by the arcs $a_i$. The
points $v_i$ are called the vertices of $P$. An arc $a_i$ is
a \emph{convex arc} if every line on the plane intersects $a_i$ at at
most two points or along a line segment.
A polygon $P$ is called a \emph{locally convex polygon}, if for every
point $p$ on the boundary of $P$, with the possible exception of $P$'s
vertices, there exists a disk centered at $p$, say $D_p$,  such that
$P\cap{}D_p$ is convex (see Fig. \ref{fig:poly-types}(left)).
A polygon $P$ is called a \emph{\pconvex polygon}, if it is
locally convex, and the portion of the boundary between every two
consecutive vertices is a convex arc (see
Fig. \ref{fig:poly-types}(right)).

Let $a_i$ be an edge of a \pconvex polygon $P$ with
endpoints $v_i$ and $v_{i+1}$. We call the convex region $r_i$
delimited by $a_i$ and $\lseg{v_iv_{i+1}}$ a \emph{room}, where
$\lseg{xy}$ denotes the line segment from $x$ to $y$. A room is called
degenerate if the arc $a_i$ is a line segment. For $p,q\in{}a_i$,
$\lseg{pq}$ is called a \emph{chord} of $a_i$; the chord
of $r_i$ is $\lseg{v_iv_{i+1}}$. An \emph{empty room} is a
non-degenerate room that does not contain any vertex of $P$ in the
interior of $r_i$ or in the interior of $\lseg{v_iv_{i+1}}$. A
\emph{non-empty room} is a non-degenerate room that contains at least
one vertex of $P$ in the interior of $r_i$ or in the interior of 
$\lseg{v_iv_{i+1}}$.

We say that a point $p$ in the interior of a \pconvex polygon $P$ is
visible from a point $q$ if $\lseg{pq}$ lies in the closure of $P$. We
say that $P$ is \emph{\guard[ed]{}} by a \emph{\gset{}} $G$ if every
point in $P$ is visible from at least one point belonging to some
guard in $G$. A \emph{diagonal} of a \pconvex polygon $P$ is a
straight-line segment in the closure of $P$ the endpoints of which are
vertices of $P$.
An \emph{edge (\resp mobile) guard} is an edge (\resp edge or
diagonal) of $P$ belonging to a \gset $G$ of $P$.
An \emph{edge (\resp mobile) \gset{}} is a \gset
that consists of only edge (\resp mobile) guards.

Let $P$ be a \pconvex polygon with $n\ge{}3$ vertices.
Consider a convex arc $a_i$ of $P$, with endpoints $v_i$ and
$v_{i+1}$, and let $r_i$ be the corresponding room. If $r_i$ is a
non-empty room, let $X_i$ be the set of vertices of $P$ that lie in
the interior of $\lseg{v_iv_{i+1}}$, and let $R_i$ be the
set of vertices of $P$ in the interior of $r_i$ or in $X_i$.
If $R_i\ne{}X_i$, let $C_i$ be the set of vertices in the convex hull
of the vertex set $(R_i\setminus{}X_i)\cup\{v_i,v_{i+1}\}$; if
$R_i=X_i$, let $C_i=X_i\cup\{v_i,v_{i+1}\}$. Finally, let
$C_i^*=C_i\setminus\{v_i,v_{i+1}\}$. If $r_i$ is an empty room, let
$C_i=\{v_i,v_{i+1}\}$ and $C_i^*=\emptyset$. 

\begin{figure}[!b]
\begin{center}
  \includegraphics[width=0.45\columnwidth]{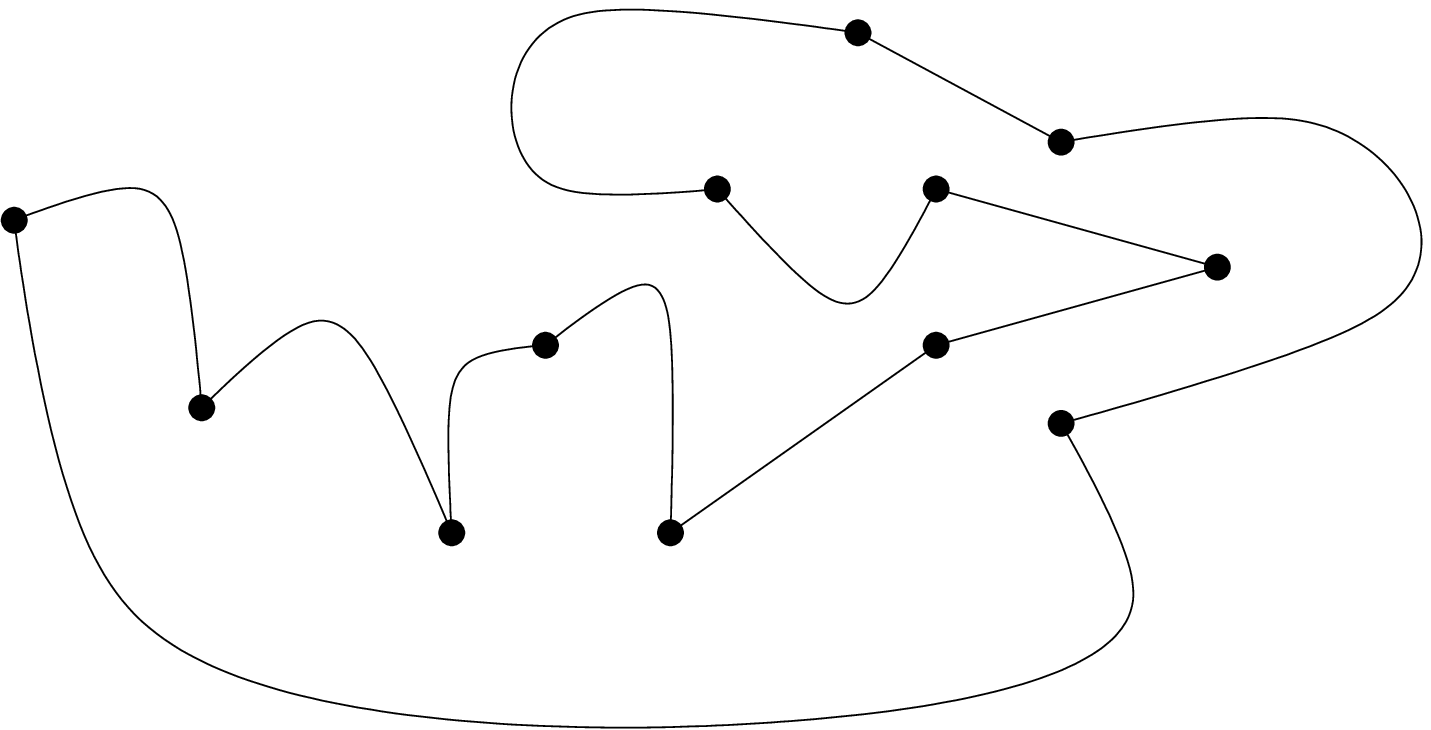}\hfill%
  \includegraphics[width=0.45\columnwidth]{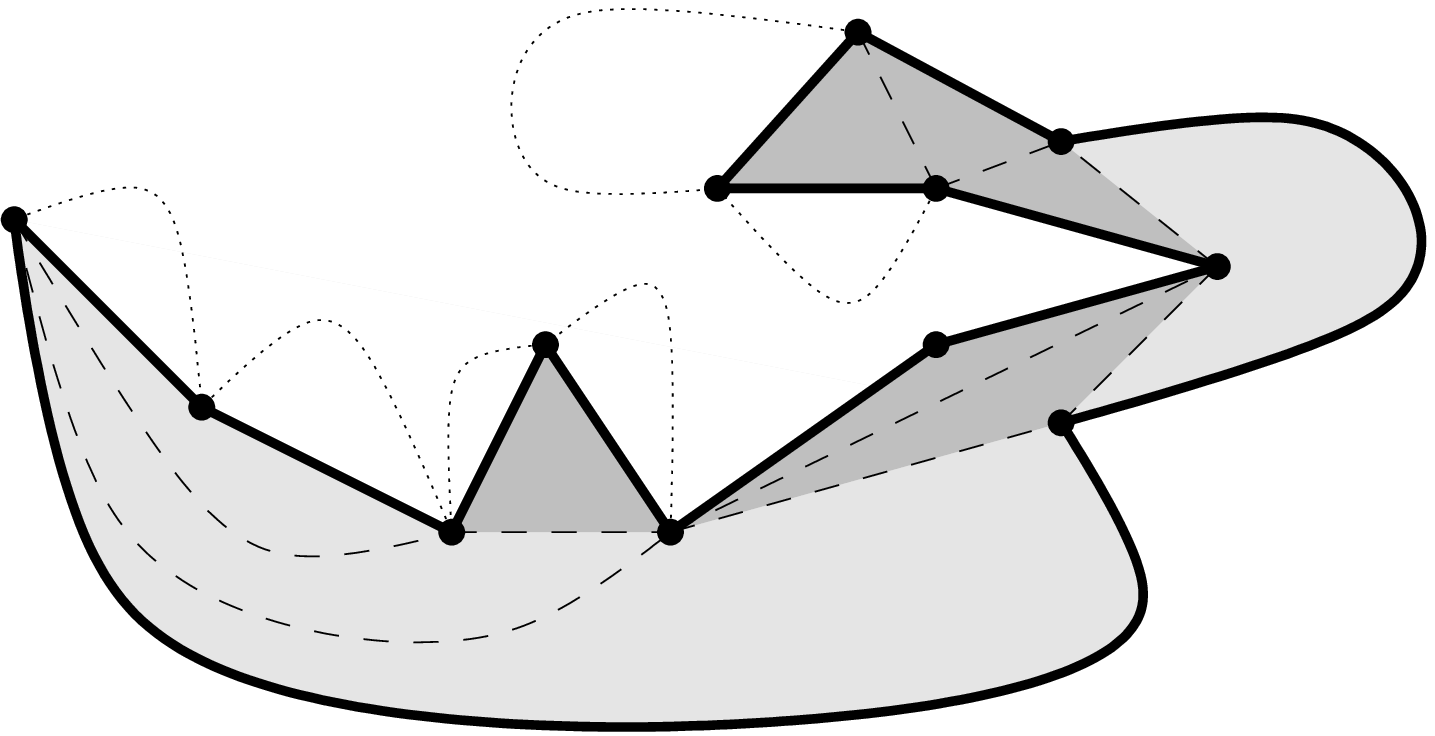}
  \caption{Left: A \pconvex polygon $P$. Right: The triangulation
    graph $\trg{P}$ of $P$. The boundary edges of $\trg{P}$ are shown
    as thick solid lines. The \new{two} crescents of $P$ are shown in light
    gray, whereas the \new{three} stars of $P$ are shown in dark gray.}
  \label{fig:poly-trgraph}
\end{center}
\end{figure}

\new{We are now going to construct a constrained triangulation graph
  $\trg{P}$ of $P$.} The vertex set of $\trg{P}$ is the set of
vertices of $P$. The edges and diagonals of $\trg{P}$, as well as
their embedding, are defined as follows (see also
Fig. \ref{fig:poly-trgraph}):
\begin{itemize}
\item[$\bullet$] If $a_i$ is a line segment or $r_i$ is an empty
  room, the edge $(v_i,v_{i+1})$ is an edge in $\trg{P}$, and is
  embedded as $\lseg{v_iv_{i+1}}$.
\item[$\bullet$] If $r_i$ is a non-empty room, the following edges or
  diagonals belong to $\trg{P}$:
  \begin{enumerate}
  \item $(v_i,v_{i+1})$,
  \item $(c_{i,j},c_{i,j+1})$, for $1\le{}j\le{}\new{|C_i|}-1$, where
    $c_{i,1}\equiv{}v_i$ and $c_{i,\new{|C_i|}}\equiv{}v_{i+1}$. The
    remaining $c_i$'s are the vertices of $P$ in $C_i^*$ as we
    encounter them when walking inside $r_i$ and on the convex hull of
    the point set $C_i$ from $v_i$ to $v_{i+1}$, and
  \item $(v_i,c_{i,j})$, for $3\le{}j\le{}\new{|C_i|}-1$, provided
    that $\new{|C_i|}\ge{}4$. We call these diagonals \emph{weak
      diagonals}.
  \end{enumerate}
  The diagonals $(c_{i,j},c_{i,j+1})$, $1\le{}j\le{}\new{|C_i|}-1$ are embedded
  as $\lseg{c_{i,j},c_{i,j+1}}$, whereas the diagonals $(v_i,c_{i,j})$,
  $3\le{}j\le{}\new{|C_i|}-1$, are embedded as curvilinear segments. Finally,
  the edges $(v_i,v_{i+1})$ are embedded as curvilinear segments,
  namely, the arcs $a_i$.
\end{itemize}

The edges $(v_i,v_{i+1})$, along with the diagonals $(c_{i,j},c_{i,j+1})$,
$1\le{}j\le{}\new{|C_i|}-1$, partition $P$ into subpolygons of two types:
(1) subpolygons that lie entirely inside a non-empty room, called
\emph{crescents}, and
(2) subpolygons delimited by edges of the polygon $P$, as well as
diagonals of the type $(c_{i,j},c_{i,j+1})$, called \emph{stars}. In
general, a \pconvex polygon may only have crescents, or only
stars, or both. The crescents are triangulated by means of the
diagonals $(v_i,c_{i,j})$, $3\le{}j\le{}\new{|C_i|}-1$. To finish the
definition of the triangulation graph $\trg{P}$, we simply need to 
triangulate all stars inside $P$. Since the delimiting edges of stars
are embedded as line segments, \ie stars are linear polygons, any
polygon triangulation algorithm may be used to triangulate them.

In direct analogy to the types of subpolygons we can have inside $P$,
we have two possible types of triangles in $\trg{P}$:
(1) triangles inside stars, called \emph{star triangles}, and 
(2) triangles inside a crescent, called \emph{crescent triangles}.
Crescent triangles have at least one edge that is a weak diagonal,
except when the number of vertices of $P$ in the interior of the
corresponding room $r$ is exactly one, in which case none of the three
edges of the unique crescent triangle in $r$ is a weak diagonal. A
crescent triangle that has at least one weak diagonal among its edges
is called a \emph{weak triangle}.


\subsection{Mobile guards}

\begin{figure}[!t]
\begin{center}
  \includegraphics[width=0.45\columnwidth]{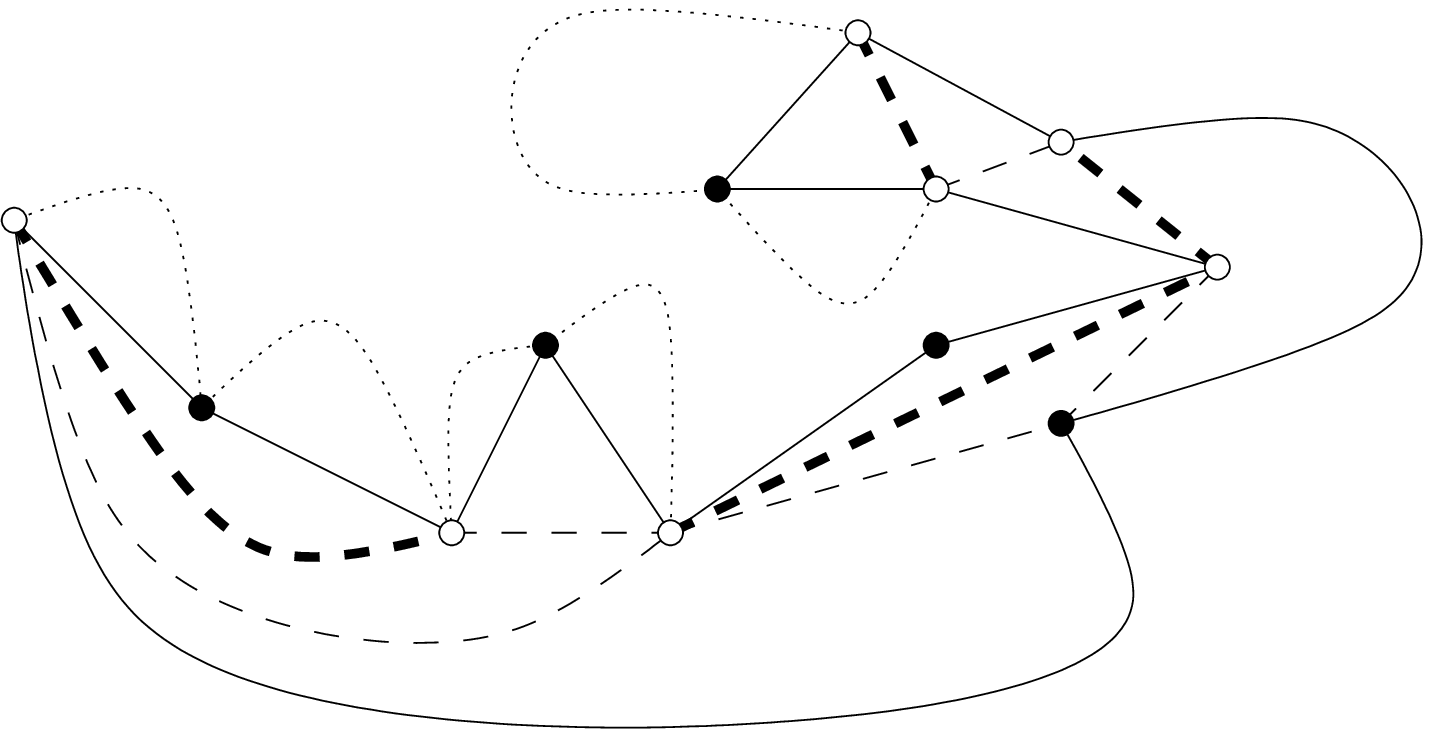}\hfill%
  \includegraphics[width=0.45\columnwidth]{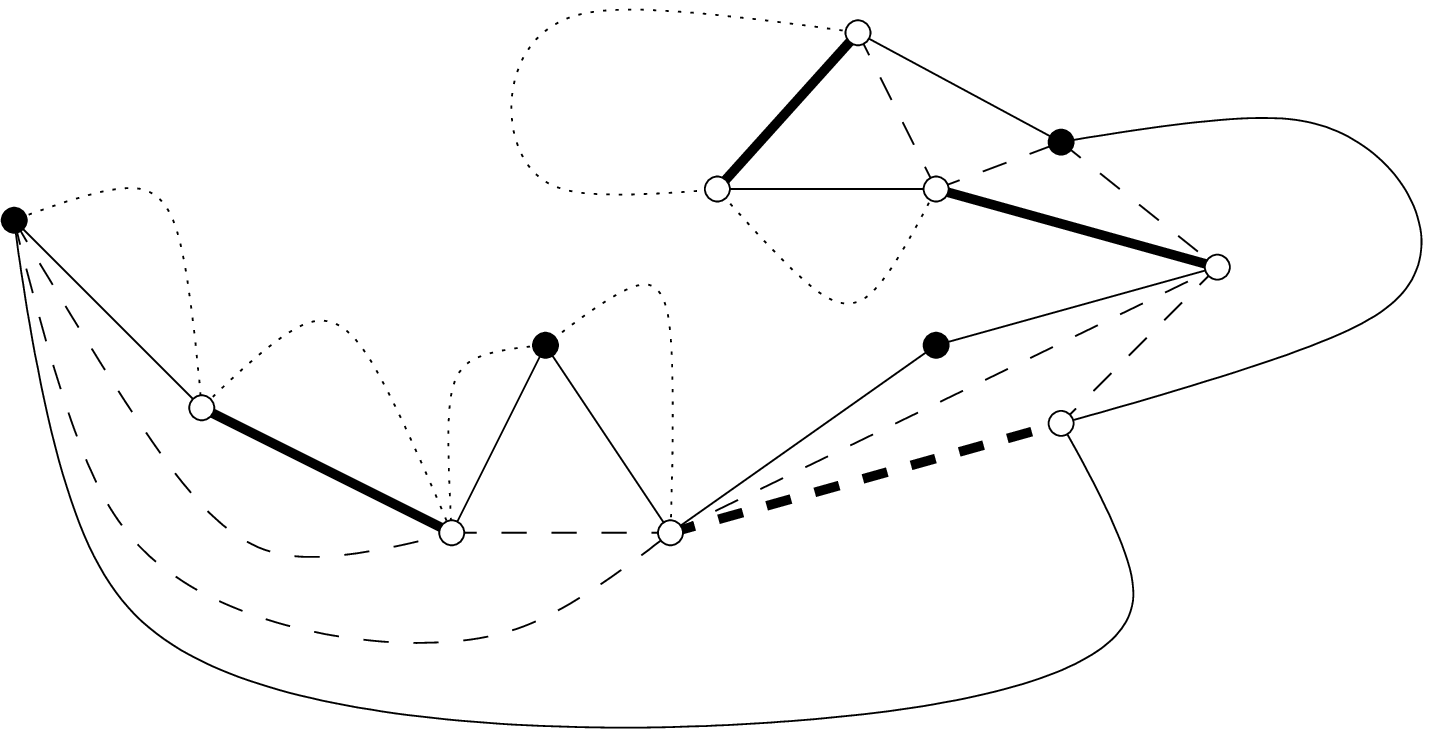}\\\vspace*{5mm}
  \includegraphics[width=0.45\columnwidth]{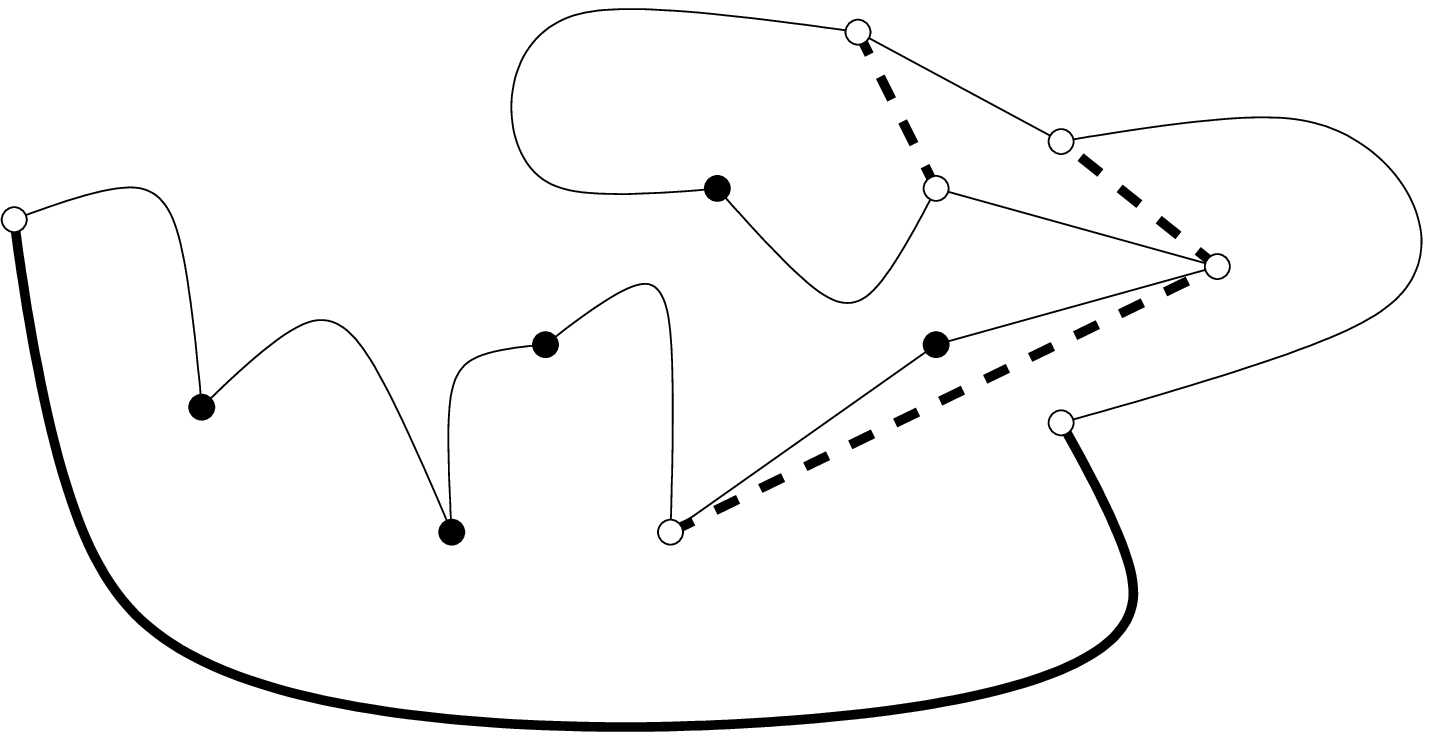}\hfill%
  \includegraphics[width=0.45\columnwidth]{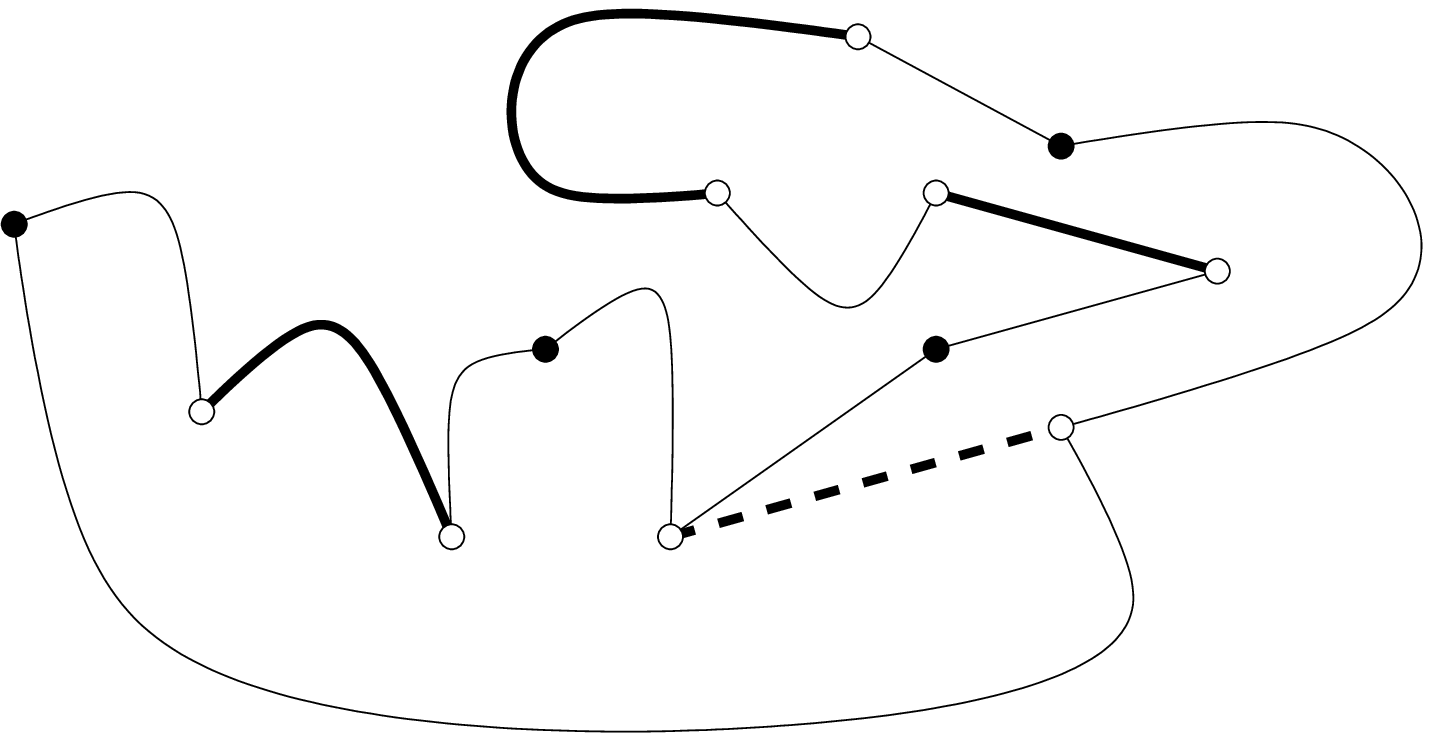}
  \caption{Top row: two diagonal 2-dominating sets for the
    triangulation graph $\trg{P}$ of $P$ from
    Fig. \ref{fig:poly-trgraph}. Bottom row: the corresponding mobile
    \gset{s} for $P$.}
  \label{fig:mobile_gset}
\end{center}
\end{figure}

Let $G_{\trg{P}}$ be a diagonal 2-dominating set of $\trg{P}$. Based
on $G_{\trg{P}}$ we define a set $G$ of edges or straight-line
diagonals of $P$ as follows (see also Fig. \ref{fig:mobile_gset}):
\new{
(1) for every edge in $G_{\trg{P}}$, add to $G$ the corresponding
convex arc of $P$,}
\new{(2)} add to $G$ every non-weak diagonal of $G_{\trg{P}}$, and
\new{(3)} for every weak diagonal in $G_{\trg{P}}$, add to $G$ the edge of
$P$ delimiting the crescent that contains the weak \new{diagonal}.
Clearly, $|G|\le|G_{\trg{P}}|$.

\begin{lemma}\label{lem:diag-guarding-set}
Let $P$ be a \pconvex polygon with $n\ge{}3$ vertices,
$\trg{P}$ its constrained triangulation graph, and $G_{\trg{P}}$ a
diagonal 2-dominating set of $\trg{P}$. The set $G$ of mobile guards,
defined by mapping \new{every edge of $G_{\trg{P}}$ to the
  corresponding convex arc of $P$},
every non-weak \new{diagonal} of $G_{\trg{P}}$ to itself, and
every weak \new{diagonal $d$} of $G_{\trg{P}}$ to the convex arc
of $P$ delimiting the crescent that contains \new{$d$}, is a mobile
\gset for $P$.
\end{lemma}

\begin{proof}
Let $q$ be a point in the interior of $P$. $q$ is either inside:
(1) an empty room $r_i$ of $P$,
(2) a star triangle $t_s$ of $\trg{P}$,
(3) a non-weak crescent triangle $t_{nw}$ of $\trg{P}$, or
(4) a weak crescent triangle $t_w$ of $\trg{P}$.
In any of the four cases, $q$ is visible from at least two
vertices $u_1$ and $u_2$ of $\trg{P}$ that are connected via an edge
or a diagonal in $\trg{P}$. In the first case, $q$ is visible from the
two endpoints $v_i$ and $v_{i+1}$ of $a_i$. In the second case, $q$ is
visible from all three vertices of $t_s$. The third case arises
when $q$ is inside a non-empty room $r_j$ with $|C_j^*|=1$
($t_{nw}$ is the unique crescent triangle in $r_j$), in which
case $q$ is visible from at least two of the three vertices $v_j$,
$v_{j+1}$ and $c_{j,1}$. Finally, in the fourth case, $q$ has to lie
inside the crescent of a non-empty room $r_j$ with $|C_j^*|\ge{}2$,
and is visible from at least two consecutive vertices $c_{j,k}$ and
$c_{j,k+1}$ of $C_j$.

Since $G$ is a diagonal 2-dominating set for $\trg{P}$, and
$(u_1,u_2)\in\trg{P}$, at least one of $u_1$ and $u_2$ belongs to
$G_{\trg{P}}$. Without loss of generality, let us assume that
$u_1\in{}G_{\trg{P}}$. If $u_1\in{}G$, $q$ is \guard[ed] by $u_1$.
If $u_1\nin{}G$, $u_1$ has to be an endpoint of a weak diagonal $d_w$
in $G_{\trg{P}}$. Let $r_\ell$ be the room, inside the crescent of
which lies $d_w$. Since $d_w\in{}G_{\trg{P}}$, we have that
$a_\ell\in{}G$. If $q$ lies inside the \new{closure of the} crescent
of the room $r_\ell$ (this can happen in case (4) above), $q$ is
visible from $a_\ell$, and thus \guard[ed] by $a_\ell$. Otherwise,
$u_1$ cannot be an endpoint of $a_\ell$ ($a_\ell\in{}G$, whereas
$u_1\nin{}G$), which implies that $u_1\in{}C_\ell^*$, \ie
$u_1\equiv{}c_{\ell,m}$, with $2\le{}m\le{}\new{|C_\ell|}-1$.
But then $q$ lies inside the cone with apex 
$c_{\ell,m}$, delimited by the rays $c_{\ell,m}c_{\ell,m-1}$ and
$c_{\ell,m}c_{\ell,m+1}$, and containing at least one of $v_\ell$ and
$v_{\ell+1}$ in its interior. Since, $q$ is visible from the
intersection point of the line $qu_1$ with $a_\ell$, $q$ is \guard[ed]
by $a_\ell$.
\end{proof}

Our approach for computing the mobile \gset $G$ of $P$ consists
of three major steps:
\begin{enumerate}
\item Construct the constrained triangulation $\trg{P}$ of $P$.
\item Compute a diagonal 2-dominating set $G_{\trg{P}}$ for the
  triangulation graph $\trg{P}$.
\item Map $G_{\trg{P}}$ to $G$.
\end{enumerate}
The sets $C_i^*$, needed in order to construct the constrained
triangulation $\trg{P}$ of $P$ can be computed in $O(n\log{}n)$ time
and $O(n)$ space (cf. \cite{ktt-gcagv-09}). Once we have the sets
$C_i^*$, the constrained triangulation $\trg{P}$ of $P$ can be
constructed in linear time and space. By Theorem \ref{thm:trg-diag-timespace},
computing $G_{\trg{P}}$ takes linear time; furthermore
$|G_{\trg{P}}|\le\bm$, which implies that $|G|\le\bm$. Finally, the
construction of $G$ from $G_{\trg{P}}$ takes $O(n)$ time and space:
\newbegin
for every edge in $G_{\trg{P}}$ we need to add to $G$ the
corresponding convex arc of $P$, while for every diagonal $d$ in
$G_{\trg{P}}$ we need to determine if it is a weak diagonal, in which
case we need to add to $G$ the edge of $P$ delimiting the crescent in
which $d$ lies, otherwise we simply add $d$ to $G$;
\newend
by appropriate bookkeeping at the time of construction of $\trg{P}$
these operations can take $O(1)$ per \new{edge or} diagonal.
Summarizing, by Theorem \ref{thm:guard-combdiag-trgraph}, Lemma
\ref{lem:diag-guarding-set} and our analysis above, we arrive at the
following theorem. The case $n=2$ can be trivially established.

\begin{theorem}\label{thm:upperbound-diags}
Let $P$ be a \pconvex polygon with $n\ge{}2$ vertices. We can
compute a mobile \gset for $P$ of size at most $\bm$ in
$O(n\log{}n)$ time and $O(n)$ space.
\end{theorem}


\subsection{Edge guards}

\begin{figure}[!b]
\begin{center}
  \includegraphics[width=0.45\columnwidth]{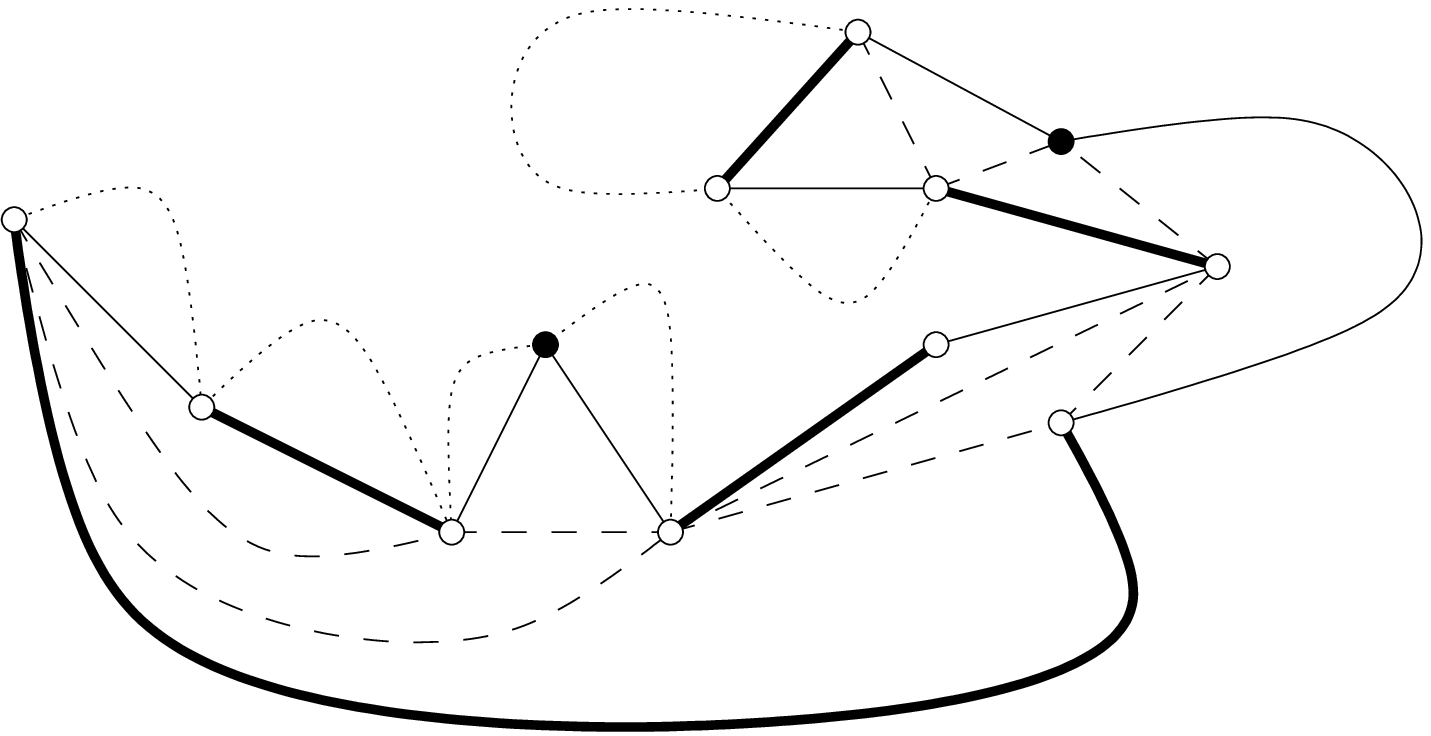}\hfill%
  \includegraphics[width=0.45\columnwidth]{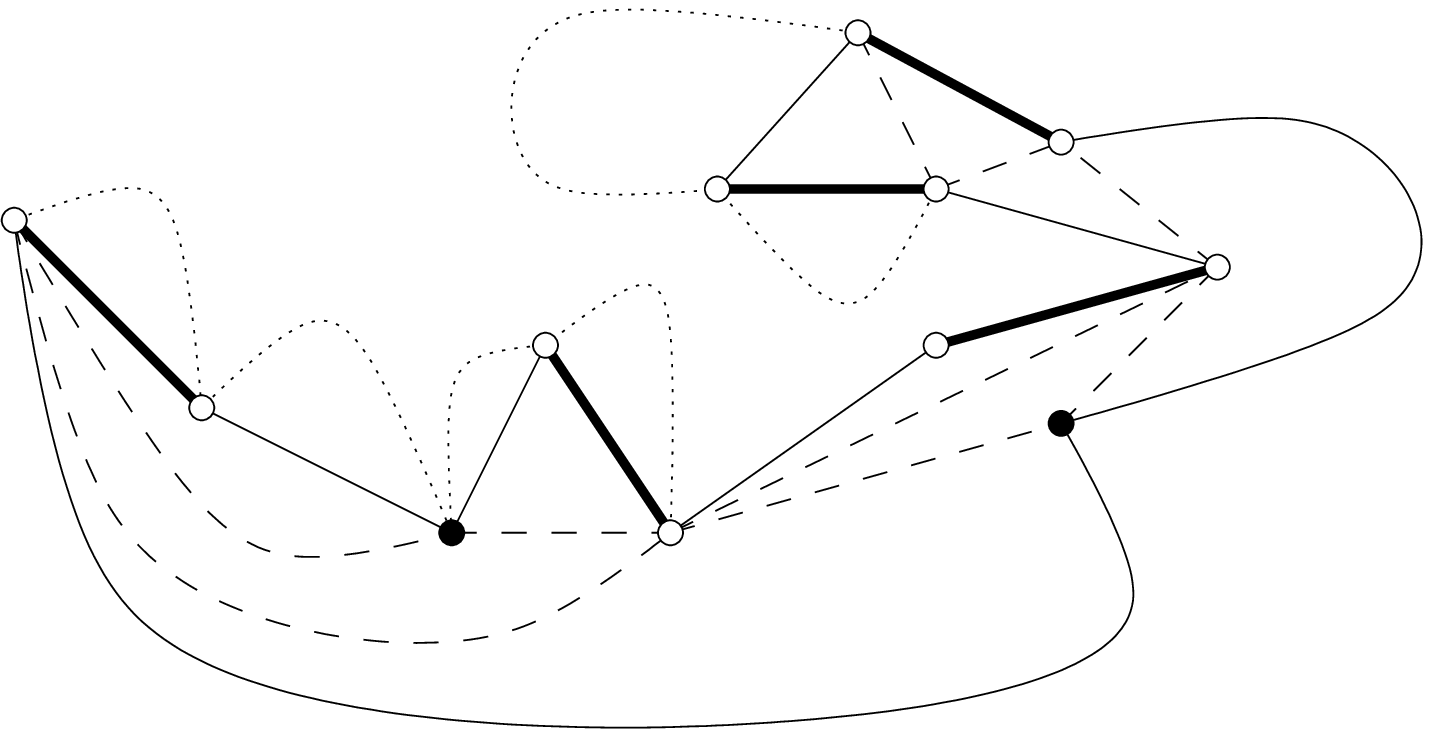}\\\vspace*{5mm}
  \includegraphics[width=0.45\columnwidth]{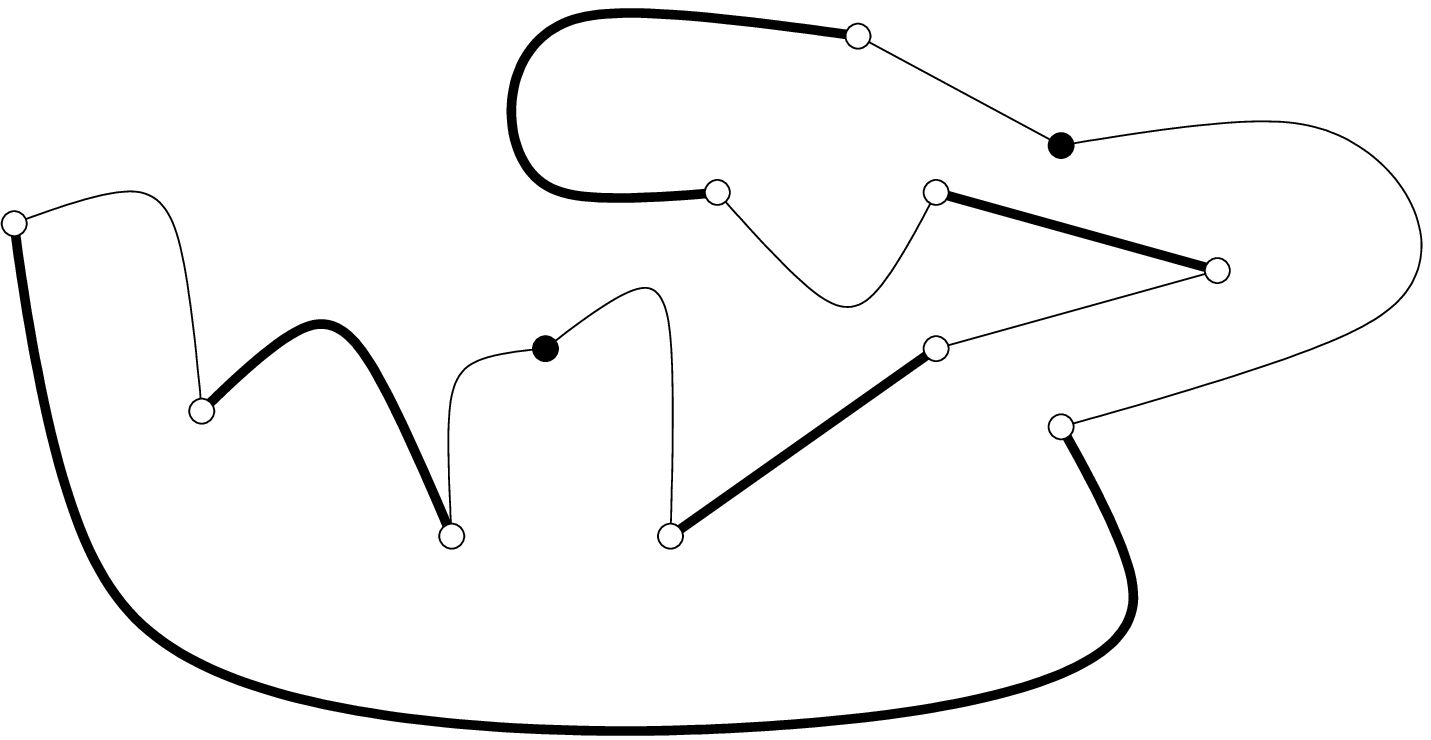}\hfill%
  \includegraphics[width=0.45\columnwidth]{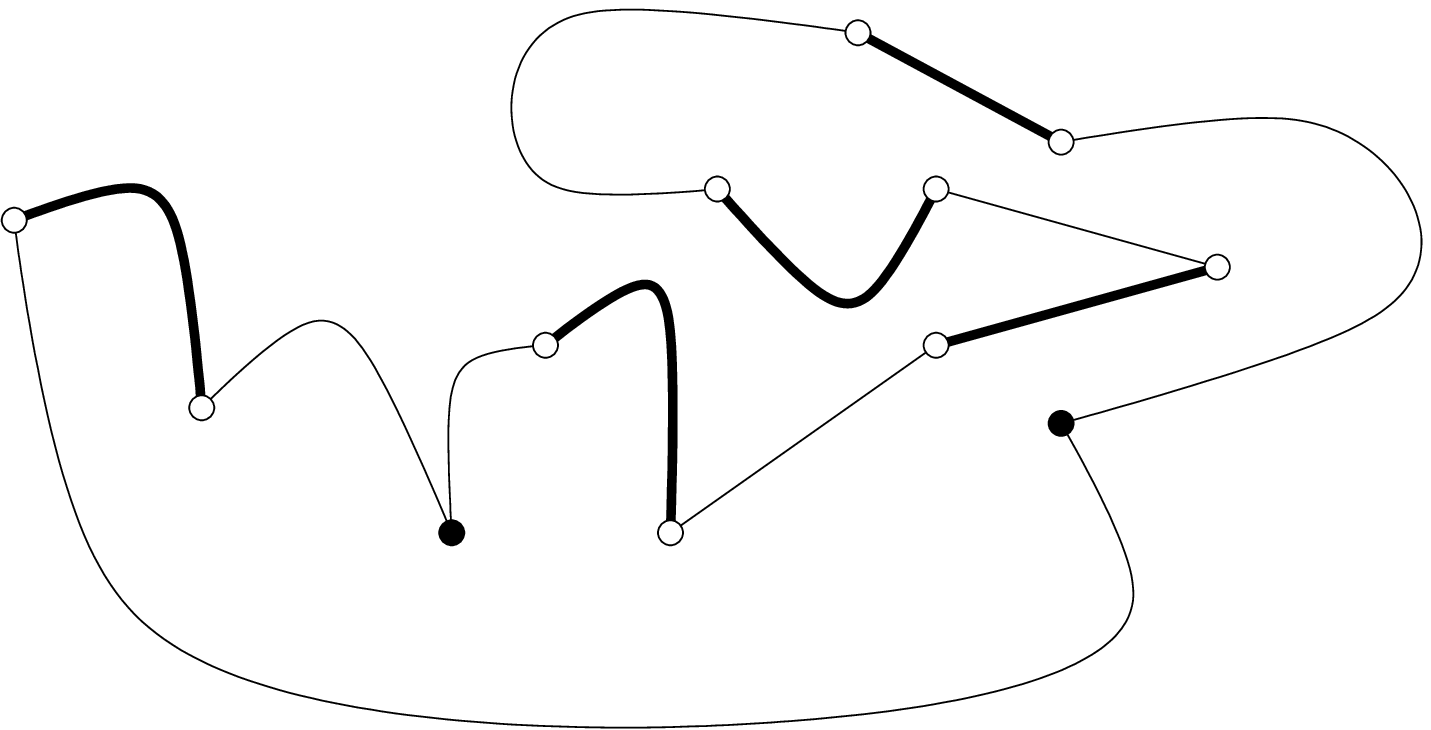}
  \caption{Top row: two edge 2-dominating sets for the triangulation
    graph $\trg{P}$ of $P$ from Fig. \ref{fig:poly-trgraph}. Bottom
    row: the corresponding edge \gset{s} for $P$.}
  \label{fig:edge_gset}
\end{center}
\end{figure}

We start by proving that an edge 2-dominating set for $\trg{P}$ is
also an edge \gset for $P$ (see also Fig. \ref{fig:edge_gset}).

\begin{lemma}\label{lem:edge-guarding-set}
Let $P$ be a \pconvex polygon with $n\ge{}3$ vertices,
$\trg{P}$ its constrained triangulation graph, and $G_{\trg{P}}$ an
edge 2-dominating set of $\trg{P}$. The set $G$ of edge guards,
defined by mapping every edge in $G_{\trg{P}}$ to the corresponding
convex arc of $P$, is an edge \gset for $P$.
\end{lemma}

\begin{proof}
Let $q$ be a point in the interior of $P$. Recall the four cases for
$q$ from the proof of Lemma \ref{lem:diag-guarding-set}. $q$ is either
inside:
(1) an empty room of $P$,
(2) a star triangle of $\trg{P}$,
(3) a non-weak crescent triangle of $\trg{P}$, or
(4) a weak crescent triangle of $\trg{P}$.
In any of the four cases, $q$ is visible from at least two
vertices $u_1$ and $u_2$ of $\trg{P}$, such that the edge or diagonal
$(u_1,u_2)$ belongs to $\trg{P}$. Let $t$ be a triangle supported by
$(u_1,u_2)$ in $\trg{P}$. At least two of the vertices of $t$ belong
to $G_{\trg{P}}$, which implies that at least one of $u_1$ and $u_2$,
belongs to \new{$G_{\trg{P}}$. Since the set of vertices that are
  endpoints of edges in $G_{\trg{P}}$ is the same as the set of
  vertices that are endpoints of edges in $G$, we conclude that $q$ is
  \guard[ed] by a vertex that is an endpoint of an edge in $G$}.
\end{proof}

By Theorems \ref{thm:guard-combedge-trgraph} and
\ref{thm:trg-edge-timespace-weak},
we can either compute an edge 2-dominating set $G_{\trg{P}}$ of size
$\be$ in $O(n^2)$ time and $O(n)$ space, or an edge 2-dominating set
$G_{\trg{P}}$ of size $\ubew$ \new{in linear time and space (except for
$n=4$ where one additional edge is needed in both cases).}
As in the case of mobile guards, the constrained triangulation graph
$\trg{P}$ of $P$ can be computed in $O(n\log{}n)$ time and $O(n)$
space. Since $|G|=|G_{\trg{P}}|$, we arrive at the following
theorem. The case $n=2$ is trivial, since in this case any of the two
edges of $P$ is an edge \gset for $P$.

\begin{theorem}\label{thm:upperbound-edges}
Let $P$ be a \pconvex polygon with $n\ge{}2$ vertices. We can
either: (1) compute an edge \gset for $P$ of size $\be$
(except for $n=4$, where one additional edge guard is required) in
$O(n^2)$ time and $O(n)$ space, or (2) compute an edge \gset
for $P$ of size $\ubew$ (except for $n=2,4$, where one additional edge
guard is required) in $O(n\log{}n)$ time and $O(n)$ space.
\end{theorem}


\subsection{Lower bound constructions}

\begin{figure}[!b]
\begin{center}
\includegraphics[width=0.75\columnwidth]{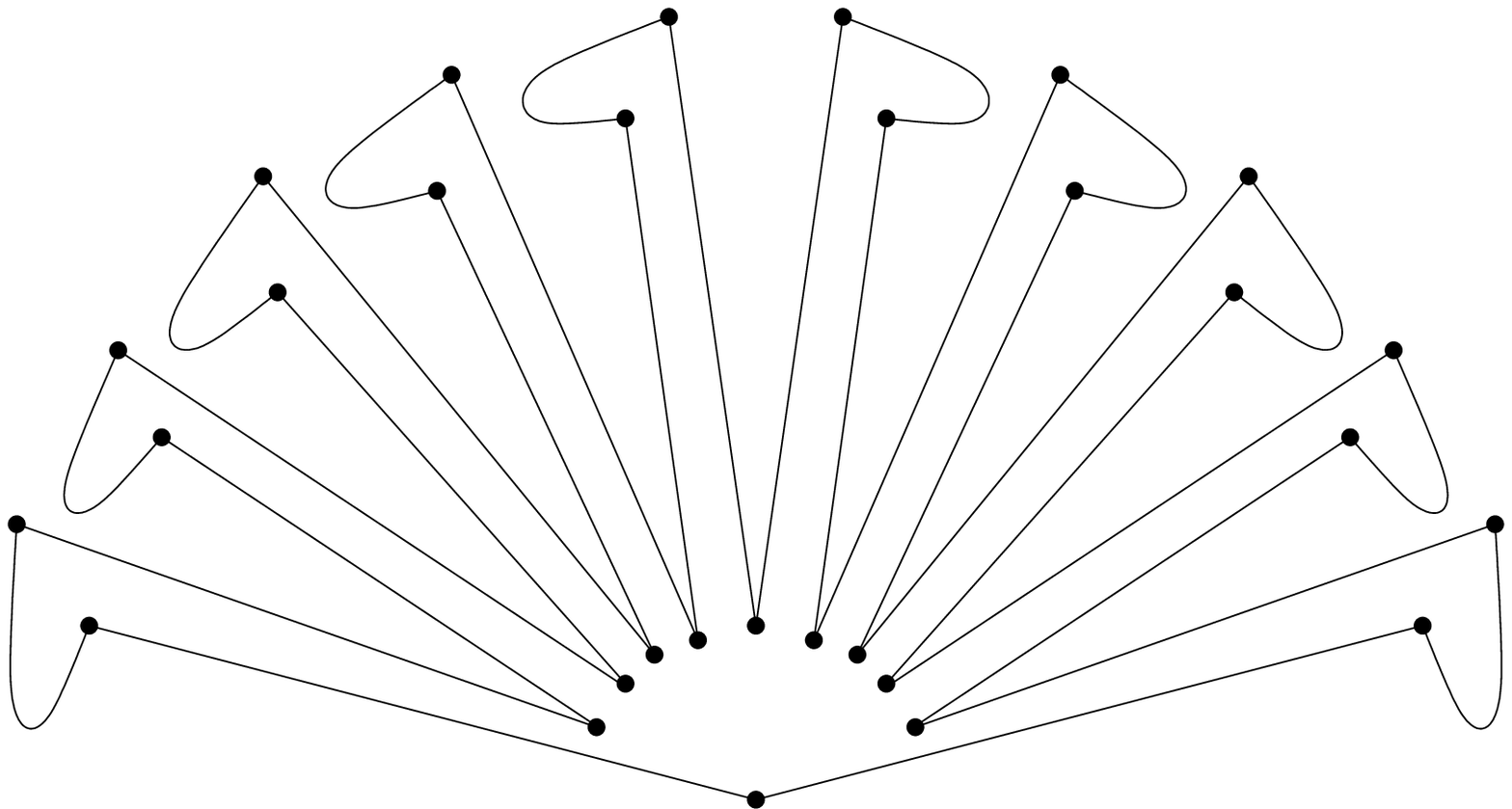}
\caption{The lower bound construction for mobile guards: the polygon
  shown contains $n=3k$ vertices, and requires $k=\lbmg$ mobile guards
  in order to be \guard[ed].}
\label{fig:lowerbound_mobile}
\end{center}
\end{figure}

Consider the \pconvex polygon $P$ of Fig. \ref{fig:lowerbound_mobile}.
Each spike consists of three edges, namely, two line segments and
a convex arc. In order for points in the non-empty room of the convex
arc to be \guard[ed], either one of the three edges of the spike, or a
diagonal at least one endpoint of which is an endpoint of the convex
arc, has to be in any \gset of $P$: the chosen edge or diagonal in a
spike cannot \guard the non-empty room inside another spike of
$P$. Since $P$ consists of $k$ spikes, yielding $n=3k$ vertices, we
need at least $k$ mobile guards to \guard $P$. We, thus, conclude that
$P$ requires at least $\lbmg$ mobile guards in order to be \guard[ed].

\begin{theorem}\label{thm:guarding-set-lb-mobile}
There exists a family of \pconvex polygons with $n\ge{}3$ vertices any
mobile \gset of which has cardinality at least $\lbmg$.
\end{theorem}

\begin{figure}[t]
\begin{center}
\includegraphics[width=0.75\columnwidth]{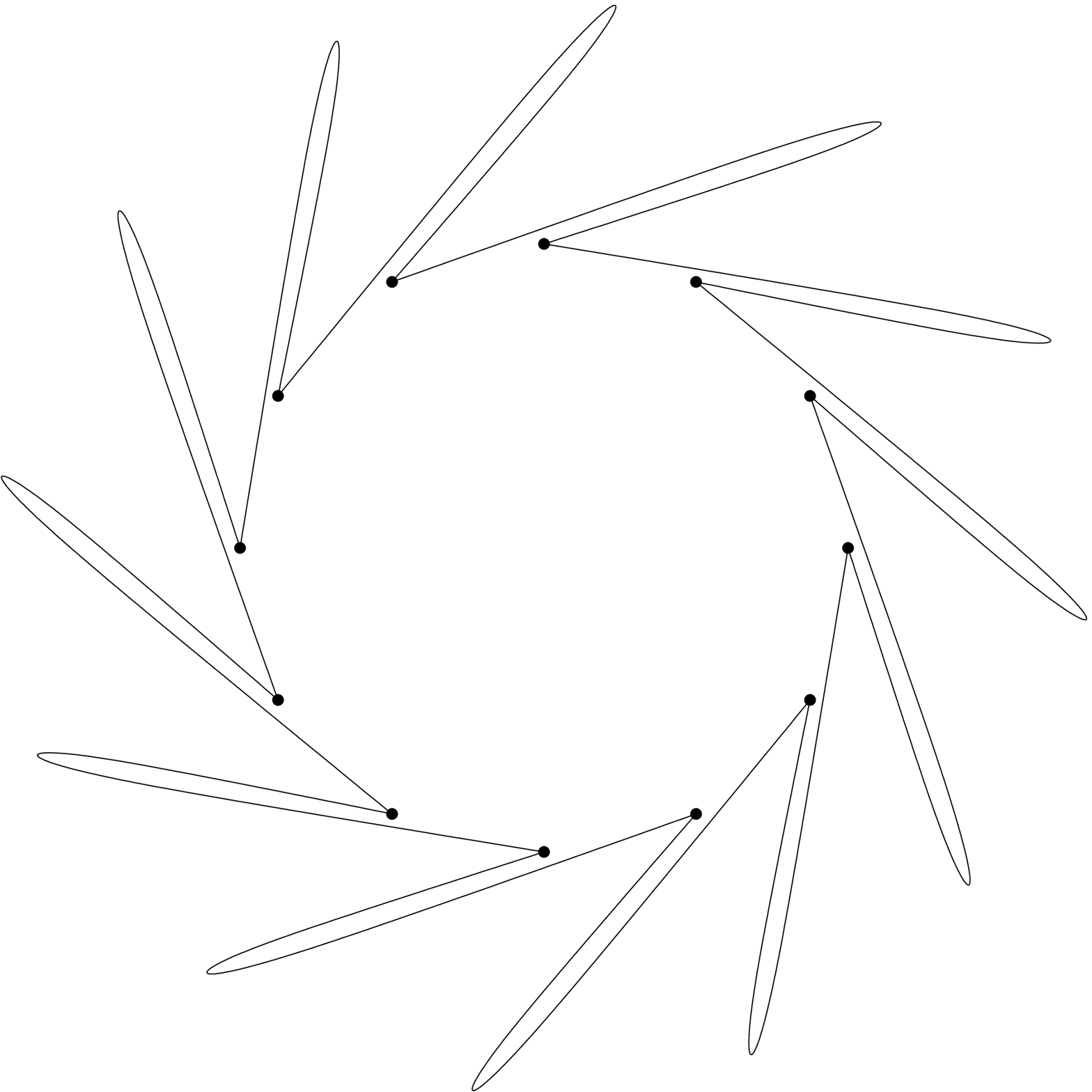}
\caption{The lower bound construction for edge guards: the polygon
  shown contains $n$ vertices, and requires $k=\lbeg$ edge guards in
  order to be \guard[ed].}
\label{fig:lowerbound_edges}
\end{center}
\end{figure}

Our lower bound for edge guards is slightly better than for
mobile guards. Consider the fan-like $n$-vertex \pconvex polygon $F$
of Fig. \ref{fig:lowerbound_edges}. $F$ is constructed from a regular
$n$-gon by replacing each edge of the $n$-gon by a highly tilted
spike. The spike $s$, bounded by the edge $e_s$ of $F$, can only be
\guard[ed] by the points of $e_s$, or some of the points of the two
neighboring edges of $e_s$. This immediately implies that in order to
\guard $F$ we need a minimum of $\lbeg$ edge guards. To see this,
assume that there exists an edge \gset $G$ for $F$ of size
$|G|<\lbeg$. Then we would be able to find three consecutive edges
$e_1$, $e_2$, $e_3$ of $F$ that do not belong to $G$, which implies
that the spike bounded by $e_2$ is not \guard[ed] by $G$, a
contradiction.

\begin{theorem}\label{thm:guarding-set-lb-edge}
There exists a family of \pconvex polygons with $n\ge{}3$ vertices any
edge \gset of which has cardinality at least $\lbeg$.
\end{theorem}


\section{Monotone piecewise-convex polygons}
\label{sec:monotone}

\begin{figure}[!b]
  \begin{center}
    \psfrag{u0}[][]{\scriptsize$u_0$}
    \psfrag{u1}[][]{\scriptsize$u_1$}
    \psfrag{u2}[][]{\scriptsize$u_2$}
    \psfrag{u3}[][]{\scriptsize$u_3$}
    \psfrag{u4}[][]{\scriptsize$u_4$}
    \psfrag{u5}[][]{\scriptsize$u_5$}
    \psfrag{u6}[][]{\scriptsize$u_6$}
    \psfrag{u7}[][]{\scriptsize$u_7$}
    \psfrag{u8}[][]{\scriptsize$u_8$}
    \psfrag{u9}[][]{\scriptsize$u_9$}
    \psfrag{u10}[][]{\scriptsize$u_{10}$}
    \psfrag{k0}[][]{\scriptsize$\kappa_0$}
    \psfrag{k1}[][]{\scriptsize$\kappa_1$}
    \psfrag{k2}[][]{\scriptsize$\kappa_2$}
    \psfrag{k3}[][]{\scriptsize$\kappa_3$}
    \psfrag{k4}[][]{\scriptsize$\kappa_4$}
    \psfrag{k5}[][]{\scriptsize$\kappa_5$}
    \psfrag{k6}[][]{\scriptsize$\kappa_6$}
    \psfrag{k7}[][]{\scriptsize$\kappa_7$}
    \psfrag{k8}[][]{\scriptsize$\kappa_8$}
    \psfrag{k9}[][]{\scriptsize$\kappa_9$}
    \psfrag{l0}[][]{\scriptsize$\ell_0$}
    \psfrag{l1}[][]{\scriptsize$\ell_1$}
    \psfrag{l2}[][]{\scriptsize$\ell_2$}
    \psfrag{l3}[][]{\scriptsize$\ell_3$}
    \psfrag{l4}[][]{\scriptsize$\ell_4$}
    \psfrag{l5}[][]{\scriptsize$\ell_5$}
    \psfrag{l6}[][]{\scriptsize$\ell_6$}
    \psfrag{l7}[][]{\scriptsize$\ell_7$}
    \psfrag{l8}[][]{\scriptsize$\ell_8$}
    \psfrag{l9}[][]{\scriptsize$\ell_9$}
    \psfrag{l10}[][]{\scriptsize$\ell_{10}$}
    \psfrag{e3r}[][]{\scriptsize$e_3^r$}
    \psfrag{e3l}[][]{\scriptsize$e_3^\ell$}
    \psfrag{e3opp}[][]{\scriptsize$e_3^{opp}$}
    \psfrag{e0r}[][]{\scriptsize$e_0^r$}
    \psfrag{e10l}[][]{\scriptsize$e_{10}^\ell$}
    \includegraphics[width=\textwidth]{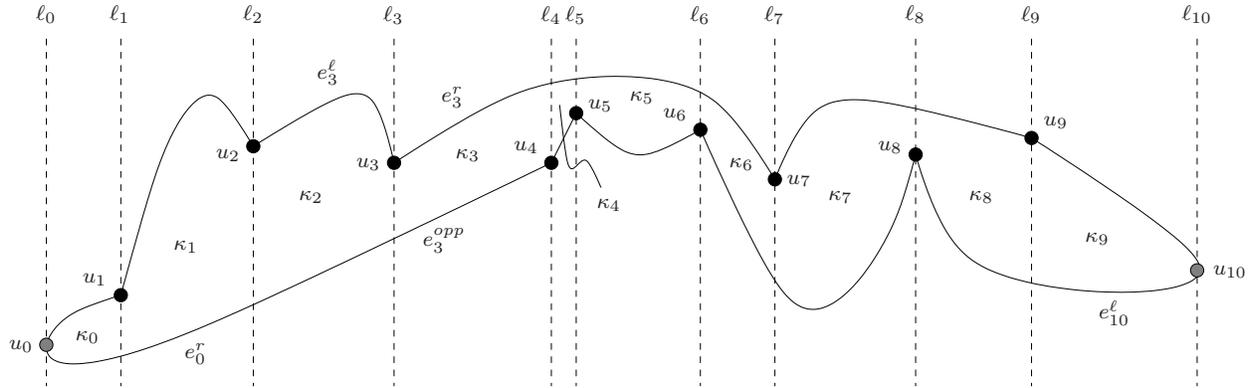}
    \caption{A monotone \pconvex polygon $P$ with 9 vertices. The
      decomposition of $P$ into the convex regions $\kappa_j$,
      $0\le{}j\le{}\new{9}$ is shown. The edges $e_3^\ell$ and $e_3^r$
      are the edges of $P$ having $u_3$ to their left and right,
      respectively. The edge $e_3^{opp}$ is the edge of $P$ opposite
      to $u_3$ (\ie the edge of $P$ intersected by $\ell_3$ lying on
      the monotone chain of $P$ not containing $u_3$). The indices of
      the vertices of $P$ are as follows: $\sigma_0=\sigma_{10}=0$;
      $\sigma_1=\sigma_2=\sigma_3=\sigma_7=\sigma_9=+1$;
      $\sigma_4=\sigma_5=\sigma_6=\sigma_8=-1$.}
    \label{fig:monpconvex}
  \end{center}
\end{figure}

In this section we consider the special case of \emph{monotone
  \pconvex polygons}. We start by restating the definition of
monotonicity: a \pconvex polygon $P$ is called \emph{monotone} if
there exists a line $L$, such that every line $L^\perp$ perpendicular
to $L$ intersects $P$ at at most two points or line segments. Without
loss of generality we may assume that the line $L$, with respect to
which $P$ is monotone, is the $x$-axis.
Let $u_j$, $1\le{}j\le{}n$, be the vertex of $P$ with the $j$-th
largest $x$-coordinate --- ties are broken lexicographically \new{(also
refer to Fig. \ref{fig:monpconvex})}. Let $u_0$ (\resp $u_{n+1}$) be
the point of $P$ of minimal (\resp maximal) $x$-coordinate. Let
$\ell_j$, $0\le{}j\le{}n+1$, be the line passing through $u_j$,
perpendicular to $L$. The collection
$\mathcal{L}=\{\ell_0,\ell_1,\ldots,\ell_{n+1}\}$
of lines decompose the interior of $P$ into $n+1$ (possibly empty)
convex regions $\kappa_j$, $0\le{}j\le{}n$, that are free of vertices
or edges of $P$. Each region $\kappa_j$, $0\le{}j\le{}n$, has on its
boundary both $u_j$ and $u_{j+1}$. Let $e_j^\ell$ (\resp $e_j^r$),
$1\le{}j\le{}n$, be the edge of $P$ that has $u_j$ as its right (\resp
left) endpoint, \ie $e_j^\ell$ (\resp $e_j^r$) lies to left (\resp
right) of $u_j$. We define $e_0^r$ (\resp $e_{n+1}^\ell$) to be the edge
containing $u_0$ (\resp $u_{n+1}$). For a vertex $u_j$,
$1\le{}j\le{}n$, let $e_j^{opp}$ be edge of $P$ opposite to $u_j$, \ie
the edge intersected by $\ell_j$ on the monotone chain on $P$ not
containing $u_j$. Finally, for each $u_j$, $0\le{}j\le{}n+1$, define
its index $\sigma_j$ to be equal to $0$ if $u_j$ lies on both the
upper and monotone chain of $P$ (this is the case for $u_0$ and
$u_{n+1}$), $+1$ if $u_j$ lies on the upper but not the lower monotone
chain of $P$, and $-1$ if $u_j$ lies on the lower but not the upper
monotone chain of $P$.

We are going to compute an edge set $G$ for $P$ of size at most
$\mbd$ as will be described below. The idea behind computing $G$ is to
split $P$ into subpieces consisting of (at most) four convex regions
$\kappa_j$ and for each such four-tuple of convex pieces choose an
edge of $P$ that \guard[s] them. The procedure for computing $G$ is as
follows. For $j>n$, set $\kappa_j=\emptyset$, and initialize $G$ to be
empty. Let

\[  K_i=\bigcup_{j=4i-4}^{4i-1}\kappa_j,\quad
    1\le{}i\le{}\left\lceil\frac{n+1}{4}\right\rceil.\]
For each $K_i$, $1\le{}i<\mbd$, we are going to add one edge of $P$ to
$G$ according to the following procedure.
\begin{enumerate}
\item If $\sigma_{4i+1}\ne\sigma_{4i+2}$, add $e_{4i+1}^r$ to $G$.
\item Otherwise, if $\sigma_{4i+2}\ne\sigma_{4i+3}$, add $e_{4i+3}^\ell$ to $G$.
\item Otherwise, if $\sigma_{4i}\ne\sigma_{4i+1}$, add $e_{4i}^r$ to $G$.
\item Otherwise, if $\sigma_{4i+3}\ne\sigma_{4i+4}$, add $e_{4i+4}^\ell$ to $G$.
\item Otherwise, add $e_{4i+2}^{opp}$ to $G$.
\end{enumerate}
The procedure for adding an edge of $P$ for $K_{\mbd}$ is \new{analogous or
simpler, since we only need to account for four or less consecutive
convex regions.}

\begin{lemma}\label{lem:mon_edge_gset}
The edge set $G$ defined via the procedure above is an edge \gset
for $P$.
\end{lemma}

\begin{proof}
We are going to show that the set $K_i$, $1\le{}i<\mbd$ is \guard[ed]
by the corresponding edge added to $G$. The argument for $K_{\mbd}$ is
\new{analogous or} simpler and is omitted.

Given a point $p\in{}P$, let $\ell^\perp(p)$ be the line passing
through $p$ that is perpendicular to $L$.

Suppose that $\sigma_{4i+1}\ne\sigma_{4i+2}$. The edge $e_{4i+1}^r$
has as right endpoint a vertex $u_\lambda$ with
$\lambda\ge{}4i+3$. Clearly, $\kappa_{4i}$ and $\kappa_{4i+1}$ are
\guard[ed] by $u_{4i+1}\in{}e_{4i+1}^r$. If $\lambda=4i+3$, then
$\kappa_{4i+2}$ and $\kappa_{4i+3}$ are \guard[ed] by
$u_{4i+3}\in{}e_{4i+1}^r$. Otherwise, $\lambda\ge{}4i+4$, in which
case for every point $p\in\kappa_{4i+2}\cup\kappa_{4i+3}$ the line
$\ell^\perp(p)$ intersects $e_{4i+1}^r$.
The argument is symmetric if $\sigma_{4i+1}=\sigma_{4i+2}$, but
$\sigma_{4i+2}\ne\sigma_{4i+3}$.

Otherwise, consider the case $\sigma_{4i}\ne\sigma_{4i+1}$. The edge
$e_{4i}^r$ has a right endpoint a vertex $u_\lambda$ of $P$, with
$\lambda\ge{}4i+3$. If $\lambda=4i+3$, both $\kappa_{4i+2}$ and
$\kappa_{4i+3}$ are \guard[ed] by $u_{4i+3}$. $\kappa_{4i}$ is
\guard[ed] by $u_{4i}$, whereas for every point $p\in\kappa_{4i+1}$,
the line $\ell^\perp(p)$ intersects $e_{4i}^r$. If $\lambda>4i+3$, \ie
$\lambda\ge{}4i+4$, then for every point $p\in{}K_i$, the line
$\ell^\perp(p)$ intersects $e_{4i}^r$.
The argument is symmetric if
$\sigma_{4i}=\sigma_{4i+1}=\sigma_{4i+2}=\sigma_{4i+3}$, but
$\sigma_{4i+3}\ne\sigma_{4i+4}$.

Finally, consider the case
$\sigma_{4i}=\sigma_{4i+1}=\sigma_{4i+2}=\sigma_{4i+3}=\sigma_{4i+4}$. In
this case for every point $p\in{}K_i$, the line $\ell^\perp(p)$ intersects
$e_{4i+2}^{opp}$.
\end{proof}

Given Lemma \ref{lem:mon_edge_gset} we can now state and prove the
main result of this section.

\begin{theorem}\label{thm:monotonepiececonvex}
Given a monotone \pconvex polygon $P$ with $n\ge{}2$,
$\mbd$ edge or mobile guards are always sufficient and sometimes
necessary in order to \guard $P$. We can compute such an edge \gset
in $O(n)$ time and $O(n)$ space.
\end{theorem}

\begin{figure}[!t]
  \begin{center}
    \psfrag{u0}[][]{\scriptsize$u_0$}
    \psfrag{u1}[][]{\scriptsize$u_1$}
    \psfrag{u2}[][]{\scriptsize$u_2$}
    \psfrag{u3}[][]{\scriptsize$u_3$}
    \psfrag{u4}[][]{\scriptsize$u_4$}
    \psfrag{u5}[][]{\scriptsize$u_5$}
    \psfrag{u6}[][]{\scriptsize$u_6$}
    \psfrag{u7}[][]{\scriptsize$u_7$}
    \psfrag{u8}[][]{\scriptsize$u_8$}
    \psfrag{u9}[][]{\scriptsize$u_9$}
    \psfrag{u10}[][]{\scriptsize$u_{10}$}
    \psfrag{u11}[][]{\scriptsize$u_{11}$}
    \psfrag{u12}[][]{\scriptsize$u_{12}$}
    \psfrag{u13}[][]{\scriptsize$u_{13}$}
    \psfrag{u14}[][]{\scriptsize$u_{14}$}
    \psfrag{s0}[][]{\scriptsize$s_0$}
    \psfrag{s1}[][]{\scriptsize$s_1$}
    \psfrag{s2}[][]{\scriptsize$s_2$}
    \psfrag{s3}[][]{\scriptsize$s_3$}
    \psfrag{s4}[][]{\scriptsize$s_4$}
    \psfrag{s5}[][]{\scriptsize$s_5$}
    \psfrag{s6}[][]{\scriptsize$s_6$}
    \psfrag{s7}[][]{\scriptsize$s_7$}
    \psfrag{s8}[][]{\scriptsize$s_8$}
    \psfrag{s9}[][]{\scriptsize$s_9$}
    \psfrag{s10}[][]{\scriptsize$s_{10}$}
    \psfrag{s11}[][]{\scriptsize$s_{11}$}
    \psfrag{s12}[][]{\scriptsize$s_{12}$}
    \psfrag{s13}[][]{\scriptsize$s_{13}$}
    \includegraphics[width=\textwidth]{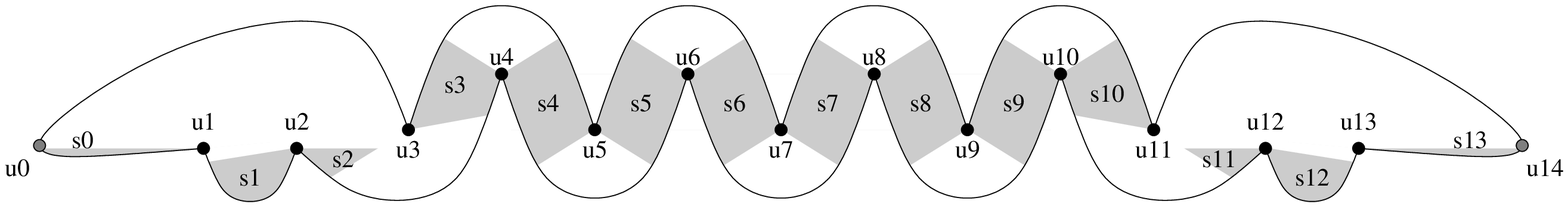}\\[20pt]
    \includegraphics[width=0.9\textwidth]{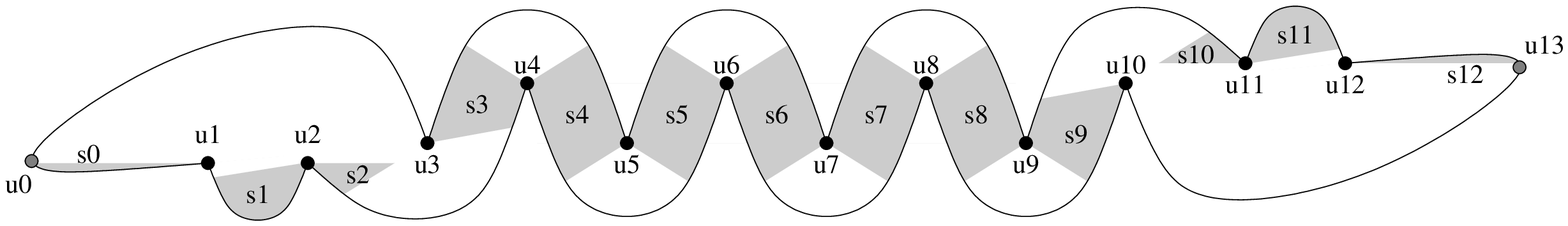}
  \end{center}
  \caption{The lower bound construction for monotone \pconvex
    polygons. The polygon $M_1$ (top) consists of $n_1=13$
    vertices, whereas $M_2$ (bottom) consists of $n_2=12$ vertices.
    Each region $s_j$, $1\le{}j\le{}n_i-1$, is only visible by $u_j$,
    $u_{j+1}$, some or all points on $e_j^r$ and $e_{j+1}^\ell$, or
    points on diagonals of $M_i$ that have either $u_j$ or $u_{j+1}$
    as one of their endpoints. The shaded region $s_0$ (\resp
    $s_{n_i}$) is only visible by $u_0$, all points on $e_0^r$ or the
    diagonals $d_{12}$, $d_{13}$, $d_{23}$ (\resp by $u_{n_i}$, all
    points on $e_{n_i+1}^\ell$ or the diagonals $d_{n_i-2,n_i}$,
    $d_{n_i-1,n_i}$, $d_{n_i-2,n_i-1}$).}
  \label{fig:monotone_lb}
\end{figure}

\begin{proof}
Lemma \ref{lem:mon_edge_gset} gives us the upper bound, since an edge
\gset is also a mobile \gset. The time and space
complexities are a result of the fact that determining whether a
\pconvex polygon is monotone can be determined in linear time
\cite{ds-cgcw-90}, and the fact that the procedure for computing an
edge \gset described above takes linear time and space.

Let us now concentrate on proving the lower bound. It suffices to
present the proof for the case of mobile guards. \new{Our claim is trivial
for $n\in\{2,3\}$.} Consider the monotone
\pconvex polygons $M_1$ (top) and $M_2$ (bottom) of 
Fig. \ref{fig:monotone_lb}. $M_1$ consists of $n_1=2m_1+5$, $m_1\ge{}0$,
vertices, whereas $M_2$ consists of $n_2=2m_2+4$, $m_2\ge{}0$, vertices
(in our example $m_1=m_2=4$). The rationale behind the construction of
$M_i$, $i=1,2$, lies in the properties of the shaded regions $s_j$,
$0\le{}j\le{}n_i$, shown in Fig. \ref{fig:monotone_lb}. Each region
$s_j$, $1\le{}j\le{}n_i-1$, is only visible by the two vertices $u_j$
and $u_{j+1}$ of $M_i$, some or all points on the edges $e_j^r$ and
$e_{j+1}^\ell$, as well as points on diagonals of $M_i$ that have
either $u_j$ or $u_{j+1}$ as one of their endpoints. Finally, 
the shaded region $s_0$ (\resp $s_{n_i}$) is only visible by $u_0$,
all points on $e_0^r$ or the diagonals $d_{12}$, $d_{13}$ and $d_{23}$
(\resp by $u_{n_i}$, all points on $e_{n_i+1}^\ell$ or the diagonals
$d_{n_i-2,n_i}$, $d_{n_i-1,n_i}$ and $d_{n_i-2,n_i-1}$).

Let $G_i$ be the mobile \gset for $M_i$, $i=1,2$. Suppose that
we can \guard $M_i$ with less than $\mbd$ mobile guards. This implies
that the number of vertices of $M_i$ in $G_i$ is less than 
$\lceil\frac{n+1}{2}\rceil$, which further implies that either: (1)
there exist two consecutive vertices of $M_i$ that do not belong to
$G_i$, or: (2) $u_1$ or $u_{n_i}$ is not incident to an edge or
diagonal of $M_i$ in $G_i$. In the former case, let $u_k$ and
$u_{k+1}$ be the two consecutive vertices of $M_i$ that are not
incident to edges that belong to $G_i$. This implies, in particular,
that neither $e_k^r$ nor $e_{k+1}^\ell$, nor any diagonal of $M_i$
incident to $u_k$ or $u_{k+1}$, belongs to $G_i$ and therefore the
shaded region $s_k$ is not \guard[ed] by the edges or diagonals in $G_i$,
a contradiction. In the latter case, $e_0$, $d_{12}$, $d_{13}$ or
$d_{23}$ (\resp $e_{n_i}$, $d_{n_i-2,n_i}$, $d_{n_i-1,n_i}$ or
$d_{n_i-2,n_i-1}$) cannot belong to $G_i$, which implies that $s_0$
(\resp $s_{n_i}$) is not \guard[ed] by any of the edges or diagonals in
$G_i$, again a contradiction. Hence our assumption that $M_1$ or $M_2$
can be \guard[ed] with less that $\mbd$ edge guards is false.
\end{proof}

\begin{remark}\sl
The results presented in this section for monotone \pconvex
polygons are also valid for monotone locally convex polygons, \ie
curvilinear polygons that are locally convex except possibly at their
vertices. The proof technique for producing the upper bound is
identical to the case of monotone \pconvex polygons. Since
monotone \pconvex polygons is a subclass of locally convex
polygons, the lower bound construction presented in Theorem
\ref{thm:monotonepiececonvex} still applies.
\end{remark}


\section{Discussion and open problems}
\label{sec:conclusion}

In this paper we have dealt with the problem of \guard[ing] \pconvex
polygons with edge or mobile guards. Our proof technique first
transforms the problem of \guard[ing] the \pconvex polygon to the
problem of 2-dominating a constrained triangulation graph.
For the problem of 2-dominance of triangulation graphs,
we have shown that $\bm$ diagonal guards are always sufficient and
sometimes necessary, while such a 2-dominating set can be computed in
$O(n)$ time and space. When edge guards are to be used in the context
of 2-dominance, $\be$ guards are always sufficient and
sometimes necessary. We have not yet found a way to compute an
edge 2-dominating set of size at most $\be$ in $o(n^2)$ time, whereas
we have shown that it is possible to compute an edge 2-dominating
set of size at most $\ubew$ in linear time and space. It, thus,
remains an open problem how to compute an edge 2-dominating set of
size at most $\be$ in $o(n^2)$ time and linear space. 

Once a 2-dominating set $D$ has been found for the constrained
triangulation graph, we either prove that $D$ is also a \gset for the
\pconvex polygon (this is the case for edge guards) or we map $D$ to a
mobile \gset for the \pconvex polygon. In the case of edge
guards, the \pconvex polygon is actually \guard[ed] by the
endpoints of the edges in the \gset. In the case of mobile
guards, interior points of edges may also be needed in order to
\guard the interior of the polygon. The latter observation should be
contrasted against the corresponding results for the class of linear
polygons, where, for both edge and mobile guards, the polygon is
essentially \guard[ed] by the endpoints of these guards
(cf. \cite{o-agta-87}). 
Based on our results on 2-dominance of triangulation graphs, we show
that a mobile \gset of size at most $\bm$ can be computed in
$O(n\log{}n)$ time and $O(n)$ space. As far as edge guards are
concerned, we can either compute an edge \gset of size at most
$\be$ in $O(n^2)$ time and $O(n)$ space, or an edge \gset of
size at most $\ubew$ in $O(n\log{}n)$ time and $O(n)$ space. Finally,
we have presented families of \pconvex polygons that require
a minimum of $\lbmg$ mobile or $\lbeg$ edge guards in order to be
\guard[ed].
An important remark, due to the lower bound \new{of Theorem
\ref{thm:guard-combedge-trgraph-lb}}, is that the proof technique of
this paper cannot possibly yield better results for the edge guarding
problem. If we are to close the gap between the upper and lower
bounds, a fundamentally different technique will have to be used.

When restricted to the subclass of monotone \pconvex polygons,
we were able to derive better bounds on the number of edge or mobile
guards that are sufficient in order to \guard these polygons. In
particular, we can \guard monotone \pconvex polygons with $\mbd$
edge or mobile guards, and this bound is tight for both types of
guards. \new{The same results apply to monotone locally convex
  polygons.}

Thus far we have limited our attention to the class of \pconvex
polygons. It would be interesting to attain similar results for
locally concave polygons (\ie curvilinear polygons that are
locally concave except possibly at the vertices), for \pconcave
polygons (\ie locally concave polygons the edges of which 
are convex arcs), or for curvilinear polygons with holes.


%% file: appendix.tex
\section{2-dominance with diagonal guards: alternative proof}
\label{sec:trg-diag-guard-alt}

The proof that follows is an alternative, much simpler proof for
Theorem \ref{thm:guard-combdiag-trgraph}. Its disadvantage is that it
makes use of edge contractions (cf. Lemma \ref{lem:contraction-gset}),
thus yielding an $O(n^2)$ time and $O(n)$ space algorithm instead of a
linear time and space algorithm, like the one provided in Section
\ref{sec:trg-diag-guards}.

\medskip

\begin{proof}
By Lemma \ref{lem:guard-combdiag-smallpoly} the theorem holds true for 
$3\le{}n\le{}7$.
Let us now assume that $n\ge{}8$ and that the theorem holds for all
$n'$ such that $3\le{}n'<n$. By means of Lemma \ref{lem:diag_existence}
with $\lambda=3$, there exists a diagonal $d$ that partitions
$\trg{P}$ into two triangulation graphs $T_1$ and $T_2$, where $T_1$
contains $k$ boundary edges of $\trg{P}$ with $3\le{}k\le{}4$. Let
$v_i$, $0\le{}i\le{}k$, be the $k+1$ vertices of $T_1$, as we
encounter them while traversing $P$ counterclockwise, and
let $v_0v_k$ be the common edge of $T_1$ and $T_2$. In what follows
$d_{ij}$ denotes the diagonal $v_iv_j$, whereas $e_i$ denotes the edge
$v_iv_{i+1}$. Consider each value of $k$ separately (see also
Fig. \ref{fig:guard-combdiag}):

\begin{mathdescription}
\item[k=3.]
  Without loss of generality let $d_{02}$ be the diagonal of the
  quadrilateral $T_1$. $T_2$ contains $n-2$ vertices. By Lemma
  \ref{lem:contraction-gset} and our induction hypothesis, we can
  2-dominate $T_2$ with $f(n-3)$ diagonal guards and $v_0$.
  $\trg{P}$ can be 2-dominated by the $f(n-3)$ diagonal guards of
  $T_2$ plus the diagonal $d_{02}$.
\item[k=4.]
  In this case $T_2$ contains $n-3$ vertices. Let $t$ be the
  triangle in $T_1$ supported by $d$, and let $v$ be the third vertex of
  $t$ besides $v_0$ and $v_4$. The presence of diagonals $d_{03}$ or
  $d_{14}$ would violate the minimality of $k$, which implies that $v$
  is actually $v_2$. By our induction hypothesis, we can 2-dominate
  $T_2$ with $f(n-3)=\bm-1$ diagonal guards. Let $D_2$ be
  the diagonal 2-dominating set of $T_2$. Notice that at least one of
  $v_0$ and $v_4$ has to be in $D_2$. Let us assume, without loss of
  generality, that $v_0$ is in $D_2$. Then the set
  $D=D_2\cup\{d_{24}\}$ is a diagonal 2-dominating set for $\trg{P}$
  of size $f(n-3)+1=\bm$.\qedhere
\end{mathdescription}

\begin{figure}[ht]
  \begin{center}
    \psfrag{0}[][]{\scriptsize$v_0$}
    \psfrag{1}[][]{\scriptsize$v_1$}
    \psfrag{2}[][]{\scriptsize$v_2$}
    \psfrag{3}[][]{\scriptsize$v_3$}
    \psfrag{4}[][]{\scriptsize$v_4$}
    \psfrag{t}[][]{\scriptsize$t$}
    \psfrag{d}[][]{\scriptsize$d$}
    \includegraphics[width=0.8\columnwidth]{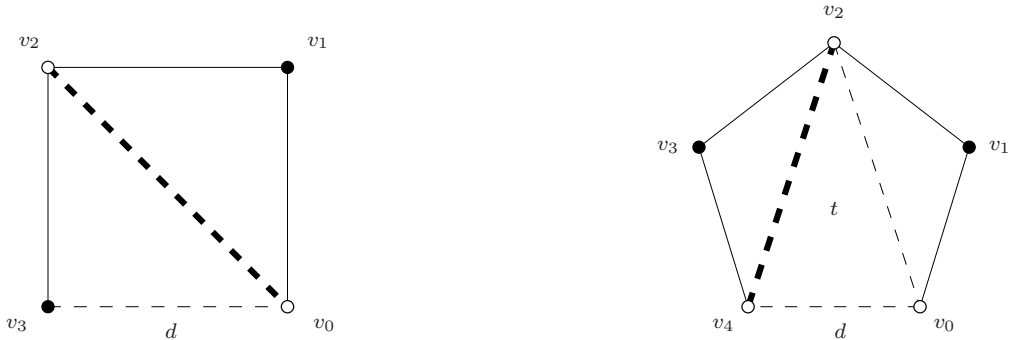}
  \end{center}
  \caption{Proof of Theorem \ref{thm:guard-combdiag-trgraph}. Left: 
    $k=3$. Right: $k=4$.}
  \label{fig:guard-combdiag}
\end{figure}
\end{proof}


%% file: mobile_guards.bbl
\begin{thebibliography}{10}

\bibitem{a-agpiv-84}
A.~Aggarwal.
\newblock {\em The art gallery problem: {Its} variations, applications, and
  algorithmic aspects}.
\newblock PhD thesis, Dept. of Comput. Sci., Johns Hopkins University,
  Baltimore, MD, 1984.

\bibitem{ae-caps-83}
D.~Avis and H.~ElGindy.
\newblock A combinatorial approach to polygon similarity.
\newblock {\em IEEE T. Inform. Theory}, IT-2:148--150, 1983.

\bibitem{b-egrc-98}
I.~Bjorling-Sachs.
\newblock Edge guards in rectilinear polygons.
\newblock {\em Comp. Geom.-Theor. Appl.}, 11(2):111--123, 1998.

\bibitem{BoiTeil:ECG:book}
J.-D. Boissonnat and M.~Teillaud, editors.
\newblock {\em Effective Computational Geometry for Curves and Surfaces}.
\newblock Mathematics and Visualization. Springer, 2007.

\bibitem{b-beddc-88}
W.~Bronsvoort.
\newblock Boundary evaluation and direct display of {CSG} models.
\newblock {\em Comput. Aided Design}, 20:416--419, 1988.

\bibitem{clu-vgac-09}
J.~Cano-Vila, J.~E. Longi, and J.~Urrutia.
\newblock Vigilancia en galer{\'i}as de arte curvil{\'i}neas.
\newblock In {\em Proc. XIII Encuentros de Geometria Computacional}, pages
  56--66, Zaragoza, Spain, June 29 -- July 1, 2009.

\bibitem{ci-tsc-84}
B.~Chazelle and J.~Incerpi.
\newblock Triangulation and shape-complexity.
\newblock {\em ACM T. Graphic.}, 3(2):135--152, 1984.

\bibitem{c-ctpg-75}
V.~Chv\'atal.
\newblock A combinatorial theorem in plane geometry.
\newblock {\em J. Comb. Theory B}, 18:39--41, 1975.

\bibitem{cgl-iscd-89}
C.~Coullard, B.~Gamble, W.~Lenhart, W.~Pulleyblank, and G.~Toussaint.
\newblock On illuminating a set of disks.
\newblock Manuscript, 1989.

\bibitem{cgru-ihcs-95}
J.~Czyzowicz, B.~Gaujal, E.~Rivera-Campo, J.~Urrutia, and J.~Zaks.
\newblock Illuminating higher-dimensionall convex sets.
\newblock {\em Geometriae Dedicata}, 56:115--120, 1995.

\bibitem{cruz-pcs-94}
J.~Czyzowicz, E.~Rivera-Campo, J.~Urrutia, and J.~Zaks.
\newblock Protecting convex sets.
\newblock {\em Graph. Combinator.}, 19:311--312, 1994.

\bibitem{ds-cgcw-90}
D.~P. Dobkin and D.~L. Souvaine.
\newblock Computational geometry in a curved world.
\newblock {\em Algorithmica}, 5:421--457, 1990.

\bibitem{ek-hstsr-89}
K.~Eo and C.~Kyung.
\newblock Hybrid shadow testing scheme for ray tracing.
\newblock {\em Comput. Aided Design}, 21:38--48, 1989.

\bibitem{egs-gpepp-07}
D.~Eppstein, M.~T. Goodrich, and N.~Sitchinava.
\newblock Guard placement for efficient point-in-polygon proofs.
\newblock In {\em Proc. 23rd Annu. ACM Sympos. Comput. Geom.}, pages 27--36,
  2007.

\bibitem{f-spcwt-78}
S.~Fisk.
\newblock A short proof of {Chv\'atal's} watchman theorem.
\newblock {\em J. Comb. Theory B}, 24:374, 1978.

\bibitem{ghks-ggprp-96}
E.~Gy{\"o}ri, F.~Hoffmann, K.~Kriegel, and T.~Shermer.
\newblock Generalized guarding and partitioning for rectilinear polygons.
\newblock {\em Comp. Geom.-Theor. Appl.}, 6:21--44, 1996.

\bibitem{ktt-gcagv-09}
M.~I. Karavelas, C.~D. T{\'o}th, and E.~P. Tsigaridas.
\newblock Guarding curvilinear art galleries with vertex or point guards.
\newblock {\em Comp. Geom.-Theor. Appl.}, 42(6--7):522--535, August 2009.

\bibitem{kt-gcagv-08}
M.~I. Karavelas and E.~P. Tsigaridas.
\newblock Guarding curvilinear art galleries with vertex or point guards, 2008.
\newblock {\tt arXiv:0802.2594v1 [cs.CG]}.

\bibitem{ks-errss}
R.~Kuc and M.~Siegel.
\newblock Efficient representation of reflecting structures for a sonar
  navigation model.
\newblock In {\em Proc. 1987 IEEE Internat. Conf. Robotics and Automation},
  pages 1916--1923, 1987.

\bibitem{ll-ccagp-86}
D.~Lee and A.~Lin.
\newblock Computational complexity of art gallery problems.
\newblock {\em IEEE T. Inform. Theory}, 32(2):276--282, 1986.

\bibitem{lw-apcfp-79}
T.~Lozano-P{\'e}rez and M.~A. Wesley.
\newblock An algorithm for planning collision-free paths among polyhedral
  obstacles.
\newblock {\em Commun. ACM}, 22(10):560--570, 1979.

\bibitem{m-wcohs-87}
M.~McKenna.
\newblock Worst-case optimal hidden-surface removal.
\newblock {\em ACM T. Graphic.}, 6:19--28, 1987.

\bibitem{m-aaspt-88}
J.~S.~B. Mitchell.
\newblock An algorithmic approach to some problems in terrain navigation.
\newblock {\em Artif. Intell.}, 37(1-3):171--201, 1988.

\bibitem{o-gnfmg-83}
J.~O'Rourke.
\newblock Galleries need fewer mobile guards: a variation on {Chv{\'a}tal}'s
  theorem.
\newblock {\em Geometriae Dedicata}, 14:273--283, 1983.

\bibitem{o-agta-87}
J.~O'Rourke.
\newblock {\em Art Gallery Theorems and Algorithms}.
\newblock The International Series of Monographs on Computer Science. Oxford
  University Press, New York, NY, 1987.

\bibitem{s-rrag-92}
T.~C. Shermer.
\newblock Recent results in art galleries.
\newblock {\em P. IEEE}, 80(9):1384--1399, Sept. 1992.

\bibitem{sc-bwfmr-86}
J.~Stenstrom and C.~Connolly.
\newblock Building wire frames for multiple range views.
\newblock In {\em Proc. 1986 IEEE Internat. Conf. Robotics and Automation},
  pages 615--650, 1986.

\bibitem{t-prgc-80}
G.~T. Toussaint.
\newblock Pattern recognition and geometrical complexity.
\newblock In {\em Proc. 5th IEEE Internat. Conf. Pattern Recogn.}, pages
  1324--1347, 1980.

\bibitem{uz-ics-89}
J.~Urrutia and J.~Zaks.
\newblock Illuminating convex sets.
\newblock Technical {Report} TR-89-31, Dept. Comput. Sci., Univ. Ottawa,
  Ottawa, ON, 1989.

\bibitem{xcb-pvicb-86}
S.~Xie, T.~Calvert, and B.~Bhattacharya.
\newblock Planning views for the incremental construction of body models.
\newblock In {\em Proc. 8th IEEE Internat. Conf. Pattern Recong.}, pages
  154--157, 1986.

\bibitem{y-3damv-86}
M.~Yachida.
\newblock 3-{D} data acquisition by multiple views.
\newblock In {\em Robotics Research: Third Internat. Sympos.}, pages 11--18,
  1986.

\end{thebibliography}
